\documentclass[a4paper,usenatbib]{mnras}
\pdfoutput=1
\usepackage{newtxtext,newtxmath}

\usepackage[T1]{fontenc}
\usepackage{ae,aecompl}

\usepackage{graphicx}	
\usepackage{amsmath}	
\usepackage{amssymb}	
\usepackage{verbatim}
\usepackage{epstopdf}
\usepackage[flushleft]{threeparttable}

\newcommand{\Jone}{SDSS\,J013644.02+044039.00}
\newcommand{\Jtwo}{SDSS\,J085859.67+174925.32}
\newcommand{\Jthree}{SDSS\,J090609.46+054818.72}
\newcommand{\Jfour}{SDSS\,J094649.47+121628.56}
\newcommand{\Jfive}{SDSS\,J114638.95+074311.28}
\newcommand{\Jsix}{SDSS\,J123602.11+001024.60}
\newcommand{\Jseven}{SDSS\,J234730.76$-$005131.68}
\newcommand{\Jones}{J0136+0440}
\newcommand{\Jtwos}{J0858+1749}
\newcommand{\Jthrees}{J0906+0548}
\newcommand{\Jfours}{J0946+1216}
\newcommand{\Jfives}{J1146+0743}
\newcommand{\Jsixs}{J1236+0010}
\newcommand{\Jsevens}{J2347$-$0051}

\newcommand{\HI}{\ion{H}{i}}
\newcommand{\CI}{\ion{C}{i}}
\newcommand{\zabs}{\ensuremath{z_{\rm abs}}}

\newcommand{\HH}{\mbox{H$_2$}}
\newcommand{\kms}{km\,s$^{-1}$}

\title[X-shooter observations of strong H$_2$-bearing DLAs]{X-shooter observations of strong H$_2$-bearing DLAs at high redshift\thanks{Based on observations carried out at the European Southern Observatory (ESO) Very Large Telescope Unit-2, Kueyen, under program ID 094.A-0362 (PI Balashev).}}

\author[S. Balashev et al.]{
S.~A.~Balashev$^{1}$\thanks{E-mail: s.balashev@gmail.com}
V.~V.~Klimenko$^{1}$, P.~Noterdaeme$^{2}$, J.-K.~Krogager$^{2}$, \newauthor
D.~A.~Varshalovich$^{1}$, A.~V.~Ivanchik$^{1}$, P.~Petitjean$^{2}$, R.~Srianand$^{3}$, 
C.~Ledoux$^{4}$
\\
$^{1}$ Ioffe Institute, Polytechnicheskaya ul. 26, Saint Petersburg, 194021, Russia\\
$^{2}$ Institut d'Astrophysique de Paris, UMR\,7095, CNRS-SU, 98bis bd Arago, 75014 Paris, France\\
$^{3}$ Inter-University Centre for Astronomy and Astrophysics, Post Bag 4, Ganeshkhind, 411 007, Pune, India \\
$^{4}$ European Southern Observatory, Alonso de C\'ordova 3107, Casilla 19001, Vitacura, Santiago, Chile 
}

\date{Accepted 2019 September 22. Received 2019 July 09}

\pubyear{2019}

\begin{document}
\label{firstpage}
\pagerange{\pageref{firstpage}--\pageref{lastpage}}
\maketitle

\begin{abstract}
We present results from spectroscopic observations with X-shooter at the Very Large Telescope of seven H$_2$-bearing DLAs at high redshifts ($\zabs\sim 2.5-3$). These DLAs were originally selected from the presence of strong H$_2$ lines directly seen at the DLA redshift in low-resolution, low S/N SDSS spectra. We confirm the detection of molecular hydrogen in all of them. We measure the column densities of H\,{\sc i}, H$_2$ in various rotational levels, and metal species, and associated dust extinction.
The metallicities, obtained from undepleted species, are in the range 
$\log Z = -0.8$ to $-0.2$. 
We discuss the chemical enrichment in these clouds and compare their properties with that of other molecular-rich systems selected by other means. In particular, we 
show that three different methods of pre-selection of H$_2$-bearing DLAs in the SDSS have their own biases but complement each other mostly in terms of chemical enrichment. 
We use the rotational excitation of H$_2$ molecules together with the fine-structure energy levels of neutral carbon to constrain the physical conditions in the gas with the help of numerical modelling as well as analytical expressions for the surface density at which atomic to molecular conversion happens. We find that the H$_2$-bearing medium revealed by the studied DLAs has typical values for the kinetic temperature, hydrogen density, and UV radiation field of, respectively, $T \sim100$\,K, $n_{\rm H} \sim100$\,cm$^{-3}$, and $I_{\rm UV}$ about twice the intensity of the Draine field.
Detailed studies combining different selections should therefore bring important clues to understand the H\,{\sc i}-H$_2$ transition at high redshift.

\end{abstract}




\section{Introduction}
\label{intro}

Molecular hydrogen (H$_2$) is by far the most abundant molecule in the Universe and predominantly resides in the cold neutral gas of galaxies. This cold phase may be subject to Jeans instability and is hence thought to be tightly linked to the onset of star formation. 
While the association between star formation and molecular gas 
is well established on kpc-scales for normal galaxies in the local Universe \citep[e.g.][]{Bigiel2008, Schruba11}, observations in emission in the distant Universe have much coarser spatial resolution and are mostly limited to the bright-end (and much probably high-metallicity end) of the galaxy population. In addition, the mass of molecular gas is estimated indirectly via tracers, like CO or dust emission, which are metallicity-dependent \citep[e.g.][]{Genzel12}.
While the link between star formation and molecular gas has been observed in a variety of environments, there are some observational indications that star formation could proceed directly from the atomic gas when the metallicity is low \citep[e.g.][]{Michalowski2015}. Indeed, in the diffuse ISM H$_2$ forms on the surface of dust grains, whose abundance scales with metallicity. Therefore, it is expected that at low metallicities, the H$_2$ formation timescale can be larger than the gravitational collapse time \citep[see e.g.][]{Krumholz2012}.

The star formation process can be studied not only in emission (where the relation between star formation and atomic gas is usually derived over the whole extent of the galaxy), but also through absorption-line studies. 
This is particularly useful for high-redshift galaxies, since the detection in absorption is independent of the luminosity of the associated object. 
Blind absorption-line studies indeed select systems owing to their cross-section only. 
The neutral gas in and around high-redshift galaxies in absorption studies is mostly detected via the \HI-Ly$\alpha$ absorption line, producing the so-called damped Lyman-$\alpha$ systems \citep[DLA,][]{Wolfe2005}, when the \ion{H}{i} column densities reaches $\log N(\ion{H}{i}) \gtrsim 20$, ensuring the gas is predominantly neutral. Statistical studies have shown that DLAs contain the bulk of the neutral atomic gas in the Universe \citep[e.g.][]{Noterdaeme2012c} and present a significant metallicity evolution with redshift \citep[e.g.][]{DeCia2018}, i.e., the smoking gun of star formation.
However, despite recent progress, it remains technically challenging to detect galaxy counterparts of DLAs in emission \citep[see][and references therein]{Krogager2017}, and therefore to understand the origin of DLAs and constrain the associated star-formation rates.
However,
absorption studies of cold and molecular gas already provide very useful information to understand the star-formation processes in high-redshift galaxies. Indeed molecular hydrogen is found in a fraction of high-$z$ DLAs \citep[$\sim 4\%$ for $\log N(\rm H_2) \gtrsim 18$,][]{Balashev2018} meaning that cold gas is intercepted by the line-of-sight. Such H$_2$-bearing DLAs potentially provide information on the transition from atomic to molecular gas and subsequently on the onset of star formation.

Locally, observations towards nearby stars have revealed that the H$_2$ column density significantly increases when the \ion{H}{i} column density reaches some threshold value.
This threshold directly traces the \ion{H}{i}-to-H$_2$ transition and is well-described by analytic and numerical modelling \citep[see][and references therein]{Sternberg2014}. The \ion{H}{i}-to-H$_2$ transition is also found to occur at higher \HI\ column density in the Magellanic clouds, owing to lower metallicities and stronger UV fields, in agreement with expectations from the models.
In turn, a global understanding of the \HI-to-H$_2$ transition at higher redshifts is still missing
\citep[see][]{Noterdaeme2015b}, mostly because of small statistical samples, which actually probe a wide range of physical conditions.
To date, there are $\sim 40$ confirmed intervening H$_2$-bearing DLAs detected at redshifts $z>2$.

Before the advent of the Sloan Digital Sky Survey \citep[SDSS;][]{York2000}, H$_2$ at high redshifts has been blindly searched, mostly in high-resolution DLA spectra from archival data. The detection rate is found to be quite low: \citealt{Ledoux2003, Noterdaeme2008} found H$_2$ in about 10\% of the DLAs surveyed using high-resolution spectra, while \citealt{Jorgenson2014} found only one H$_2$ in 86 blindly-selected DLAs, using mid-resolution data. \citet{Petitjean2006} proposed that 
above a metallicity $Z>0.2Z_{\rm \odot}$, the probability to detect H$_2$ increases 
to about 50\%. However, this is of little practical use since the metallicity is known accurately only once a high-resolution spectrum is obtained.
With a huge increase in the number of quasar spectra available, the SDSS spectroscopic database has presented a unique opportunity to search for H$_2$ systems. Different techniques have been proposed to pre-select these from low-resolution, low-S/N spectra. 

A first efficient pre-selection was based on the presence of neutral carbon lines \citep[e.g.,][]{Srianand2008, Noterdaeme2010}. Indeed, \CI\ is a fragile species that survives more easily in the same conditions as H$_2$ requires \citep[see e.g.][]{Liszt1981,Ge2001}. Since the strongest \ion{C}{i} absorption lines are usually redshifted outside of the Lyman-$\alpha$ forest, it was possible to systematically search for \CI\ in SDSS spectra without any prior on \HI\ content \citep{Ledoux2015}. 
While the sample of \CI-selected molecular systems is continuously increasing thanks to the SDSS \citep[e.g.][]{Ma2015, Noterdaeme2017}, the majority of the systems have $z\lesssim2$, so that H$_2$ (and sometimes \HI) lines are inaccessible from the ground, due to the atmospheric cutoff at $\sim3000~${\AA}. For those systems where H$_2$ is covered, \citet{Noterdaeme2018} find a 100\% detection rate of H$_2$ molecules.
However, it has been found that a selection based on \CI\ only identifies metal-rich systems (with $Z\sim Z_{\odot}$, see \citealt{Zou2018} and Ledoux et al., in prep.).

Another method to pre-select possible H$_2$-bearing systems is to consider extremely-strong DLAs (ESDLAs, with $\log N(\HI) > 21.7$, \citealt{Noterdaeme2014}), since a large critical column density for atomic-to-molecular conversion is expected for clouds at the typical metallicity of DLAs.
Such a conversion could furthermore explain the steeping slope of the observed $N(\HI)$-distribution function at high $N(\HI)$ \citep{Zwaan2006,Altay2013}. From a galactic-scale point-of-view, we also expect 
the corresponding lines-of-sight to probe gas at small galacto-centric distances, as confirmed by the detection of Lyman-$\alpha$ emission in composite ESDLA spectra \citep{Noterdaeme2014} and in individual systems \citep{Noterdaeme2012a, Ranjan2018}.
This picture is corroborated by the possible increase of thermal pressure with increasing \HI\ column density \citep{Balashev2017}, indicating that such lines-of-sight probe more central regions of galaxies.
An enhanced H$_2$ detection rate in ESDLAs has indeed been confirmed through follow-up of ESDLAs \citep[][]{Noterdaeme2015b,Ranjan2019} as well as in stacked SDSS spectra \citep{Balashev2018}. 

Finally, \citet{Balashev2014} proposed to directly identify H$_2$ Lyman-Werner bands in SDSS spectra. However, this is a complex task because of the low spectral resolution, the typically low signal-to-noise ratio, and the presence of Ly$\alpha$ forest absorption lines.
Due to these limitations, only a few percent of SDSS spectra are suitable for the detection of H$_2$, and only systems with high H$_2$ column densities ($N(\HH) \gtrsim 10^{19}$~cm$^{-2}$) can be detected.
In our original search of SDSS DR9, we have found about 25 good H$_2$ candidates (with low false detection probability (FDP), see \citealt{Balashev2014}); A number that has since then been increased to about 100 using SDSS DR14.

In order to study the H$_2$ absorbers selected directly from SDSS in greater detail, we have initiated a follow-up program using the Very Large Telescope (VLT). The higher spectral resolution and signal-to-noise ratio allow us to confirm the presence of H$_2$ in the candidate systems and to measure the gas properties.
One of the H$_2$-bearing DLAs at $z=2.786$ towards J\,0843+0221 has already been published owing to its unique and extremely high H$_2$ column density of $\log N(\rm H_2)\sim 21.2$ \citep{Balashev2017}.
In the present paper, we describe the full analysis of the remaining 7 systems from this follow-up sample.

The paper is structured as follows. First, we describe the observations and the methodology that we have used to analyze the H$_2$-bearing DLAs in Sect.~\ref{sect:observations} and \ref{sect:analysis}, respectively. In Sect.~\ref{sect:discussion}, we discuss the properties of the sample, which we compare to other H$_2$ samples arising from different selection methods.
Lastly, in Sect.~\ref{sect:conclusion}, we offer a summary of the findings presented in this paper.

\section{Observations and data reduction}
\label{sect:observations}
Since the main goal of our pilot observations was to confirm the reality of the H$_2$ absorption systems detected by \citet{Balashev2014}, we first selected the targets with lowest false H$_2$-detection probability from our list of DLA systems (extended to DR10) and that were accessible from the Paranal observatory during the semester of observations.
The sample was observed with the X-shooter spectrograph \citep{Vernet2011} mounted at the Unit Telescope 2 of the VLT (VLT-UT2, Kueyen). The observations were performed in service mode under program ID 094.A-0362 (PI: Balashev) between the end of 2014 and beginning of 2015.  The log of observations is given in Table~\ref{table:jour_obs}. For all observations, we used a slit width of 1 arcsec in the UVB arm, and 0.9 arcsec in the VIS and NIR arms, leading to a resolving power of $R\sim5400$, $\sim$8900, and $\sim$5600\footnote{The actual value of the spectral resolution depends on the atmospheric seeing value, which in some cases was smaller than the slit width, see Table~\ref{table:jour_obs}.} in the UVB, VIS, and NIR arms, respectively. For each target, we obtained two 1-hour exposures except for \Jseven, which was observed only once before the end of the observations.

\begin{table}
	\centering
	\addtolength{\tabcolsep}{-2pt}
	\caption{Journal of X-shooter observations. 
	}
	\label{table:jour_obs}
	\begin{tabular}{lcccc}
		\hline
		Quasar & Date & Seeing & Airmass & Exp. times$\dagger$ \\
		       &      & ($\arcsec$) &    & (ks)                \\
		\hline
		J\,013644.02+044039.00 & 27.11.2014 & 1.15 & 1.16 & 3.6, 3.6, 3$\times$1.2\\
        J\,013644.02+044039.00 & 28.11.2014 & 1.15 & 1.26 & 3.6, 3.6, 3$\times$1.2\\
        J\,085859.67+174925.32 & 24.12.2014 & 0.73 & 1.36 & 3.6, 3.6, 3$\times$1.2\\
        J\,085859.67+174925.32 & 15.02.2015 & 1.38 & 1.37 & 3.6, 3.6, 3$\times$1.2\\
        J\,090609.46+054818.72 & 23.12.2014 & 0.79 & 1.26 & 3.6, 3.6, 3$\times$1.2\\
        J\,090609.46+054818.72 & 25.12.2014 & 0.67 & 1.22 & 3.6, 3.6, 3$\times$1.2\\
        J\,094649.47+121628.56 & 18.02.2015 & 0.83 & 1.27 & 3.5, 3.5, 3$\times$0.9\\
        J\,094649.47+121628.56 & 18.02.2015 & 0.99 & 1.29 & 3.5, 3.5, 3$\times$0.9\\
        J\,114638.95+074311.28 & 18.02.2015 & 0.79 & 1.19 & 3.6, 3.6, 3$\times$1.2\\
        J\,114638.95+074311.28 & 24.02.2015 & 0.77 & 1.25 & 3.6, 3.6, 3$\times$1.2\\
        J\,123602.11+001024.60 & 14.02.2015 & 0.87 & 1.13 & 3.6, 3.6, 3$\times$1.2\\
        J\,123602.11+001024.60 & 18.02.2015 & 1.15 & 1.12 & 3.6, 3.6, 3$\times$1.2\\
        J\,234730.76$-$005131.68 & 20.12.2014 & 0.52 & 1.39 & 3.6, 3.6, 3$\times$1.2\\
		\hline
	\end{tabular}
		\addtolength{\tabcolsep}{+2pt}
	\begin{tablenotes}
	\item $\dagger$ Exposure times given for the UVB, VIS, and NIR arms, in this order.
	\end{tablenotes}
\end{table}

We first reduced the spectra using the ESO Reflex pipeline \citep{Freudling2013}.
However, the unavailability of atmospheric dispersion correction at the time of observations led to a significant wavelength-dependent trace offset in the UVB, which was not properly taken into account by the pipeline.
Therefore, we used our own code to extract the 1-dimensional (1D) spectra along the trace starting from the 2D pipeline product (which incorporates dark/bias correction, flat-field correction, 2D geometry and arc-line solutions).
In addition, we improved the removal of cosmic-ray hits by flagging the corrupted pixels on the 2D exposures interactively. The 2D sky model was constructed using polynomial fitting along the spatial axis for each dispersion column and taking into account the rejection of outliers and correction of 2D geometrical issues at the edge of bright sky lines.
We used the optimal-extraction method with a Moffat profile to extract the 1D spectrum \citep{Horne1986}. Finally, we corrected the flux error in each pixel of each spectra by a wavelength-dependent factor so that it matches the observed dispersion relative to the quasar continuum. To do this, we identified unabsorbed regions in the continuum and estimated a smooth average of the dispersion of pixels using 5-10~\AA-wide sliding windows. This extraction resulted in an overall $\sim$25\% improvement in the S/N ratio and a better handling of lower quality regions, which were otherwise hard to manage through the pipeline.

\section{Analysis}\label{sect:analysis}

Before presenting the analysis of the different systems, we note that two out of seven DLAs studied in this paper have redshifts close to that of the quasar ($\Delta v \lesssim 3000$\,km/s) and could hence be considered as proximate. This is quite surprising at the first glance since the parent sample used by \citet{Balashev2014}, i.e., the DLA catalogue from 
\citet{Noterdaeme2012c}, was based on a search optimised for intervening DLAs and hence likely biased {\sl against} 
proximate DLAs. 
Recently, \citet{Noterdaeme2019} found that the incidence rate of proximate H$_2$-bearing DLAs (within $\sim 2000$~\kms from the quasar) per redshift path unit can be a factor of 5 higher than that of intervening systems. However, the observed number of systems depends on the total redshift path probed by each sample and therefore intervening \HH\ systems should still dominate in numbers over proximate ones, due to the much longer total pathlength. The relatively high number of proximate DLAs here is more likely to originate from selection method used for this follow-up sample. Indeed, the objects were selected upon their probability of false \HH-detection. This probability is strongly correlated with the number of the band used and therefore, the closer DLA to the quasar, the more H$_2$ bands are available in the spectrum before blue cutoff, and the smaller can be the false detection probability. In this work, as we concentrate on the chemical and physical properties of the systems and found no striking difference between those within a few thousands \kms\ from the quasar and those located further in velocity space, therefore we consider them indistinctly.

\subsection{Absorption lines}
\label{sect:alf}
We analysed all DLA systems in our sample in the same manner. We used a custom routine\footnote{We used this routine in recent analyses of absorption-line systems \citep[e.g.,][]{Balashev2016, Balashev2017} and also, partly, in \citet{Noterdaeme2018, Krogager2018b}, where we found that it usually is in good agreement with other codes used for Voigt-profile fitting, like vpfit \citep{Carswell2014} and VoigtFit \citep{Krogager2018a}.} for Voigt-profile fitting to derive the redshifts, column densities and Doppler parameters of the velocity components for \ion{H}{I}, H$_2$, \ion{C}{I}, and  various low-ionization metal absorption lines (e.g. \ion{Si}{ii}, \ion{Fe}{ii}, \ion{S}{ii}, \ion{Zn}{ii}). 
For each absorption line (except \ion{H}{I} Ly$\alpha$, described below) we derived the local continuum using a B-spline interpolation matched on the adjacent unabsorbed spectral regions. 
The spectral pixels used to constrain the fit were selected visually and iteratively. For each quasar we fit available exposures jointly without combining them.
The number of velocity components was also defined visually and increased in case of remaining structure in the residuals.
We used Monte-Carlo Markov Chain calculations with an affine invariant sampler \citep{Goodman2010} to obtain the posterior probability density function (PDF) of fitting parameters (assuming flat uninformative priors). While we directly obtained the PDF, in the following we use the standard way of reporting the best fitting parameters and uncertainties. The best fitting value corresponds to the maximum posterior probability and the estimated uncertainties correspond to the 68.3\% confidence interval around this value (corresponding to formal 1$\sigma$ uncertainty for Gaussian PDF). 

We measured the \HI\ column density from the Ly$\alpha$ line. However, since this line is strongly damped and 
frequently located close to the Ly$\alpha$ emission line, we simultaneously fitted the absorption profile and the unabsorbed continuum modelled usually by 4-8 order Chebyshev polynomials, depending on the wavelength range.
The values all agree within 0.2~dex from those measured at lower S/N and resolution power from the SDSS data.

\begin{figure}
    \centering
    \includegraphics[width=\hsize]{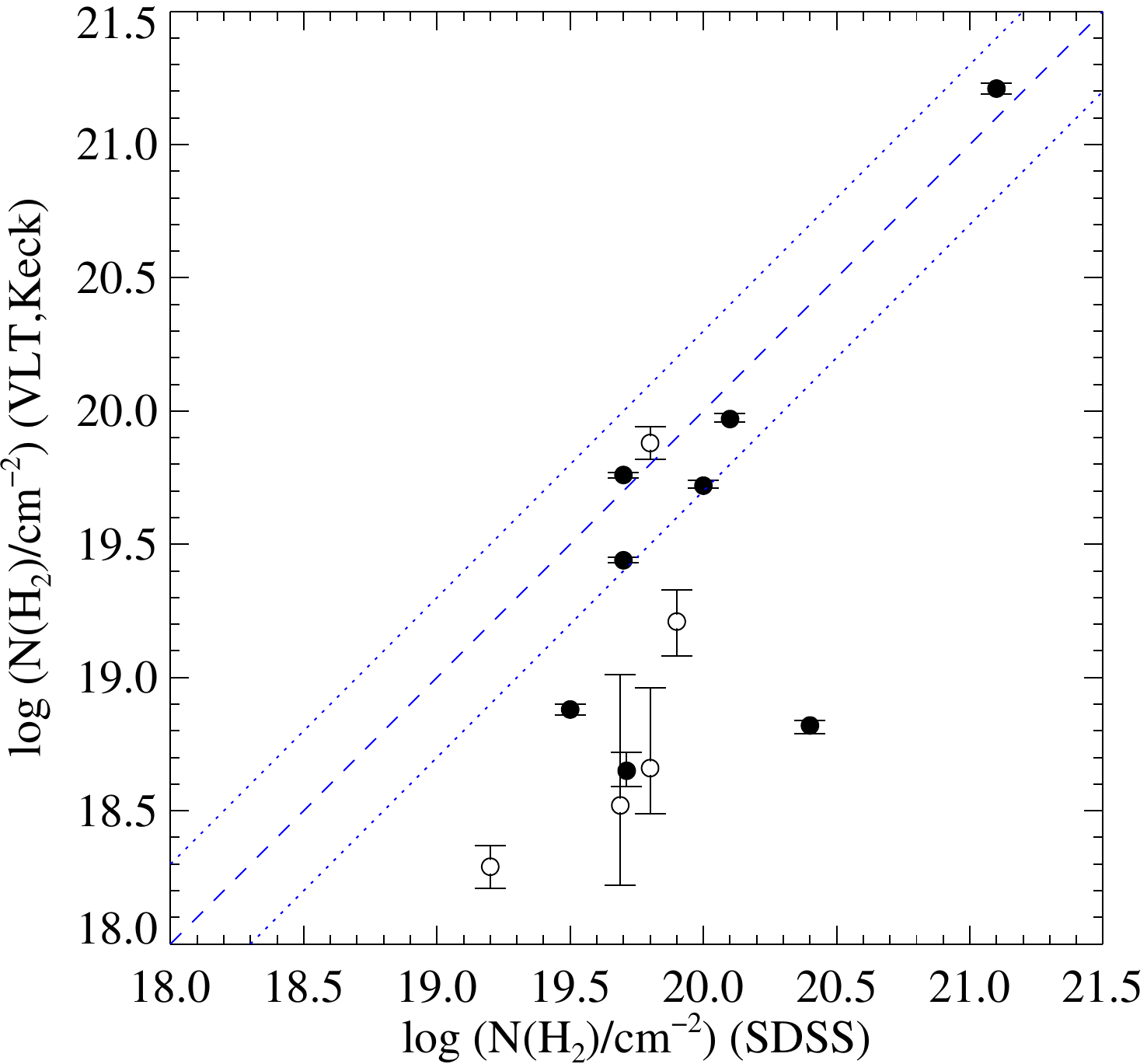}
    \caption{Comparison of \HH\ column densities automatically determined from the low-resolution low S/N SDSS data with those obtained from medium/high-resolution  observations. The sample studied in this paper is shown as filled symbols (all using X-shooter but J\,0843+0221 observed with UVES), while open symbols represent other known \HH\ systems (all observed with UVES but towards J81240.69+320808.52, observed with HIRES on Keck, see \citealt{Jorgenson2009}).}
    \label{compNH2}
\end{figure}

We confirm the presence of H$_2$ in all systems through the detection of Lyman- and Werner-band absorption lines. {We only detect lines from several low rotational levels of the $\nu=0$ vibrational ground state. Therefore, in the following, we use an abbreviated notation for those levels, as follows: $\rm H_2Ji$, where index $i$ is the rotational quantum number.} While an increase of the Doppler parameter with increasing rotational level has been observed for several H$_2$ systems 
\citep{Lacour2005, Noterdaeme2007, Balashev2009, Ivanchik2010}, the spectral resolution of X-shooter is not high enough to study this effect. Therefore, we tied the Doppler parameters of different rotational levels when fitting H$_2$ lines. 
This implies that the reported column densities for high ($J>2$) rotational levels should be considered with caution, in particular because the corresponding lines are generally in the intermediate regime. 
Notwithstanding, this has a little effect on the measurement of the total $N($H$_2)$ which is dominated by the first rotational levels whose lines are damped and hence not sensitive to the exact Doppler parameter. Additionally, for each exposure we included two additional independent parameters when fitting H$_2$ lines, to linearly correct for the wavelength distortions caused by the unavailability of the ADC \citep[see e.g.][]{Noterdaeme2017}. We then used the obtained values of the zero point and the slope of distortion to correct the spectrum before fitting the \ion{C}{i} and low-ionization metal lines. Finally, since the seeing was narrower than the slit width for some exposures, we also varied spectral resolution for each exposure as independent fitting parameter. 
We found lower H$_2$ column densities than initially estimated from the SDSS in all cases, see Fig.~\ref{compNH2}. The reasons for this is that H$_2$ profiles are unresolved in SDSS, the blends with Ly$\alpha$ forest lines are hard to take into account and the signal to noise ratio are low. For these reasons, the SDSS-based \HH\ column densities 
were considered as indicative only by \citet{Balashev2014}. 
Notwithstanding, the column densities agree within 0.3~dex for 4 out of 7 DLAs (5/8 including J0843+0221). From the X-shooter data, we could see that these are indeed the cases where the H$_2$ profiles dominated by one main component, while for the other three DLAs with larger differences, there are at least two \HH\ components with comparable column densities. 

{}

To fit the low-ionization metal lines, we tied the redshifts and Doppler parameters of various species since the broadening is dominated by turbulence which is independent of the atomic mass. We also varied spectral resolution for each exposure as independent fitting parameter. 
%
For all DLAs we detected associated \ion{C}{I} absorption lines at the position of H$_2$ components. 
Therefore, we fitted the \ion{C}{I} lines (and the associated fine-structure levels) independently from other metals.

The detailed fitting results, the corresponding figures and specific comments for each DLA in our sample are presented in 
Supplementary materials.
The total column densities of \ion{H}{i},  H$_2$, \ion{C}{i}, the average molecular fraction, metallicity, and Fe depletion are summarised in Table~\ref{table:fit_summary}.
We also add the H$_2$-bearing DLAs towards J\,0843$+$0221 to our sample, since it was selected from SDSS using the same technique. However it was obtained with separate high resolution UVES observations, which confirmed the exceptionally high H$_2$ column density \citep{Balashev2017}.

\subsection{Dust-extinction measurements}

We obtained the estimates of $E(B-V)$ by fitting a quasar spectral template to the data assuming a fixed extinction curve. Since we observe no indications of the prominent 2175~\AA\ dust feature, we used the extinction curve observed towards the Small Magellanic Cloud (SMC) as parametrized by \citet{Gordon2003}, however, we modified the far UV curvature part using $c_4 = 0$ since this provided a better description of the data. The dust reddening is assumed to be caused solely by the H$_2$-bearing DLA. We used the quasar template derived by \citet{Selsing2016} and fitted only the data in bona fide continuum regions, i.e., regions which are not strongly affected by absorption or broad emission lines. In the near-infrared, we performed a 5-$\sigma$ clipping in order to discard outlying pixels introduced by the removal of skylines in the data reduction. Moreover, we masked out pixels that are affected by telluric absorption.

Before fitting, the template by Selsing et al. was smoothed with a Gaussian kernel ($\sigma$ = 7~pixels) to prevent the noise in the template to falsely fit noise peaks in the real data. In order to take into account the uncertainty on the template, we subsequently convolved the errors on the observed data with the uncertainty estimate for the template.

We then fitted the template to the data using 3 free parameters: the amount of dust in the absorber, $A_V$, and the change in the intrinsic power-law index, $\Delta\beta$, as well as a flux scaling, $f_0$, since we do not know the intrinsic brightness prior to reddening. The best-fit $A_V$ are given in Table~\ref{table:fit_summary}. We found that power law indices, $\Delta\beta$, to be between $-0.8$ and $0.2$ for this sample. The typical statistical uncertainties on $A_V$ of such measurements are $\sim 0.01$\,mag. However, there is a strong degeneracy between $A_V$ and the intrinsic power-law shape of the quasar spectrum. 
We estimated the systematic uncertainty related to the unknown spectral shape by assuming the inferred Gaussian dispersion for $\beta$ of $\sigma_{\beta} = 0.19$ \citep{Krawczyk2015}. We drew a random $\Delta\beta$ from a Gaussian centred at $\Delta\beta = 0$ with dispersion $\sigma_{\beta}$ and change the quasar template accordingly. We then refitted the spectra with the newly altered, random template to obtain a distribution of $A_V$.  Based on this distribution, we inferred an estimate of the systematic uncertainty of $\sigma_{\rm sys} \sim 0.1$\,mag. We therefore consider this systematic uncertainty as a conservative estimate of the error on individual measurements.

\subsection{Physical conditions} 
\label{sect:phys_cond}
In this section, we derive the temperature and number density of the H$_2$-bearing gas in our sample as well as the intensity of the UV irradiation onto the clouds. In cases where more than one component is detected in H$_2$, we use only the strongest component since \ion{C}{i} is not always detected for the weaker component. We note however that for one absorber (toward J1146+0743) we are able to properly constrain \ion{C}{i} and H$_2$ in both components, and for both components, we infer consistent results for temperature, density and UV intensity. Also for J\,0843+0221 we used the total column densities of H$_2$ and \CI, since absorption lines from $J=0,1$ H$_2$ rotational levels damped and two main components are not resolved.  

We used the $J=0$ to $J=1$ rotational excitation temperature of H$_2$, $T_{01}$, as a proxy for the kinetic temperature $T_k$ in the cold diffuse medium \citep[see][]{Roy2006}. The inferred temperatures for the individual systems are given in Table~\ref{table:fit_summary}. We note that these estimates of the temperature only corresponds to the temperature of the H$_2$-bearing gas and not the overall DLA gas, for which there most likely is a contribution from a warm neutral phase with $T \gtrsim 10^3$\,K \citep[e.g.,][]{Srianand2012, Neeleman2015}.

In order to constrain the number density and the UV flux in the H$_2$-bearing medium, we used the relative population of \ion{C}{I} fine structure levels \citep[see][]{Silva2002}, H$_2$ rotational levels and the analytical theory of \HI-H$_2$ transition from \citet{Sternberg2014}. For \ion{C}{I}, we used the same calculations and code as in \citep{Balashev2017, Ranjan2018}. Here we briefly outline the assumptions used in the analysis. For \ion{C}{i} excitation, we assumed a homogeneous medium, where \ion{C}{i} fine-structure levels are populated by the Cosmic Microwave Background (CMB) radiation, UV pumping and collisions. The CMB temperature at the absorbers redshift is fixed to be $T_{\rm CMB}(z) = T_{\rm CMB}(z=0)(1+z)$ \citep[see, e.g.,][]{Srianand2008,Noterdaeme2011}.
Since the observed H$_2$ column densities are high, we assume that the gas phase containing the bulk of \ion{C}{i} is fully molecularized \citep[e.g.,][]{Balashev2015}, i.e. the main collisional partners are H$_2$ and He. 
Due to the relatively low measured \ion{C}{i} column densities we neglect self-shielding in \ion{C}{i} lines in which UV pumping occurs. 
{Since the collisional coefficients for \ion{C}{i} depend on temperature, we fixed the temperature to the measured $T_{01}$ value, reducing the free parameters to be the hydrogen number density ($n_{\rm H}$) and the intensity of the UV field ($I_{\rm UV}$, that we express relative to the Draine field). The exact shape of the spectrum cannot be constrained, since it depends on the unknown galaxy's stellar population and ambient conditions. Therefore, we assumed Draine shape for the UV background spectrum. We also added the UV metagalactic background in the calculation. However, at redshifts $z\sim 2-3$, it is $\sim$10$\div$30 times lower than the Draine field in the considered wavelength range \citep[see e.g.][]{Khaire2019}.}

{The population of the H$_2$ rotational levels also provides constraints on the ($I_{\rm UV}$, $n_{\rm H}$)-plane, and were previously used to estimate the physical conditions in a few high-$z$ H$_2$-bearing DLAs \citep[e.g.,][]{Klimenko2016, Shaw2016, Noterdaeme2017, Rawlins2018}.}
While UV pumping plays a direct role in the excitation of high rotational level, we are not confident that $J\ge3$ rotational levels are well constrained (see note in Sect.~\ref{sect:alf}). We hence here focus on the excitation of the $J=0,1,2$ levels only. These levels are predominantly thermalised and their excitation typically close to the kinetic temperature \citep[see][]{LePetit2006}, which is set 
by the thermal balance, itself being a function of density and UV field. 
Notwithstanding, UV pumping could also contribute to the excitation of these low-J levels, 
so that the assumption of a homogeneous cloud is not correct anymore (the resonant H$_2$ lines --at which pumping occur -- quickly saturate inside the cloud). We therefore used the 
PDR Meudon code \citep{LePetit2006} which performs full radiative transfer calculations in H$_2$ lines. 
We used grids of pseudo-spherical constant-density models with the only parameters being the metallicity (assuming dust scales to metallicity), density and intensity of UV field. {The temperature profile in the cloud was calculated from the solution of thermal balance for each model.}
We then ran grids of models with ten logarithmically-spaced points for both the number density and UV field intensity in the respective ranges of $0 \leq \log \rm n_{H} / cm^{-3} \leq 4$ and $-1.5 \leq \log I_{\rm UV} \leq 2.5$ (where $I_{\rm UV}$ is measured in Draine field units). For each system, we chose the appropriate grid corresponding to the observed metallicity. 
We used models assuming a slab of gas with thickness corresponding to $A_V=0.5$ irradiated from both sides. To compare the modeled and observed column densities of different H$_2$ rotational levels, we cut the model at a depth in the cloud where half of the total H$_2$ column density equals half the observed one. This approach mimics a cloud illuminated on both sides with a total column density of H$_2$ equal to the measured one.

We then compared the modelled excitation of H$_2$ rotational levels to the observed ones, including 
a factor of 2 (0.3 dex) uncertainty for the observed H$_2$ column densities to take account for unknown geometry (the H$_2$ cloud is likely not exactly spherical). In order to obtain a smooth representation of the results, we interpolated the calculated H$_2$ excitation on a denser grid in the $n_{\rm H}$--$I_{\rm UV}$ parameter space.

\begin{figure}
\centering
\includegraphics[trim={15 0 0 0},clip,width=\hsize]{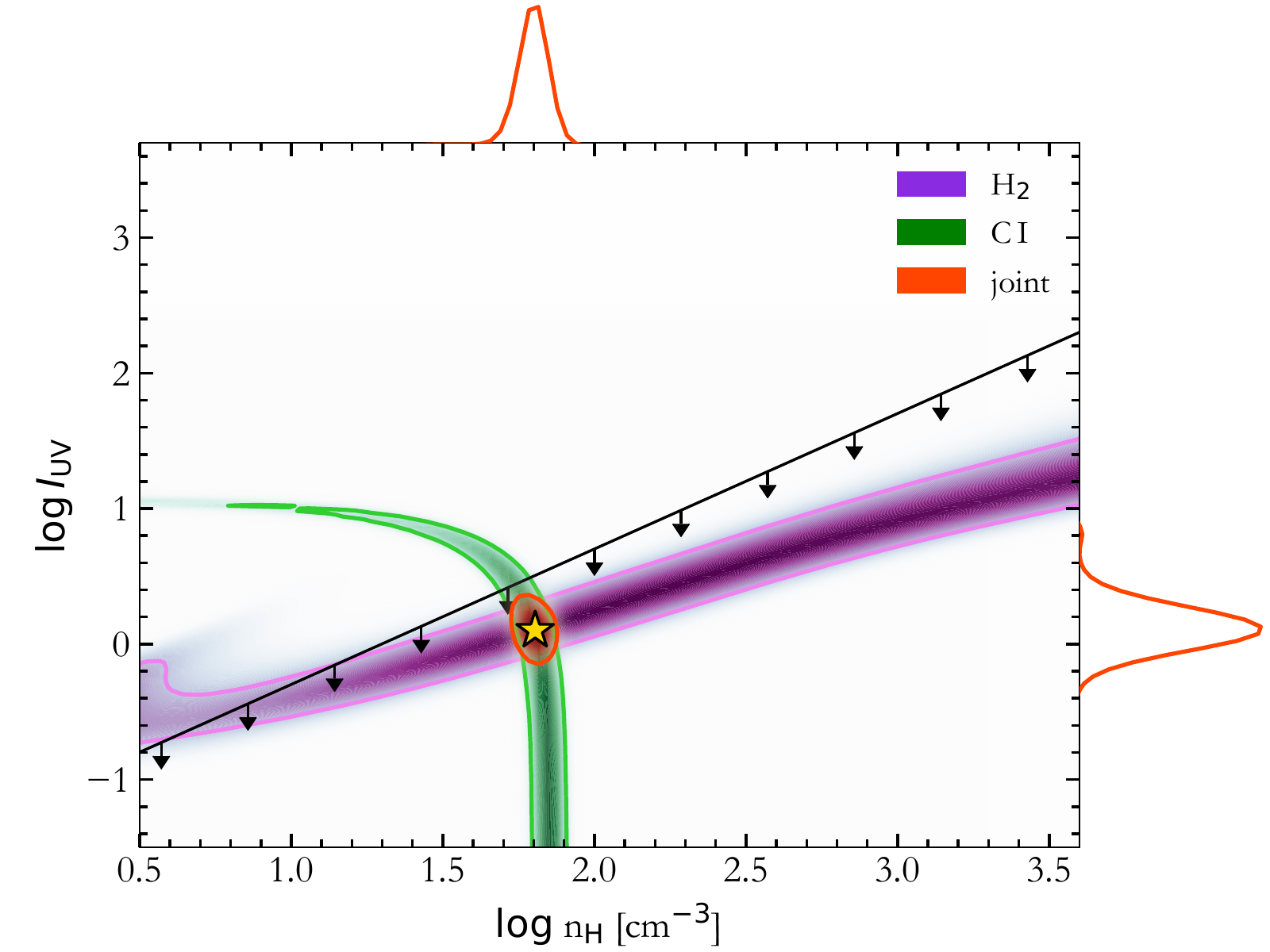}
		\caption{
		Constraints on the hydrogen number density $n_{\rm H}$ and UV field strength ($I_{\rm UV}$, in Draine field units) in the H$_2$-bearing medium toward \Jtwo. 
		The green and violet color gradients represent the obtained probability density function (with the contour lines encompassing 68.3\% of it) using excitation of \CI\ fine-structure levels and H$_2$ rotational levels, respectively. The red contour shows the joint constraint with the star located at the most probable value. The red lines on the top and right axis indicate the marginalized distributions of respectively, 
		$n_{\rm H}$ and $I_{\rm UV}$ from the joint fit. The black line with downward arrows show the constraint obtained from the theoretical 
		\HI-H$_2$ transition in the medium.
		}
		\label{fig:J0858_phys}
\end{figure}

In Fig.~\ref{fig:J0858_phys}, we show an example of the constrained regions from \CI\ and H$_2$ for the DLA towards \Jtwos\ using green and purple colors, respectively. The figures for the remaining targets in our sample are shown in the Appendix. Although independently \CI\ and H$_2$ provide degenerate constraints in the $\rm n_{\rm H}$-$I_{\rm UV}$ parameter space, their constraints are nearly orthogonal allowing us to break the degeneracy. Therefore, a joint fit (shown by the red contour in Fig.~\ref{fig:J0858_phys}) provides significantly tighter constraints as evidenced by the marginalized joint probability densities shown on the right and top axes of Fig.~\ref{fig:J0858_phys}.
The black line in Fig.~\ref{fig:J0858_phys} indicates the constraint on the parameters using the analytical theoretical expression from \citet{Bialy2017}, based on 
the \HI-H$_2$ theory from \citet{Sternberg2014}, which relates the \HI\ column density in the envelope of H$_2$ cloud to the ratio of the UV intensity to density ($I_{\rm UV}/n$) for a given dust abundance. The latter is assumed here to scale with metallicity. 
Since neutral gas not directly related to the H$_2$-bearing cloud also contributes to the observed $N(\HI)$, the constraint following \citet{Bialy2017} results in an upper limit on $I_{\rm UV}$ for a given $n_{\rm H}$. For all DLAs in our sample, the \HI-H$_2$ transition theory provides constraints in agreement with the joint measurements from \ion{C}{i} and H$_2$.

The estimates of $n_{\rm H}$ and $I_{\rm UV}$ for our sample are given in Table~\ref{table:fit_summary}.

\begin{table*}
	\centering
	\caption{Summary of column densities, chemical enrichment, and physical conditions in the sample. 
	}
	\addtolength{\tabcolsep}{-1pt}
	\label{table:fit_summary}
	\begin{tabular}{lccccccccccr}
		\hline
		Quasar & \zabs & $\log N(\rm H_2)$ & $\log N(\rm \HI)$ & $\log f$ & $\log N(\CI)$ & [X/H]$^{\rm a}$ & [Fe/X] & $A_{\rm V}$\,$^{\rm b}$ & $T_{01}$ & $\log n_{\rm H}$ & ${\log I_{\rm UV}}^{\rm c}$\\
		       &       &   [cm$^{-2}$]     &  [cm$^{-2}$]    &          &  [cm$^{-2}$]  &           &        &             &  [K]     & [cm$^{-3}$] & \\
		\hline
		{\Large \strut}J\,0136+0440 & 2.779 & $18.65^{+0.06}_{-0.07}$ & $20.73\pm0.01$ &  $-1.79^{+0.06}_{-0.07}$ & $14.67^{+0.17}_{-0.15}$ & $-0.58^{+0.03}_{-0.03}$ & $-0.70^{+0.03}_{-0.04}$ & $0.16$ & $77^{+7}_{-6}$ & $2.2^{+0.1}_{-0.2}$ & $-0.2^{+0.2}_{-0.1}$ \\
        {\Large \strut}J\,0858+1749  & 2.625 & $19.72^{+0.01}_{-0.02}$ & $20.40\pm0.01$ &  $-0.53^{+0.01}_{-0.02}$ & $14.34^{+0.04}_{-0.04}$ & $-0.63^{+0.02}_{-0.02}$ & $-1.70^{+0.03}_{-0.03}$ & $0.29$ & $102^{+2}_{-2}$ & $1.8^{+0.1}_{-0.1}$ & $0.1^{+0.2}_{-0.2}$  \\
        {\Large \strut}J\,0906+0548  & 2.567 & $18.88^{+0.02}_{-0.02}$ & $20.13\pm0.01$ &  $-1.00^{+0.02}_{-0.02}$ & $13.86^{+0.06}_{-0.05}$ & $-0.18^{+0.05}_{-0.08}$ & $-1.28^{+0.08}_{-0.06}$ & $0.21$ & $116^{+26}_{-4}$ & $2.6^{+0.1}_{-0.1}$ & $0.7^{+0.1}_{-0.2}$ \\
        {\Large \strut}J\,0946+1216  & 2.607 & $19.97^{+0.01}_{-0.02}$ & $21.15\pm0.02$ &  $-0.93^{+0.02}_{-0.03}$ & $14.43^{+0.05}_{-0.05}$ & $-0.48^{+0.01}_{-0.01}$ & $-0.77^{+0.01}_{-0.01}$ & $0.25$ & $118^{+9}_{-8}$ & $2.3^{+0.1}_{-0.1}$ & $0.5^{+0.2}_{-0.2}$ \\
        {\Large \strut}J\,1146+0743  & 2.840 & $18.82^{+0.03}_{-0.02}$ & $21.54\pm0.01$ &  $-2.42^{+0.03}_{-0.02}$ & $13.91^{+0.09}_{-0.07}$ & $-0.57^{+0.02}_{-0.02}$ & $-0.67^{+0.02}_{-0.02}$ & $0.17$ & $101^{+12}_{-9}$ & ${2.6^{+0.2}_{-0.2}}^c$  &  ${0.6^{+0.2}_{-0.2}}^d$ \\
        {\Large \strut}J\,1236+0010  & 3.033 & $19.76^{+0.01}_{-0.01}$ & $20.78\pm0.01$ &  $-0.80^{+0.01}_{-0.01}$ & $14.11^{+0.08}_{-0.07}$ & $-0.58^{+0.04}_{-0.03}$ & $-1.09^{+0.04}_{-0.04}$ & $0.07$ & $118^{+7}_{-6}$ & $1.5^{+0.6}_{-0.6}$ & $0.4^{+0.4}_{-0.4}$ \\
        {\Large \strut}J\,2347$-$0051 & 2.588 & $19.44^{+0.01}_{-0.01}$ & $20.47\pm0.01$ &  $-0.80^{+0.01}_{-0.01}$ & $14.05^{+0.37}_{-0.25}$ & $-0.82^{+0.05}_{-0.03}$ & $-0.71^{+0.03}_{-0.05}$ & $0.04$ & $76^{+2}_{-2}$ & $2.0^{+0.4}_{-0.5}$ & $-0.4^{+0.3}_{-0.4}$ \\
		\hline
		{\Large \strut}J\,0843+0221 & 2.786 & $21.21^{+0.02}_{-0.02}$ & $21.82\pm0.11$ &  $-0.48^{+0.08}_{-0.08}$ & $13.98^{+0.04}_{-0.04}$ & $-1.52^{+0.08}_{-0.10}$ & $-1.01^{+0.07}_{-0.05}$ & $<0.1$ & $123^{+9}_{-8}$ & $2.82^{+0.05}_{-0.06}$ & $1.18^{+0.24}_{-0.17}$ \\
		\hline
	\end{tabular}
		\addtolength{\tabcolsep}{+1pt}
    \begin{tablenotes}
    \item (a) For all systems the metallicity given here is based on the abundance of sulphur, except for that towards J\,1146+0743 for which we used zinc because \ion{S}{ii} lines are blended.
    \item (b) The typical statistical uncertainty on $A_{\rm V}$ is 0.01~mag, but the systematic uncertainty due to intrinsic spectral slope variations is 0.1~mag (see text).
    \item (c) In units of the Draine field.
    \item (d) For the system towards J\,1146+0743, the estimated number density in the two components is similar. Here we give the value corresponding to the strongest H$_2$ component.
    \end{tablenotes}
\end{table*}

\section{Discussion}\label{sect:discussion}

\subsection{Comparison between different H$_2$-absorber samples}\label{sect:properties}

In this section, we compare the chemical properties of various samples of H$_2$-bearing DLAs (and sub-DLAs) in the distant Universe: 
this includes (1) high-redshift systems from this paper based on direct H$_2$ identification in SDSS spectra of DLAs \citep{Balashev2014}; (2) systems pre-selected using \ion{C}{I} without any prior on \HI\ content \citep[][]{Ledoux2015, Noterdaeme2018}; (3) systems pre-selected based on very high \ion{H}{i} column densities \citep[see][]{Noterdaeme2015b, Ranjan2019}; and (4) regular DLAs systems without any further or particular selection (mostly from the VLT/UVES database, see \citealt{Noterdaeme2008}, \citealt{Balashev2017} and references therein, but also from \citealt{Muzahid2015, Muzahid2016} for low-$z$ systems). 
For convenience, in the following we will refer to 
these samples as, respectively, 
{\sl direct-H$_2$}, {\sl \CI}, {\sl strong-\HI} and {\sl blind}.

We note that the H$_2$-bearing DLAs towards J\,0843$+$0221\footnote{This system has the second highest H$_2$ column density to date ($\log N(\HH)=21.2$, \citealt{Balashev2017}), only slightly below that towards J\,1513+0352, selected for its high $N(\HI)$ ($\log N(\HH)=21.3$,  \citealt{Ranjan2018}).} and J\,1146$+$0743 belong to both the high-$N(\HI)$ and the direct-H$_2$ selections \citep{Noterdaeme2014,Balashev2014}. 

We also compare the properties of these samples with 
those observed at $z=0$ in the local group \citep{Savage1977, Rachford2009, Gillmon2006, Welty2012}. By construction (due to various selection effects) the different samples probe different regions in terms of the redshift, metallicity, \HI, and H$_2$ abundance, which we discuss in the following.

\subsubsection{Hydrogen content of molecular absorbers}

\begin{figure}
\centering
\includegraphics[width=\hsize]{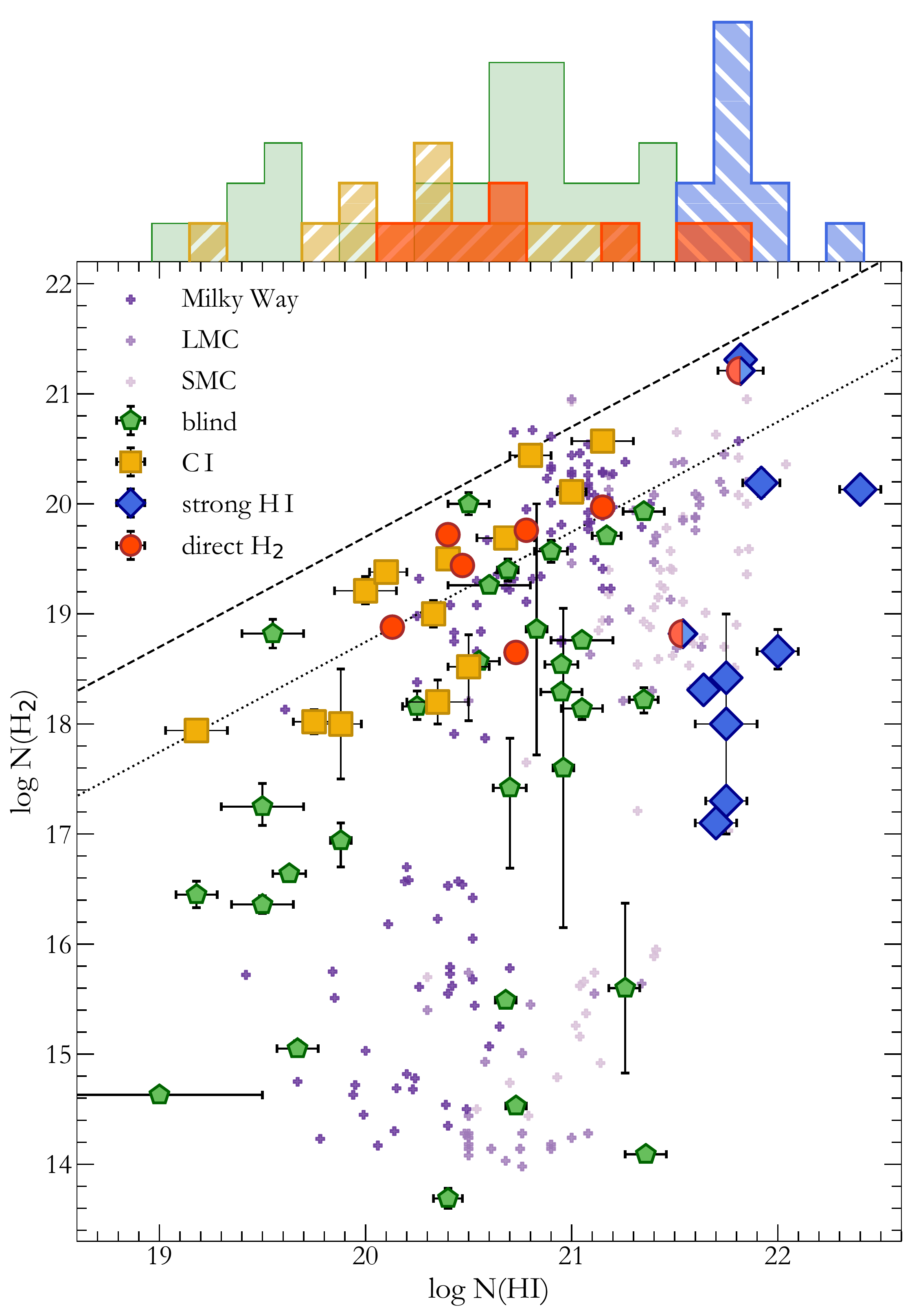}
		\caption{The dependence of $\log N(\rm H_2)$ on $\log N(\ion{H}{i})$ measured in absorption at high redshift and locally. The green pentagons, blue diamonds and yellow squares represent H$_2$-bearing DLAs from the `blind', \ion{C}{I} and strong-\ion{H}{I} selections, respectively. The sample presented in this paper and concerned with direct-H$_2$ detection is shown by red circles. Note that two systems, towards J\,0843+0221 and J\,1146+0743, belong to both 
		the H$_2$ and the \HI-selections and the symbols are then half red circle, half blue diamond. 
		The measurements in the local galaxies (Milky Way, LMC and SMC) are shown by purple crosses with slightly different hue. The dashed and dotted lines depict constant average molecular fractions of 0.5 and 0.1, respectively. The histograms 
	    on top show the $N(\HI)$-distributions for the different samples using the same colour coding.
		\label{fig:H2vsHI}}
\end{figure}

In Fig.~\ref{fig:H2vsHI}, we plot the dependence of H$_2$ on \ion{H}{i} column densities measured in different samples. 
It can be seen that, for a given $H_2$ column density, \CI-selected systems tend to probe lower 
\HI\ column densities than the other samples. 
This can be explained by the absence of any prior on \HI\ for the detection of these systems \citep{Ledoux2015}, and the corresponding sightlines may statistically intercept less unrelated \HI\ gas compared to DLAs. Notwithstanding, when compared to the cosmological \HI\ frequency distribution,
the $N(\HI)$-distribution of the \CI-selected sample remains skewed towards high values. In other words, the probability to find \CI\ remains higher at higher \HI\ column densities, but the overwhelming incidence of neutral absorption systems with low $N(\HI)$ \citep[e.g.,][]{Noterdaeme2012c}
makes them over-represented in the \CI\ sample.

The {\sl blind} sample should in principle be more representative of the overall DLA population. However, while the corresponding H$_2$-bearing systems indeed cover a wide range of \HI\ column densities, there is an excess of high-\HI\ systems. This reflects not only the increasing H$_2$-detection probability with increasing $N(\HI)$ but also a bias towards high-$N(\HI)$ in the UVES DLA/sub-DLA database \citep{Noterdaeme2008}. Since these systems were mostly observed in high spectral resolution, high-S/N data, the corresponding detection limits can indeed be as low as $N(\HH)\sim 10^{14}~$cm$^{-2}$, so that systems with low molecular content \citep[e.g.,][]{Noterdaeme2007} are also present in the figure.

The strong-\HI\ sample, by construction, probes the high end of \HI\ column densities. Although the incidence rate of H$_2$ in that sample is quite high, $\sim 50$ per cent, the actual H$_2$ column densities correspond to a very wide range\footnote{Note that most ESDLAs were followed up at intermediate resolution using the X-shooter spectrograph, thus formally reported non-detections of H$_2$ actually correspond to upper limits of $\log N(\rm H_2)\gtrsim 16-17$.}, including the two highest H$_2$ column densities found in ESDLAs, towards J\,0843+0221 and J\,1513+0352 \citep{Balashev2017, Ranjan2018}.

Finally, the direct H$_2$-selection in SDSS is, by construction, mostly limited to high \HH\ column densities, typically $\log N(\HH)>19$, that produce detectable lines at low spectral resolution. In addition, the search was limited to DLAs and strong sub-DLAs ($\log N(\HI)\ge 20$) from the SDSS catalogue \citep{Noterdaeme2012c}.
{We note that six out of eight H$_2$-selected systems have 
high average molecular fractions, as seen for \CI-selected systems. In the case of the latter, the presence of \CI\ depends on the abundance of carbon (hence the metallicity) and the shielding from UV photons, both favouring high molecular fractions, as discussed in next section. In the case of the former, by selection we restrict to high \HH-column density systems when the \HI\ frequency distribution decreases with increasing \HI\ column density, also favouring high-molecular fraction systems. 
}

\subsubsection{Dependence on metallicity}

\begin{figure*}
\centering
\setlength{\tabcolsep}{2pt}
    \begin{tabular}{cc}
    \includegraphics[trim={0 0 0 0},clip,width=0.5\textwidth]{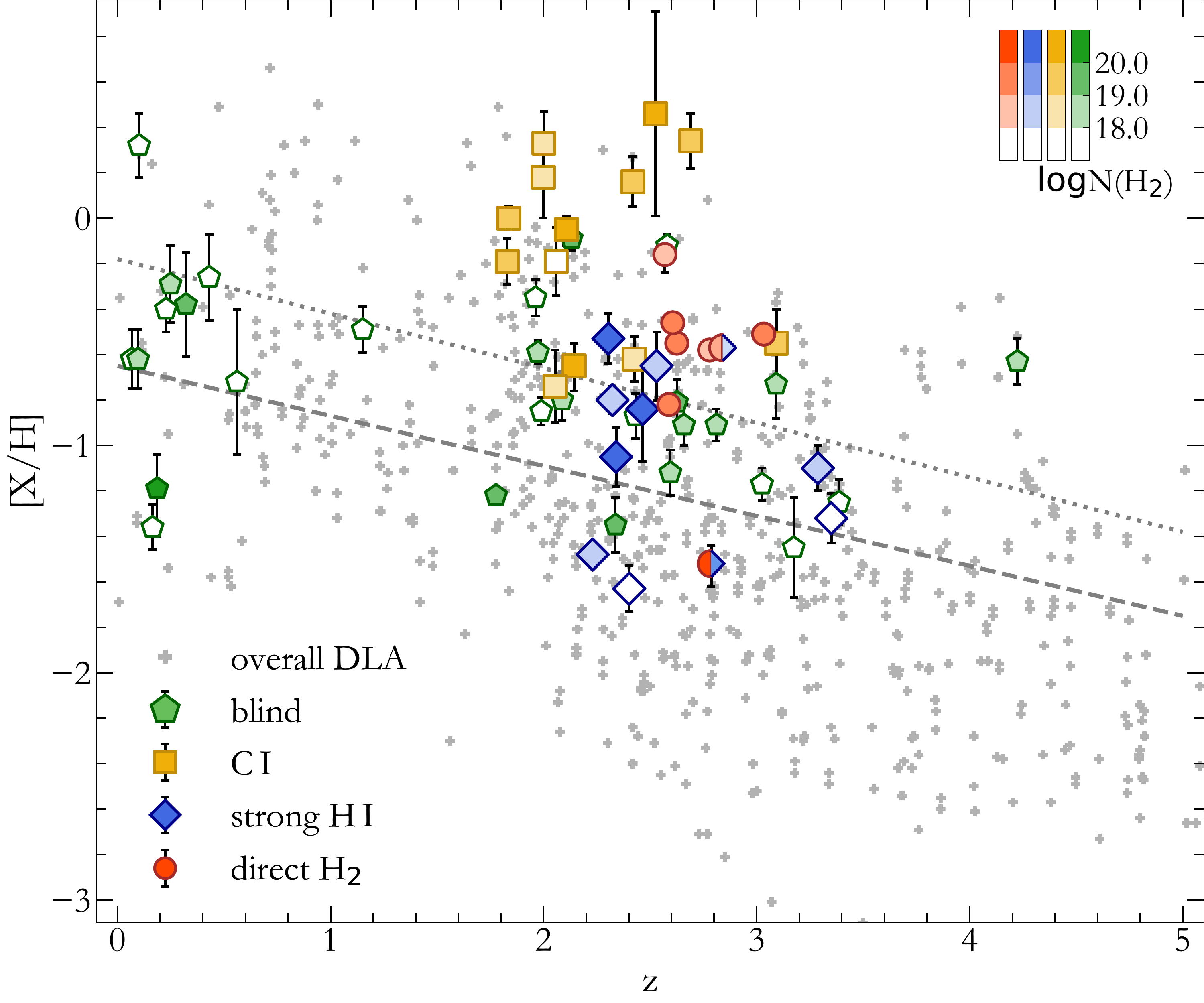}
    & \includegraphics[trim={0 0 0 0},clip,width=0.5\textwidth]{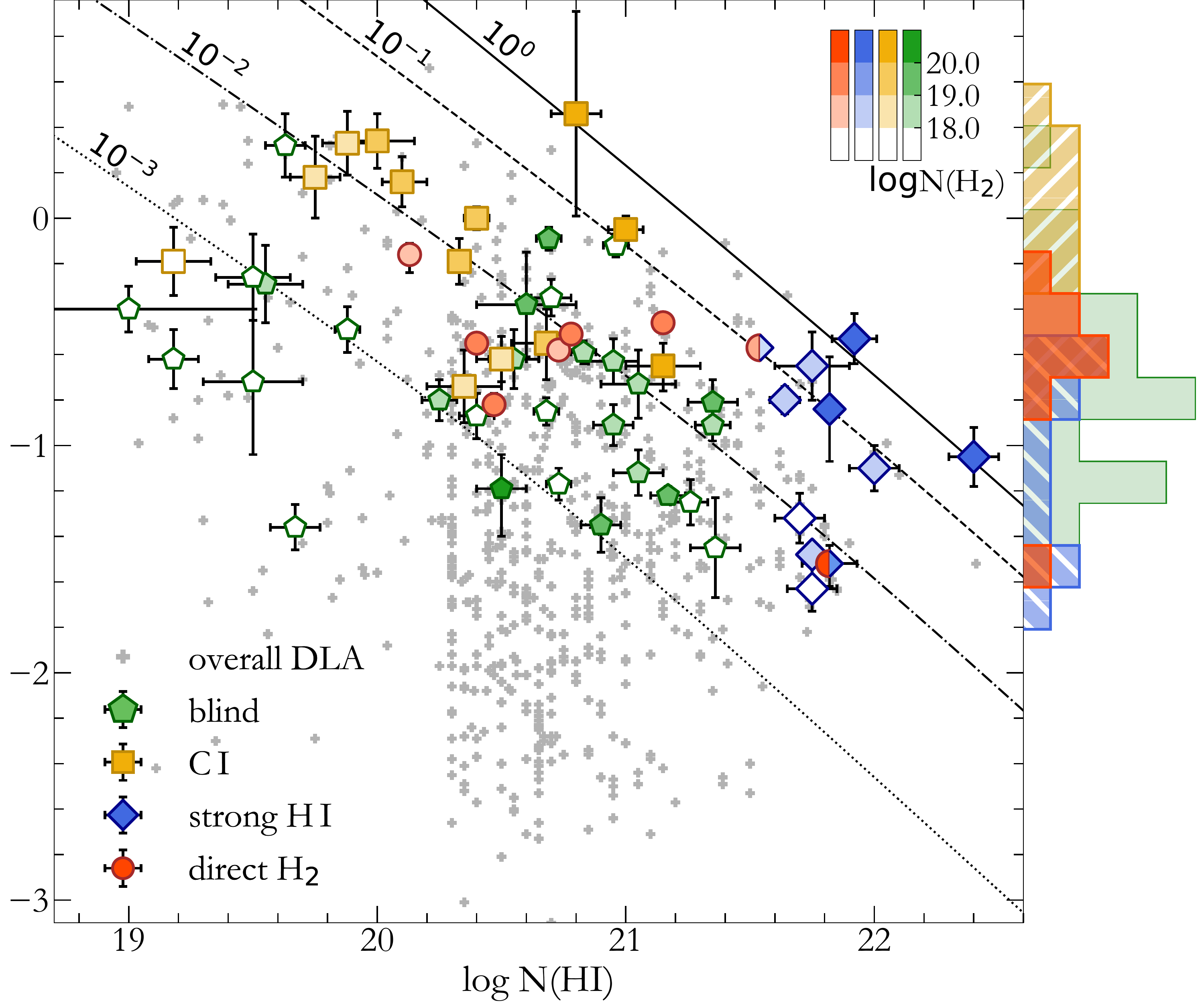} \\
    \end{tabular}
\setlength{\tabcolsep}{6pt}
\caption{The metallicity as a function of the redshift (left panel) and $\log N(\HI)$ (right panel) measured in DLAs. The symbols are the same as in Fig.~\ref{fig:H2vsHI} with colour saturation representing the H$_2$ column density. The mean metallicity in DLAs as a function of redshift is shown on the left panel by dash \citep{Rafelski2012} and dotted \citep[$N(\HI)$ weighted and depletion corrected, see][]{DeCia2018} lines. The different lines in the right panel correspond to the $N(\HI)$ in the envelope of a single H$_2$ cloud \citep{Bialy2017} for the different ratios of the  UV field to the number density, $I_{\rm UV} / n$, where $I_{\rm UV}$ is expressed in units of the Draine field and the $n$ in units of cm$^{-3}$.} 
\label{fig:MevsHI}
\label{fig:Mevsz}
\end{figure*}

Aside from the obvious differences in H$_2$ and \HI\ column densities, the different pre-selection techniques also exhibit prominent differences in the probed metallicities and redshifts. In Fig.~\ref{fig:MevsHI} we show metallicity as a function of redshift and \HI\ column density. We here use metallicities based on volatile species, i.e. either sulphur or zinc \citep[e.g.,][]{Vladilo2011}  depending on the line coverage and possible blends. 

From the left panel of Fig.~\ref{fig:Mevsz}, it can be seen that different samples probe slightly different redshifts. The obvious reason for this is that each selection utilizes a different redshift cutoff depending on the instrument used, the atmospheric cutoff in the blue, and the rest-frame wavelength of the transitions used for detection. Indeed, \CI\ lines are detectable down to lower redshifts than \HI, while the observation of H$_2$ lines in SDSS spectra requires even higher-redshifts than for \HI. Note that the sensitivity functions, $g(z)$, of the samples increases with decreasing redshift, and therefore the density of points is enhanced  close to their respective redshift cutoff.

H$_2$-bearing DLAs mainly occupy the upper half of the overall observed DLA metallicity distribution, consistent with previous works \citep[e.g.,][]{Petitjean2006}. 
While the \CI-selected systems probe mostly the high-end of the H$_2$-bearing DLA metallicity distribution (and consequently of the overall DLA distribution, Ledoux et al. in prep.), the strong-\HI-selected sample corresponds to low metallicity H$_2$-bearing systems. The direct H$_2$ selection tends to probe metallicities in between those of \CI\ and ESDLAs and this remains evident even after taking into account the small redshift evolution of metallicity.

The metallicity actually plays a dominant role in the detectability of \CI\ lines for a given \HI\ column density.
Not only does a high metallicity imply a high carbon abundance, but it also promotes the survival of \CI\ through shielding by dust and H$_2$, both of which scale with metallicity. Therefore the \CI-selection predominantly picks out high metallicity systems, in particular at low $N(\HI)$ \citep[][Ledoux et al. (in prep)]{Noterdaeme2018,Zou2018}.
Interestingly, for a given metallicity, the \HH\ column densities in \CI-selected systems increases with increasing $N(\HI)$. Since theoretical models of \HI-H$_2$ conversion indicate that the \HI\ envelope of a single cloud has almost constant column density whatever the $N(\HH)$ (i.e. more gas added to a cloud is converted to H$_2$), this could indicate the interception of multiple \HI-H$_2$ layers. This is consistent with a higher probability to find cold gas close to a given cold gas cloud, either along the same line of sight (appearing as multiple components, see Ledoux et al. in prep), or transversely \citep{Krogager2018}.

The chemical conditions for the presence of H$_2$ at detectable levels are less restrictive than those for \CI, allowing the direct H$_2$ and strong \HI-selections to probe lower metallicities. Furthermore, in the high \HI\ column density regime, the presence of \HH\ in a given system appears to be little dependent on the metallicity. Indeed, the DLA towards J\,0843+0221 has a very high \HI\ and H$_2$ content despite its low metallicity, and the metallicity distribution of H$_2$-bearing ESDLAs is not significantly different from that of non H$_2$-bearing ESDLAs (see Fig.~\ref{fig:MevsHI}). 
The local physical conditions (density and UV field) seem to play a more important role for these systems. Indeed, these systems are expected to probe small galacto-centric distances \citep{Noterdaeme2014,Ranjan2019}, where the pressure could be increased \citep[in theoretical models of warm and cold gas distribution in galaxies][]{Wolfire2003}. 
We also note that high-$N(\HI)$ DLAs with high metallicities could be under-represented among them due to the associated dust reddening. This could exclude the background quasars from optically-selected samples \citep[see, e.g.,][]{Geier2019, Krogager2019} and lead to infeasible amounts of observing time required for spectroscopic follow-up.

The intermediate metallicities seen for the H$_2$-selected sample suggest that the intrinsic metallicity distribution of H$_2$-bearing systems is shifted compared to that of the overall DLA population.
However, the paucity of low-metallicity systems in the direct H$_2$ sample {\sl in comparison to H$_2$-bearing ESDLAs}, if real, can be due also to differences in physical conditions. The presence of H$_2$ depends not only on the metallicity and column density but also on the number density ($n$) and intensity of the UV field ($I_{\rm UV}$). To understand this further, we compare the observed $N(\HI)$ and metallicities with the theoretical expectation from the work by \citet{Bialy2016}.
We caution that the theoretical predictions are only valid for a single cloud whereas the actual situation is more likely a concatenation of H$_2$ clouds along the line of sight embedded in a diffuse and warm \HI\ phase that also contributes to the measured \HI\ column density. In that sense, all measured values of $N(\HI)$ are upper limits to the \HI\ column density associated with the envelope of the H$_2$ cloud.
The different theoretical lines should therefore be regarded as the minimal conditions for \HI/H$_2$ transition at a given metallicity and \HI\ column density. That is, the calculated \HI\ column densities are only applicable for systems where a transition from \HI\ to \HH\ has occurred. For the typical ranges of the physical conditions in the cold ISM (mainly, the metallicity, number density and UV flux) 
this corresponds to a \HH\ column density larger than $\log N (\rm H_2) \sim 18-20$.
Therefore we highlight the points correspond to different ranges of \HH\ column densities in Fig.~\ref{fig:H2vsHI} using color saturation. The blank points represent the lowest \HH\ column densities ($\log N (\rm H_2) < 18$) for which the model lines are most likely not applicable. Almost all saturated H$_2$-bearing DLAs, regardless of their selection, are located in between the $I_{\rm UV}/n_{\rm H}\sim 10^{-3}$ and $I_{\rm UV}/n_{\rm H}\sim 1$ lines. Systems with low metallicity and \HI\ column density would require very low $I_{\rm UV}/n_{\rm H}$ ratios (below $\lesssim10^{-3}$) in order to efficiently form H$_2$. High density systems with low total column density must have very small physical dimensions leading to a low absorption cross-section. This is consistent with the absence of H$_2$ among such absorption-selected samples.
For example, for a given UV field typical of our Galaxy (i.e. $I_{\rm UV} = 1$) this correspond to $n\gtrsim10^3$\,cm$^{-3}$. Therefore at column density at the DLA threshold, and metallicity typical of DLAs ([X/H]$\sim-1.3$), a cloud should be compressed less than $<0.1$\,pc in order to H$_2$ to form. For a temperature of $T\sim 100$~K, this also means a thermal pressure $P\sim10^5$\,K\,cm$^{-3}$, which is not very common, especially at low N(\HI). 

To summarize, the three preselection methods presented above complement each other and enable us to study the physical conditions and the H$_2$/\ion{H}{I} transition in a wide range of environments.

\subsubsection{Fe depletion and dust}

\begin{figure}
\centering
\includegraphics[trim={0.0cm 0.0cm 0.0cm 0.0cm},clip,width=\hsize]{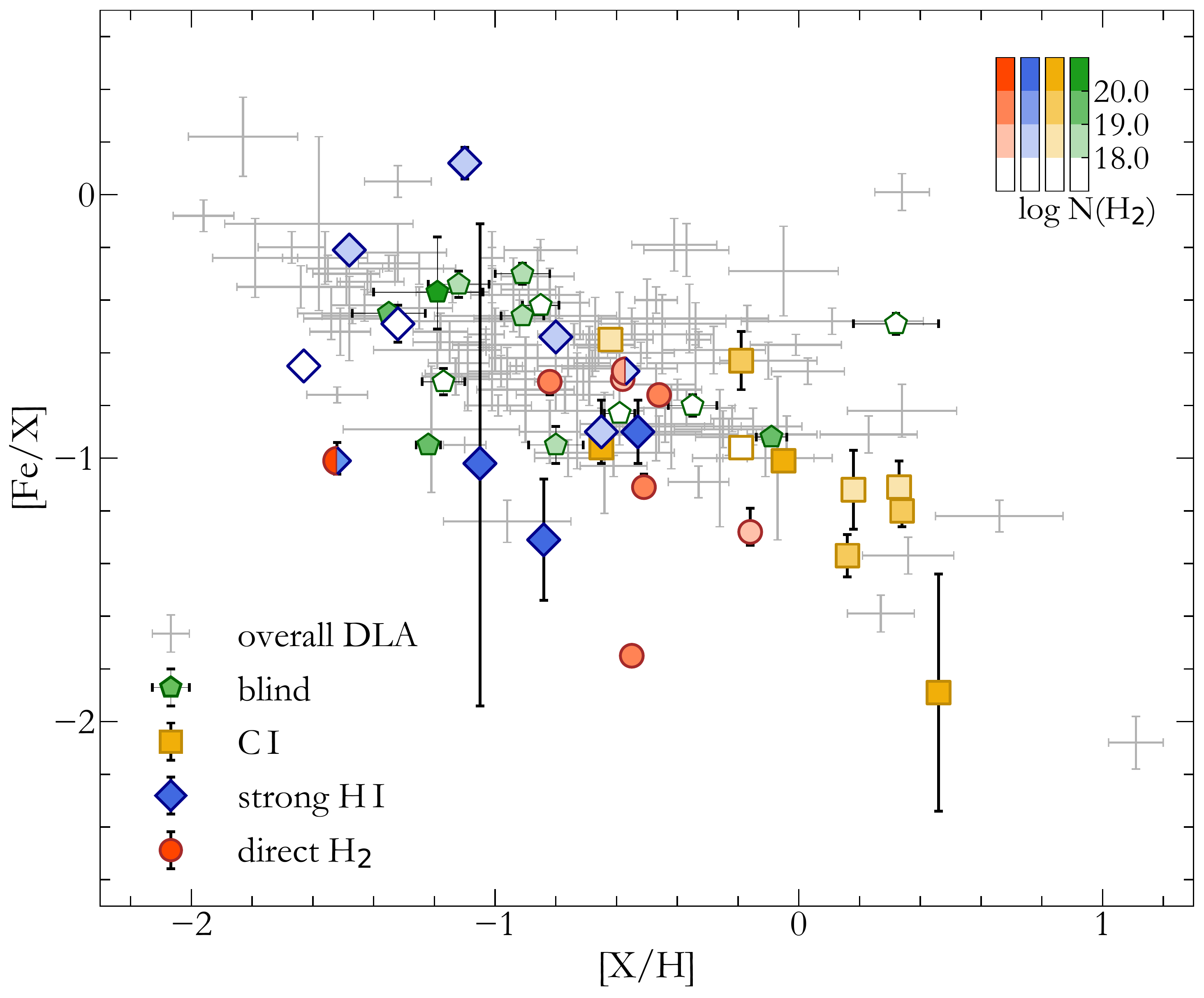}
		\caption{The depletion versus metallicity in DLA at high redshifts. Symbols and color encoding are the same as in Fig.~\ref{fig:Mevsz}.\label{fig:DeptoMe}}
\end{figure}

\begin{figure}
\centering
\includegraphics[trim={0.0cm 0.0cm 0.0cm 0.0cm},clip,width=\hsize]{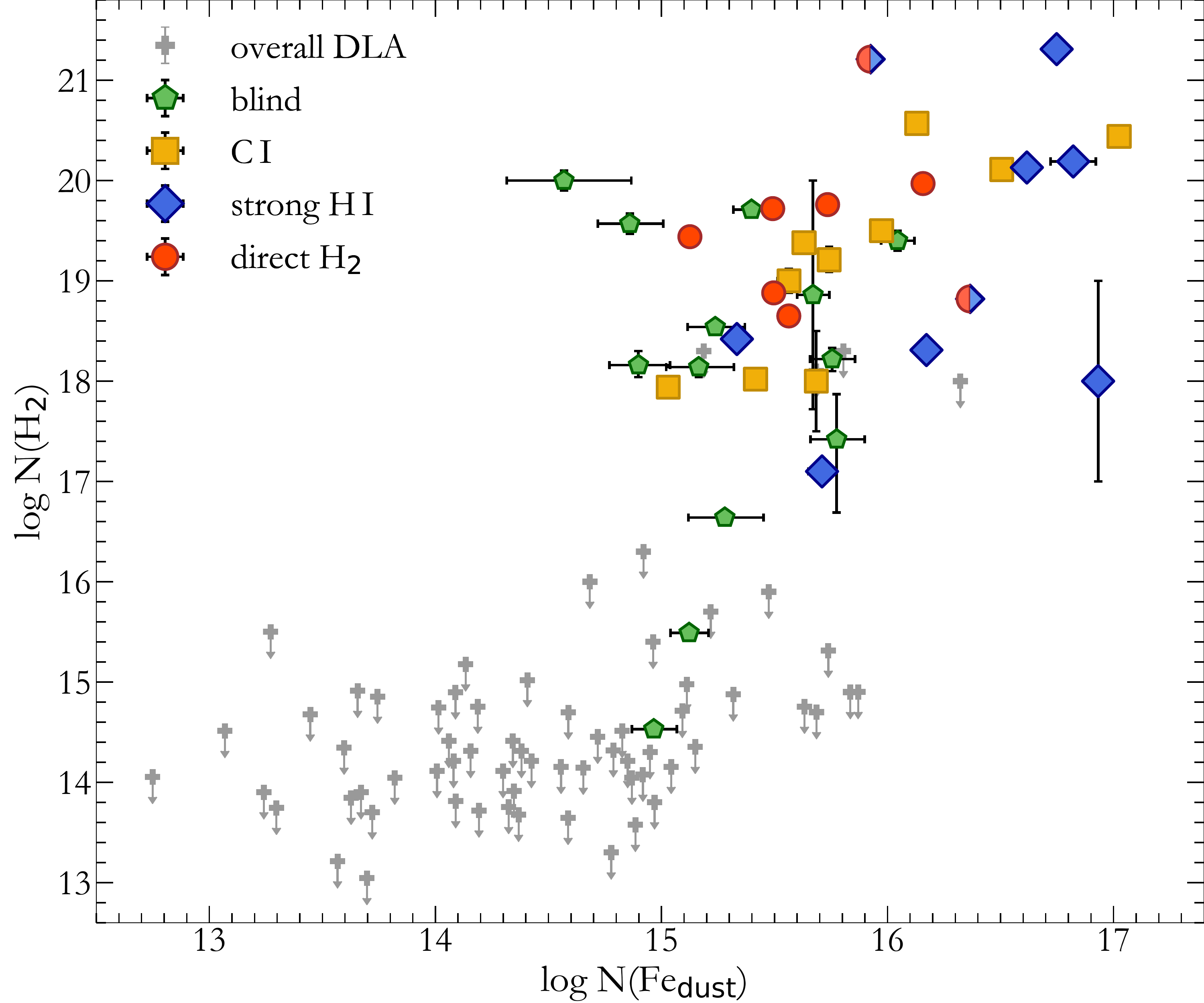}
		\caption{H$_2$ column density versus column density of Iron in the dust. Symbols and color encoding are the same as in Fig.~\ref{fig:Mevsz}. The grey crosses show the upper limits for DLAs taken from \citealt{Noterdaeme2008, Ranjan2019}.
		\label{fig:H2vsFedust}}
\end{figure}

In previous discussions, we did not differentiate between the metallicity and dust content, assuming that the dust content directly scales with the metallicity. However, the presence of H$_2$ is directly linked to the dust content, since the formation of molecules occurs on the surface of dust grains that also participate to the shielding of UV photons.
Unfortunately, the large systematic uncertainties of the $A_V$ measurements, compared to the actual (low) values, prevent us from quantifying the dust content directly from the measured extinction. Instead, we can use the depletion of refractory elements (like iron) compared to volatile ones (zinc or sulphur).
All samples follow a similar trend of increasing Fe depletion with increasing metallicity, as illustrated in Fig.~\ref{fig:DeptoMe}, reflecting the scaling of dust with metallicity \citep[see][]{Ledoux2003}.
Since the H$_2$ formation rate is roughly proportional to the dust abundance, we might statistically expect a enhanced depletion for H$_2$-bearing DLAs at a given metallicity. 
However, we do not clearly observe such a behaviour: only two out of eight DLAs in our sample have a significant stronger depletion (i.e. lower [Fe/X]) than seen from the overall DLA population at their given metallicity. This can be 
due to a fact that the quantities used for comparison are averaged throughout the absorption complexes \footnote{See, e.g., the H$_2$-bearing DLA towards J\,0000+0048, where the detailed analysis of high-resolution data indicates that the H$_2$-bearing component is significantly more depleted than the average over all DLA components.}. 
More importantly, the dependence on average depletion shown in Fig.~\ref{fig:DeptoMe} doesn't take into account the column densities. Observationally, it is already known that the total column of the dust probably plays a major role in the presence of H$_2$ at detectable levels \citep[see][]{Noterdaeme2008}. The physical reason is that larger dust column imply larger shielding from UV photons, either directly from the dust grains, or indirectly provided by H$_2$ self-shielding (since the more dust, the higher H$_2$ formation rate and the larger H$_2$ column density). 
In Fig.~\ref{fig:H2vsFedust} we show the dependence of the H$_2$ column density on the column density of iron locked into dust, which provides a way to quantify the column density of dust \citep{Vladilo2006}. Overall, the different samples are not distinguishable in this Figure and confirm the neat increase in the H$_2$ column densities at $\log N(\rm Fe_{\rm dust})\gtrsim15$. This suggests that the physical conditions (ratio of UV to number density) should not be drastically different for different H$_2$ samples, since the critical $N(\rm Fe_{\rm dust})$ is mostly defined by this ratio \citep[see, e.g.,][]{Bialy2017}.
For given physical conditions, an increasing $N(\HH)$ with increasing $N(Fe_{\rm dust})$ could then be expected at higher dust column, regardless of whether this is due to a concatenation of several transition layers seen along a line of sight, or single clouds with larger column densities. Such a correlation may possibly be seen in the \CI-selected sample which possibly share more homogeneous conditions.


\subsection{Physical conditions}

\begin{figure}
\centering
\includegraphics[trim={0.0cm 0.0cm 0.0cm 0.0cm},clip,width=\hsize]{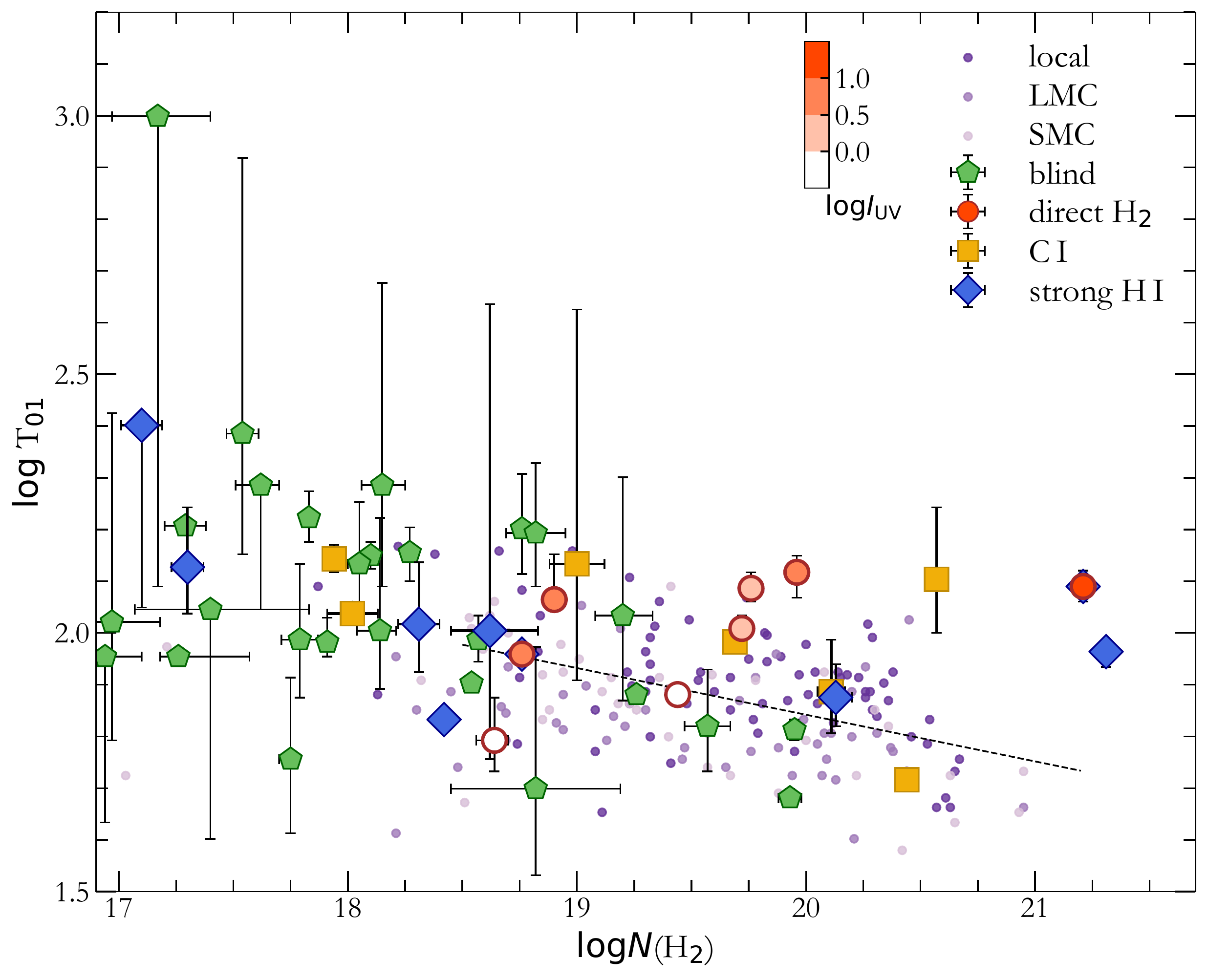}
		\caption{The excitation temperature T$_{01}$ versus column density of molecular hydrogen. Symbols are as per Fig.~\ref{fig:H2vsHI}, with colour-filling for the sample presented here (red-circles) representing the inferred intensity of the UV field (see enclosed colour bar). 
		\label{fig:T01}}
\end{figure}

We find values of the temperature in the range 75-120~K, with a median of $T\sim 100$~K\footnote{We note that this value corresponds to the temperature of the H$_2$-bearing gas and not the overall DLA system, which is more likely characterized by a warm neutral phase with $T \gtrsim 10^3$\,K.}. 
These values are slightly higher than what is measured towards nearby stars \citep[$T\sim 80$~K, see, e.g.,][]{Savage1977}, but lower than previously seen in the overall population of high-$z$ DLAs \citep[$T\sim 150$~K;][]{Srianand2005}. However,
$T_{01}$ temperatures in H$_2$-bearing DLAs are known to decrease with increasing H$_2$ column density \citep{Muzahid2015, Balashev2017}, a trend also seen in the local ISM.
The systems in our sample have typically higher \HH\ column densities than those from \citet{Srianand2005}, but lower than towards nearby stars, that can explain the difference. However, from our sample alone, we do not see any trend in T$_{01}$ vs. $\log N (\rm H_2)$, though our measurements are still within the measured dispersion of T$_{01}$. 
Metallicity could also play a role to explain the differences in the temperatures since it plays an important role in 
the thermal balance \citep[e.g.,][]{Wolfire2003}. 
{Indeed, the cooling is dominated by line emission (in particular [C\,{\sc ii}] at $158\mu\,{\rm m}$) while the heating is dominated by photoelectric effect on dust grains. Both these processes are directly linked to the gas metallicity.}  
Observationally, a low H$_2$ temperature of $T\sim 50$~K was measured in the super-Solar metallicity DLA towards J\,0000+0048 \citep{Noterdaeme2017} while higher temperatures where measured at low metallicities, $\rm [X/H]\lesssim -1$ towards  J\,0843+0221 and J\,1513+0352 \citep{Balashev2017, Ranjan2018}. 
{However, since the photo-electric heating involves UV photons, 
the intensity of the UV field directly plays an important role.} Indeed, one can see in Fig.~\ref{fig:T01}, where the estimated UV flux is coded by the color saturation, that our sample (for which most systems span a narrow range of the metallicity) supports such a dependence. 
We can further speculate that, knowing the gas metallicity, the relative $J=1/J=0$ population of H$_2$ could give a useful diagnostic of the UV to density ratio, without the need of detailed modelling.

\begin{figure}
\centering
\includegraphics[trim={0.0cm 0.0cm 0.0cm 0.0cm},clip,width=\hsize]{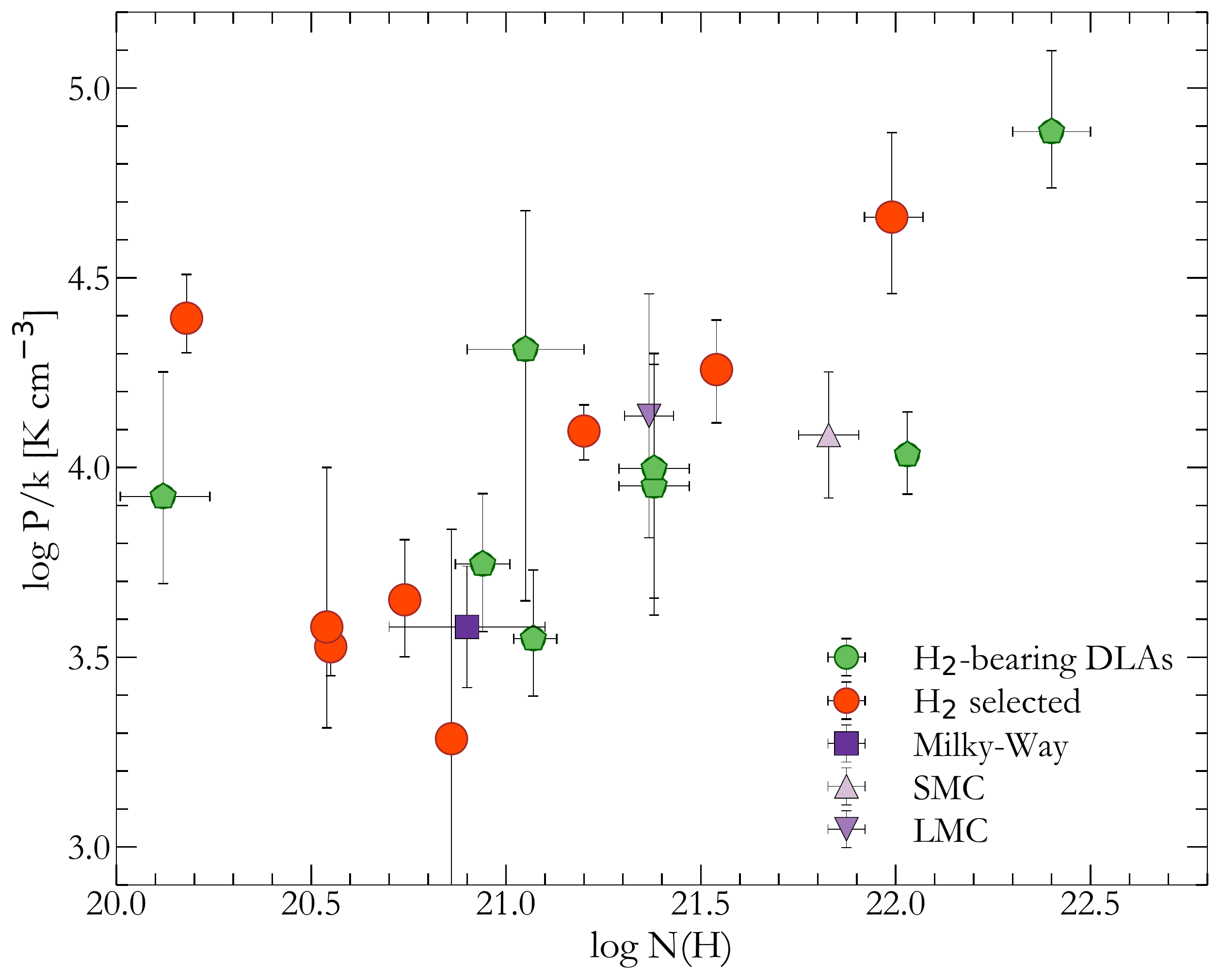}
		\caption{The dependence of the  thermal pressure ${\rm P}/k=nT$ on $\log N(\rm H_{\rm tot})$ measured in DLAs (circles) and local galaxies (square and triangles). Green and red circles correspond to the known H$_2$/\CI\ bearing DLAs at high redshift and our sample, respectively.
		\label{fig:PHI}}
\end{figure}
We found a typical hydrogen number density of $n \gtrsim 100$ cm$^{-3}$ and typical UV field on average twice higher than the Draine field. The obtained number densities and kinetic temperatures agree well with the canonical values for the cold ISM in the two-phase model from \citet{Wolfire2003}. This shows that H$_2$ is a good tracer of the CNM in DLAs. This is particularly useful since 21-cm absorption studies remain limited at high redshift \citep{Srianand2012}, due to the paucity of strong compact radio sources observable with current facilities \citep[but see][for future prospects]{GuptaMALS}.

Using our estimates of the number density and temperature, we compare the thermal pressures (${\rm P}/ k = n T$) to those seen in other H$_2$-bearing DLAs as well as in the local group in Fig.~\ref{fig:PHI}. 
Our sample probes a wide range of hydrogen column densities, from values similar to that of the Magellanic clouds \citep{Welty2016} down to about an order of magnitude lower than in the Milky Way \citep{Jenkins2011}. The pressure in our sample is about $\sim10^4$\,K\,cm$^{-3}$ with 0.5 dex dispersion around this value, similar to what is seen locally. Our data further strengthen the suggestion from \citet{Balashev2017} that the thermal pressure increases with the hydrogen column density. This is likely the consequence of both the pressure and the column density to increase with decreasing galacto-centric radius.

\section{Conclusion}
\label{sect:conclusion}

We presented follow-up observations with the X-shooter spectrograph on the Very Large Telescope of seven strong H$_2$-bearing DLAs at high redshifts. These DLAs were selected from the SDSS DLA catalogue \citep[see][]{Noterdaeme2012c} directly upon the presence of H$_2$ Lyman bands seen in the SDSS spectra of their background quasars, see \citet{Balashev2014}.
Our X-shooter observations confirm the detection of H$_2$ in all systems and hence the high efficiency of the direct H$_2$ detection method even at low spectral resolution. We measured the neutral, molecular, and metal column densities of the systems and found typical H$_2$ column densities in the range $\log N(\rm H_2) = 18.5-20$, when the \HI\ column density span a range $\log N(\HI) \sim 20.1-21.5$. The metallicities, obtained from undepleted species, are in the range 
$\log Z = -0.8$ to $-0.2$. 
Finally, in all systems, we detected absorption lines from \CI\ in various fine-structure energy levels associated to the H$_2$ components.
We used the rotational excitation of H$_2$, the fine-structure excitation of \CI, and theoretical relations for the \HI-H$_2$ transition to estimate the physical conditions in the H$_2$-bearing gas. 
We found a typical kinetic temperature of $\sim100$\,K and hydrogen densities $n_{\rm H}\sim100$\,cm$^{-3}$, within clouds embedded in a UV field about twice higher than the Draine field.
This confirms that H$_2$ is a good tracer of the cold neutral medium\footnote{However, \citet{Srianand2012} noted there is no one-to-one correspondence between the population of cold gas traced by H$_2$ and that traced by \ion{H}{i} 21-cm absorption.}. 

Comparing our sample with other samples of H$_2$-bearing DLAs and sub-DLAs selected differently, we discuss how the selection methods probe different regimes of column density and chemical enrichment and how
the presence of H$_2$ depends on the physical conditions. 
We show that all these methods complement each other and by the SDSS limitation tend to select high H$_2$ column-density systems that can be efficiently used to study the \ion{H}{i}-H$_2$ transition.

\section*{Acknowledgements}
We thank the referee for useful comments and suggestions.
This paper was supported by RSF grant 18-12-00301.
PN and JKK acknowledge support from the \textit{French Agence Nationale de la Recherche}, under grant ANR-17-CE31-0011-01 (Projet "HIH2", PI Noterdaeme) and from the CNRS-RFBR Russian-French grant PRC No\,1845 (PI Noterdaeme). We are also grateful to the Institut d'Astrophysique de Paris (SB and VK) and to the Ioffe Institute in Saint Petersburg (PN and JKK) for hospitality when part of this work was achieved.




\bibliographystyle{mnras}
\bibliography{bib} 

\begin{thebibliography}{}
\makeatletter
\relax
\def\mn@urlcharsother{\let\do\@makeother \do\$\do\&\do\#\do\^\do\_\do\%\do\~}
\def\mn@doi{\begingroup\mn@urlcharsother \@ifnextchar [ {\mn@doi@}
  {\mn@doi@[]}}
\def\mn@doi@[#1]#2{\def\@tempa{#1}\ifx\@tempa\@empty \href
  {http://dx.doi.org/#2} {doi:#2}\else \href {http://dx.doi.org/#2} {#1}\fi
  \endgroup}
\def\mn@eprint#1#2{\mn@eprint@#1:#2::\@nil}
\def\mn@eprint@arXiv#1{\href {http://arxiv.org/abs/#1} {{\tt arXiv:#1}}}
\def\mn@eprint@dblp#1{\href {http://dblp.uni-trier.de/rec/bibtex/#1.xml}
  {dblp:#1}}
\def\mn@eprint@#1:#2:#3:#4\@nil{\def\@tempa {#1}\def\@tempb {#2}\def\@tempc
  {#3}\ifx \@tempc \@empty \let \@tempc \@tempb \let \@tempb \@tempa \fi \ifx
  \@tempb \@empty \def\@tempb {arXiv}\fi \@ifundefined
  {mn@eprint@\@tempb}{\@tempb:\@tempc}{\expandafter \expandafter \csname
  mn@eprint@\@tempb\endcsname \expandafter{\@tempc}}}

\bibitem[\protect\citeauthoryear{{Altay}, {Theuns}, {Schaye}, {Booth}  \&
  {Dalla Vecchia}}{{Altay} et~al.}{2013}]{Altay2013}
{Altay} G.,  {Theuns} T.,  {Schaye} J.,  {Booth} C.~M.,   {Dalla Vecchia} C.,
  2013, \mn@doi [\mnras] {10.1093/mnras/stt1765}, \href
  {https://ui.adsabs.harvard.edu/\#abs/2013MNRAS.436.2689A} {436, 2689}

\bibitem[\protect\citeauthoryear{{Balashev} \& {Noterdaeme}}{{Balashev} \&
  {Noterdaeme}}{2018}]{Balashev2018}
{Balashev} S.~A.,  {Noterdaeme} P.,  2018, \mn@doi [\mnras]
  {10.1093/mnrasl/sly067}, \href
  {https://ui.adsabs.harvard.edu/#abs/2018MNRAS.478L...7B} {478, L7}

\bibitem[\protect\citeauthoryear{{Balashev}, {Varshalovich}  \&
  {Ivanchik}}{{Balashev} et~al.}{2009}]{Balashev2009}
{Balashev} S.~A.,  {Varshalovich} D.~A.,   {Ivanchik} A.~V.,  2009, \mn@doi
  [Astronomy Letters] {10.1134/S1063773709030025}, \href
  {https://ui.adsabs.harvard.edu/#abs/2009AstL...35..150B} {35, 150}

\bibitem[\protect\citeauthoryear{{Balashev}, {Klimenko}, {Ivanchik},
  {Varshalovich}, {Petitjean}  \& {Noterdaeme}}{{Balashev}
  et~al.}{2014}]{Balashev2014}
{Balashev} S.~A.,  {Klimenko} V.~V.,  {Ivanchik} A.~V.,  {Varshalovich} D.~A.,
  {Petitjean} P.,   {Noterdaeme} P.,  2014, \mn@doi [\mnras]
  {10.1093/mnras/stu275}, \href
  {https://ui.adsabs.harvard.edu/#abs/2014MNRAS.440..225B} {440, 225}

\bibitem[\protect\citeauthoryear{{Balashev}, {Noterdaeme}, {Klimenko},
  {Petitjean}, {Srianand}, {Ledoux}, {Ivanchik}  \& {Varshalovich}}{{Balashev}
  et~al.}{2015}]{Balashev2015}
{Balashev} S.~A.,  {Noterdaeme} P.,  {Klimenko} V.~V.,  {Petitjean} P.,
  {Srianand} R.,  {Ledoux} C.,  {Ivanchik} A.~V.,   {Varshalovich} D.~A.,
  2015, \mn@doi [\aap] {10.1051/0004-6361/201425553}, \href
  {https://ui.adsabs.harvard.edu/#abs/2015A&A...575L...8B} {575, L8}

\bibitem[\protect\citeauthoryear{{Balashev}, {Zavarygin}, {Ivanchik},
  {Telikova}  \& {Varshalovich}}{{Balashev} et~al.}{2016}]{Balashev2016}
{Balashev} S.~A.,  {Zavarygin} E.~O.,  {Ivanchik} A.~V.,  {Telikova} K.~N.,
  {Varshalovich} D.~A.,  2016, \mn@doi [Monthly Notices of the Royal
  Astronomical Society] {10.1093/mnras/stw356}, \href
  {https://ui.adsabs.harvard.edu/abs/2016MNRAS.458.2188B} {458, 2188}

\bibitem[\protect\citeauthoryear{{Balashev} et~al.,}{{Balashev}
  et~al.}{2017}]{Balashev2017}
{Balashev} S.~A.,  et~al., 2017, \mn@doi [\mnras] {10.1093/mnras/stx1339},
  \href {https://ui.adsabs.harvard.edu/#abs/2017MNRAS.470.2890B} {470, 2890}

\bibitem[\protect\citeauthoryear{{Bialy} \& {Sternberg}}{{Bialy} \&
  {Sternberg}}{2016}]{Bialy2016}
{Bialy} S.,  {Sternberg} A.,  2016, \mn@doi [\apj]
  {10.3847/0004-637X/822/2/83}, \href
  {https://ui.adsabs.harvard.edu/\#abs/2016ApJ...822...83B} {822, 83}

\bibitem[\protect\citeauthoryear{{Bialy}, {Bihr}, {Beuther}, {Henning}  \&
  {Sternberg}}{{Bialy} et~al.}{2017}]{Bialy2017}
{Bialy} S.,  {Bihr} S.,  {Beuther} H.,  {Henning} T.,   {Sternberg} A.,  2017,
  \mn@doi [\apj] {10.3847/1538-4357/835/2/126}, \href
  {https://ui.adsabs.harvard.edu/\#abs/2017ApJ...835..126B} {835, 126}

\bibitem[\protect\citeauthoryear{{Bigiel}, {Leroy}, {Walter}, {Brinks}, {de
  Blok}, {Madore}  \& {Thornley}}{{Bigiel} et~al.}{2008}]{Bigiel2008}
{Bigiel} F.,  {Leroy} A.,  {Walter} F.,  {Brinks} E.,  {de Blok} W.~J.~G.,
  {Madore} B.,   {Thornley} M.~D.,  2008, \mn@doi [\aj]
  {10.1088/0004-6256/136/6/2846}, \href
  {https://ui.adsabs.harvard.edu/#abs/2008AJ....136.2846B} {136, 2846}

\bibitem[\protect\citeauthoryear{{Carswell} \& {Webb}}{{Carswell} \&
  {Webb}}{2014}]{Carswell2014}
{Carswell} R.~F.,  {Webb} J.~K.,  2014, {VPFIT: Voigt profile fitting program}
  (\mn@eprint {ascl} {1408.015})

\bibitem[\protect\citeauthoryear{{De Cia}, {Ledoux}, {Petitjean}  \&
  {Savaglio}}{{De Cia} et~al.}{2018}]{DeCia2018}
{De Cia} A.,  {Ledoux} C.,  {Petitjean} P.,   {Savaglio} S.,  2018, \mn@doi
  [\aap] {10.1051/0004-6361/201731970}, \href
  {https://ui.adsabs.harvard.edu/\#abs/2018A&A...611A..76D} {611, A76}

\bibitem[\protect\citeauthoryear{{Freudling}, {Romaniello}, {Bramich},
  {Ballester}, {Forchi}, {Garc{\'{\i}}a-Dabl{\'o}}, {Moehler}  \&
  {Neeser}}{{Freudling} et~al.}{2013}]{Freudling2013}
{Freudling} W.,  {Romaniello} M.,  {Bramich} D.~M.,  {Ballester} P.,  {Forchi}
  V.,  {Garc{\'{\i}}a-Dabl{\'o}} C.~E.,  {Moehler} S.,   {Neeser} M.~J.,  2013,
  \mn@doi [\aap] {10.1051/0004-6361/201322494}, \href
  {http://adsabs.harvard.edu/abs/2013A%26A...559A..96F} {559, A96}

\bibitem[\protect\citeauthoryear{{Ge}, {Bechtold}  \& {Kulkarni}}{{Ge}
  et~al.}{2001}]{Ge2001}
{Ge} J.,  {Bechtold} J.,   {Kulkarni} V.~P.,  2001, \mn@doi [\apj]
  {10.1086/318890}, \href
  {https://ui.adsabs.harvard.edu/\#abs/2001ApJ...547L...1G} {547, L1}

\bibitem[\protect\citeauthoryear{{Geier} et~al.,}{{Geier}
  et~al.}{2019}]{Geier2019}
{Geier} S.~J.,  et~al., 2019, \mn@doi [\aap] {10.1051/0004-6361/201935108},
  \href {https://ui.adsabs.harvard.edu/abs/2019A&A...625L...9G} {625, L9}

\bibitem[\protect\citeauthoryear{{Genzel} et~al.,}{{Genzel}
  et~al.}{2012}]{Genzel12}
{Genzel} R.,  et~al., 2012, \mn@doi [\apj] {10.1088/0004-637X/746/1/69}, \href
  {http://adsabs.harvard.edu/abs/2012ApJ...746...69G} {746, 69}

\bibitem[\protect\citeauthoryear{{Gillmon}, {Shull}, {Tumlinson}  \&
  {Danforth}}{{Gillmon} et~al.}{2006}]{Gillmon2006}
{Gillmon} K.,  {Shull} J.~M.,  {Tumlinson} J.,   {Danforth} C.,  2006, \mn@doi
  [\apj] {10.1086/498053}, \href
  {https://ui.adsabs.harvard.edu/\#abs/2006ApJ...636..891G} {636, 891}

\bibitem[\protect\citeauthoryear{{Goodman} \& {Weare}}{{Goodman} \&
  {Weare}}{2010}]{Goodman2010}
{Goodman} J.,  {Weare} J.,  2010, \mn@doi [Communications in Applied
  Mathematics and Computational Science] {10.2140/camcos.2010.5.65}, \href
  {https://ui.adsabs.harvard.edu/#abs/2010CAMCS...5...65G} {5, 65}

\bibitem[\protect\citeauthoryear{{Gordon}, {Clayton}, {Misselt}, {Land olt}  \&
  {Wolff}}{{Gordon} et~al.}{2003}]{Gordon2003}
{Gordon} K.~D.,  {Clayton} G.~C.,  {Misselt} K.~A.,  {Land olt} A.~U.,
  {Wolff} M.~J.,  2003, \mn@doi [\apj] {10.1086/376774}, \href
  {https://ui.adsabs.harvard.edu/abs/2003ApJ...594..279G} {594, 279}

\bibitem[\protect\citeauthoryear{{Gupta} et~al.,}{{Gupta}
  et~al.}{2016}]{GuptaMALS}
{Gupta} N.,  et~al., 2016, in Proceedings of MeerKAT Science: On the Pathway to
  the SKA. 25-27 May, 2016 Stellenbosch, South Africa (MeerKAT2016).. p.~14
  (\mn@eprint {arXiv} {1708.07371})

\bibitem[\protect\citeauthoryear{{Horne}}{{Horne}}{1986}]{Horne1986}
{Horne} K.,  1986, \mn@doi [\pasp] {10.1086/131801}, \href
  {http://adsabs.harvard.edu/abs/1986PASP...98..609H} {98, 609}

\bibitem[\protect\citeauthoryear{{Ivanchik}, {Petitjean}, {Balashev}, {Srianand
  }, {Varshalovich}, {Ledoux}  \& {Noterdaeme}}{{Ivanchik}
  et~al.}{2010}]{Ivanchik2010}
{Ivanchik} A.~V.,  {Petitjean} P.,  {Balashev} S.~A.,  {Srianand } R.,
  {Varshalovich} D.~A.,  {Ledoux} C.,   {Noterdaeme} P.,  2010, \mn@doi
  [\mnras] {10.1111/j.1365-2966.2010.16382.x}, \href
  {https://ui.adsabs.harvard.edu/abs/2010MNRAS.404.1583I} {404, 1583}

\bibitem[\protect\citeauthoryear{{Jenkins} \& {Tripp}}{{Jenkins} \&
  {Tripp}}{2011}]{Jenkins2011}
{Jenkins} E.~B.,  {Tripp} T.~M.,  2011, \mn@doi [\apj]
  {10.1088/0004-637X/734/1/65}, \href
  {https://ui.adsabs.harvard.edu/abs/2011ApJ...734...65J} {734, 65}

\bibitem[\protect\citeauthoryear{{Jorgenson}, {Wolfe}, {Prochaska}  \&
  {Carswell}}{{Jorgenson} et~al.}{2009}]{Jorgenson2009}
{Jorgenson} R.~A.,  {Wolfe} A.~M.,  {Prochaska} J.~X.,   {Carswell} R.~F.,
  2009, \mn@doi [\apj] {10.1088/0004-637X/704/1/247}, \href
  {https://ui.adsabs.harvard.edu/\#abs/2009ApJ...704..247J} {704, 247}

\bibitem[\protect\citeauthoryear{{Jorgenson}, {Murphy}, {Thompson}  \&
  {Carswell}}{{Jorgenson} et~al.}{2014}]{Jorgenson2014}
{Jorgenson} R.~A.,  {Murphy} M.~T.,  {Thompson} R.,   {Carswell} R.~F.,  2014,
  \mn@doi [\mnras] {10.1093/mnras/stu1314}, \href
  {https://ui.adsabs.harvard.edu/abs/2014MNRAS.443.2783J} {443, 2783}

\bibitem[\protect\citeauthoryear{{Khaire} \& {Srianand}}{{Khaire} \&
  {Srianand}}{2019}]{Khaire2019}
{Khaire} V.,  {Srianand} R.,  2019, \mn@doi [Monthly Notices of the Royal
  Astronomical Society] {10.1093/mnras/stz174}, \href
  {https://ui.adsabs.harvard.edu/abs/2019MNRAS.484.4174K} {484, 4174}

\bibitem[\protect\citeauthoryear{{Klimenko}, {Balashev}, {Ivanchik}  \&
  {Varshalovich}}{{Klimenko} et~al.}{2016}]{Klimenko2016}
{Klimenko} V.~V.,  {Balashev} S.~A.,  {Ivanchik} A.~V.,   {Varshalovich} D.~A.,
   2016, \mn@doi [Astronomy Letters] {10.1134/S1063773716030038}, \href
  {https://ui.adsabs.harvard.edu/\#abs/2016AstL...42..137K} {42, 137}

\bibitem[\protect\citeauthoryear{{Krawczyk}, {Richards}, {Gallagher},
  {Leighly}, {Hewett}, {Ross}  \& {Hall}}{{Krawczyk}
  et~al.}{2015}]{Krawczyk2015}
{Krawczyk} C.~M.,  {Richards} G.~T.,  {Gallagher} S.~C.,  {Leighly} K.~M.,
  {Hewett} P.~C.,  {Ross} N.~P.,   {Hall} P.~B.,  2015, \mn@doi [\aj]
  {10.1088/0004-6256/149/6/203}, \href
  {http://adsabs.harvard.edu/abs/2015AJ....149..203K} {149, 203}

\bibitem[\protect\citeauthoryear{{Krogager}}{{Krogager}}{2018}]{Krogager2018a}
{Krogager} J.-K.,  2018, {VoigtFit: Absorption line fitting for Voigt profiles}
  (\mn@eprint {ascl} {1811.016})

\bibitem[\protect\citeauthoryear{{Krogager}, {M{\o}ller}, {Fynbo}  \&
  {Noterdaeme}}{{Krogager} et~al.}{2017}]{Krogager2017}
{Krogager} J.~K.,  {M{\o}ller} P.,  {Fynbo} J.~P.~U.,   {Noterdaeme} P.,  2017,
  \mn@doi [\mnras] {10.1093/mnras/stx1011}, \href
  {https://ui.adsabs.harvard.edu/#abs/2017MNRAS.469.2959K} {469, 2959}

\bibitem[\protect\citeauthoryear{{Krogager} et~al.,}{{Krogager}
  et~al.}{2018a}]{Krogager2018b}
{Krogager} J.~K.,  et~al., 2018a, \mn@doi [\aap] {10.1051/0004-6361/201833608},
  \href {https://ui.adsabs.harvard.edu/abs/2018A&A...619A.142K} {619, A142}

\bibitem[\protect\citeauthoryear{{Krogager} et~al.,}{{Krogager}
  et~al.}{2018b}]{Krogager2018}
{Krogager} J.~K.,  et~al., 2018b, \mn@doi [\aap] {10.1051/0004-6361/201833608},
  \href {https://ui.adsabs.harvard.edu/abs/2018A&A...619A.142K} {619, A142}

\bibitem[\protect\citeauthoryear{{Krogager}, {Fynbo}, {M{\o}ller},
  {Noterdaeme}, {Heintz}  \& {Pettini}}{{Krogager} et~al.}{2019}]{Krogager2019}
{Krogager} J.-K.,  {Fynbo} J. P.~U.,  {M{\o}ller} P.,  {Noterdaeme} P.,
  {Heintz} K.~E.,   {Pettini} M.,  2019, \mn@doi [\mnras]
  {10.1093/mnras/stz1120}, \href
  {https://ui.adsabs.harvard.edu/abs/2019MNRAS.486.4377K} {486, 4377}

\bibitem[\protect\citeauthoryear{{Krumholz}}{{Krumholz}}{2012}]{Krumholz2012}
{Krumholz} M.~R.,  2012, \mn@doi [\apj] {10.1088/0004-637X/759/1/9}, \href
  {http://adsabs.harvard.edu/abs/2012ApJ...759....9K} {759, 9}

\bibitem[\protect\citeauthoryear{{Lacour}, {Ziskin}, {H{\'e}brard}, {Oliveira},
  {Andr{\'e}}, {Ferlet}  \& {Vidal-Madjar}}{{Lacour} et~al.}{2005}]{Lacour2005}
{Lacour} S.,  {Ziskin} V.,  {H{\'e}brard} G.,  {Oliveira} C.,  {Andr{\'e}}
  M.~K.,  {Ferlet} R.,   {Vidal-Madjar} A.,  2005, \mn@doi [\apj]
  {10.1086/430291}, \href
  {https://ui.adsabs.harvard.edu/abs/2005ApJ...627..251L} {627, 251}

\bibitem[\protect\citeauthoryear{{Le Petit}, {Nehm{\'e}}, {Le Bourlot}  \&
  {Roueff}}{{Le Petit} et~al.}{2006}]{LePetit2006}
{Le Petit} F.,  {Nehm{\'e}} C.,  {Le Bourlot} J.,   {Roueff} E.,  2006, \mn@doi
  [\apjs] {10.1086/503252}, \href
  {http://adsabs.harvard.edu/abs/2006ApJS..164..506L} {164, 506}

\bibitem[\protect\citeauthoryear{{Ledoux}, {Petitjean}  \& {Srianand}}{{Ledoux}
  et~al.}{2003}]{Ledoux2003}
{Ledoux} C.,  {Petitjean} P.,   {Srianand} R.,  2003, \mn@doi [\mnras]
  {10.1046/j.1365-2966.2003.07082.x}, \href
  {https://ui.adsabs.harvard.edu/\#abs/2003MNRAS.346..209L} {346, 209}

\bibitem[\protect\citeauthoryear{{Ledoux}, {Noterdaeme}, {Petitjean}  \&
  {Srianand}}{{Ledoux} et~al.}{2015}]{Ledoux2015}
{Ledoux} C.,  {Noterdaeme} P.,  {Petitjean} P.,   {Srianand} R.,  2015, \mn@doi
  [\aap] {10.1051/0004-6361/201424122}, \href
  {https://ui.adsabs.harvard.edu/#abs/2015A&A...580A...8L} {580, A8}

\bibitem[\protect\citeauthoryear{{Liszt}}{{Liszt}}{1981}]{Liszt1981}
{Liszt} H.~S.,  1981, \mn@doi [\apjl] {10.1086/183572}, \href
  {https://ui.adsabs.harvard.edu/abs/1981ApJ...246L.147L} {246, L147}

\bibitem[\protect\citeauthoryear{{Ma} et~al.,}{{Ma} et~al.}{2015}]{Ma2015}
{Ma} J.,  et~al., 2015, \mn@doi [\mnras] {10.1093/mnras/stv2073}, \href
  {https://ui.adsabs.harvard.edu/\#abs/2015MNRAS.454.1751M} {454, 1751}

\bibitem[\protect\citeauthoryear{{Micha{\l}owski} et~al.,}{{Micha{\l}owski}
  et~al.}{2015}]{Michalowski2015}
{Micha{\l}owski} M.~J.,  et~al., 2015, \mn@doi [\aap]
  {10.1051/0004-6361/201526542}, \href
  {https://ui.adsabs.harvard.edu/#abs/2015A&A...582A..78M} {582, A78}

\bibitem[\protect\citeauthoryear{{Muzahid}, {Srianand}  \&
  {Charlton}}{{Muzahid} et~al.}{2015}]{Muzahid2015}
{Muzahid} S.,  {Srianand} R.,   {Charlton} J.,  2015, \mn@doi [\mnras]
  {10.1093/mnras/stv133}, \href
  {https://ui.adsabs.harvard.edu/#abs/2015MNRAS.448.2840M} {448, 2840}

\bibitem[\protect\citeauthoryear{{Muzahid}, {Kacprzak}, {Charlton}  \&
  {Churchill}}{{Muzahid} et~al.}{2016}]{Muzahid2016}
{Muzahid} S.,  {Kacprzak} G.~G.,  {Charlton} J.~C.,   {Churchill} C.~W.,  2016,
  \mn@doi [\apj] {10.3847/0004-637X/823/1/66}, \href
  {https://ui.adsabs.harvard.edu/\#abs/2016ApJ...823...66M} {823, 66}

\bibitem[\protect\citeauthoryear{{Neeleman}, {Prochaska}  \&
  {Wolfe}}{{Neeleman} et~al.}{2015}]{Neeleman2015}
{Neeleman} M.,  {Prochaska} J.~X.,   {Wolfe} A.~M.,  2015, \mn@doi [\apj]
  {10.1088/0004-637X/800/1/7}, \href
  {http://adsabs.harvard.edu/abs/2015ApJ...800....7N} {800, 7}

\bibitem[\protect\citeauthoryear{{Noterdaeme}, {Ledoux}, {Petitjean}, {Le
  Petit}, {Srianand}  \& {Smette}}{{Noterdaeme} et~al.}{2007}]{Noterdaeme2007}
{Noterdaeme} P.,  {Ledoux} C.,  {Petitjean} P.,  {Le Petit} F.,  {Srianand} R.,
    {Smette} A.,  2007, \mn@doi [\aap] {10.1051/0004-6361:20078021}, \href
  {https://ui.adsabs.harvard.edu/#abs/2007A&A...474..393N} {474, 393}

\bibitem[\protect\citeauthoryear{{Noterdaeme}, {Ledoux}, {Petitjean}  \&
  {Srianand}}{{Noterdaeme} et~al.}{2008}]{Noterdaeme2008}
{Noterdaeme} P.,  {Ledoux} C.,  {Petitjean} P.,   {Srianand} R.,  2008, \mn@doi
  [\aap] {10.1051/0004-6361:20078780}, \href
  {https://ui.adsabs.harvard.edu/\#abs/2008A&A...481..327N} {481, 327}

\bibitem[\protect\citeauthoryear{{Noterdaeme}, {Petitjean}, {Ledoux},
  {L{\'o}pez}, {Srianand}  \& {Vergani}}{{Noterdaeme}
  et~al.}{2010}]{Noterdaeme2010}
{Noterdaeme} P.,  {Petitjean} P.,  {Ledoux} C.,  {L{\'o}pez} S.,  {Srianand}
  R.,   {Vergani} S.~D.,  2010, \mn@doi [\aap] {10.1051/0004-6361/201015147},
  \href {https://ui.adsabs.harvard.edu/#abs/2010A&A...523A..80N} {523, A80}

\bibitem[\protect\citeauthoryear{{Noterdaeme}, {Petitjean}, {Srianand},
  {Ledoux}  \& {L{\'o}pez}}{{Noterdaeme} et~al.}{2011}]{Noterdaeme2011}
{Noterdaeme} P.,  {Petitjean} P.,  {Srianand} R.,  {Ledoux} C.,   {L{\'o}pez}
  S.,  2011, \mn@doi [\aap] {10.1051/0004-6361/201016140}, \href
  {https://ui.adsabs.harvard.edu/abs/2011A&A...526L...7N} {526, L7}

\bibitem[\protect\citeauthoryear{{Noterdaeme} et~al.,}{{Noterdaeme}
  et~al.}{2012a}]{Noterdaeme2012a}
{Noterdaeme} P.,  et~al., 2012a, \mn@doi [\aap] {10.1051/0004-6361/201118691},
  \href {https://ui.adsabs.harvard.edu/\#abs/2012A&A...540A..63N} {540, A63}

\bibitem[\protect\citeauthoryear{{Noterdaeme} et~al.,}{{Noterdaeme}
  et~al.}{2012b}]{Noterdaeme2012c}
{Noterdaeme} P.,  et~al., 2012b, \mn@doi [\aap] {10.1051/0004-6361/201220259},
  \href {https://ui.adsabs.harvard.edu/\#abs/2012A&A...547L...1N} {547, L1}

\bibitem[\protect\citeauthoryear{{Noterdaeme}, {Petitjean}, {P{\^a}ris}, {Cai},
  {Finley}, {Ge}, {Pieri}  \& {York}}{{Noterdaeme}
  et~al.}{2014}]{Noterdaeme2014}
{Noterdaeme} P.,  {Petitjean} P.,  {P{\^a}ris} I.,  {Cai} Z.,  {Finley} H.,
  {Ge} J.,  {Pieri} M.~M.,   {York} D.~G.,  2014, \mn@doi [\aap]
  {10.1051/0004-6361/201322809}, \href
  {https://ui.adsabs.harvard.edu/#abs/2014A&A...566A..24N} {566, A24}

\bibitem[\protect\citeauthoryear{{Noterdaeme}, {Petitjean}  \&
  {Srianand}}{{Noterdaeme} et~al.}{2015}]{Noterdaeme2015b}
{Noterdaeme} P.,  {Petitjean} P.,   {Srianand} R.,  2015, \mn@doi [\aap]
  {10.1051/0004-6361/201526018}, \href
  {https://ui.adsabs.harvard.edu/#abs/2015A&A...578L...5N} {578, L5}

\bibitem[\protect\citeauthoryear{{Noterdaeme} et~al.,}{{Noterdaeme}
  et~al.}{2017}]{Noterdaeme2017}
{Noterdaeme} P.,  et~al., 2017, \mn@doi [\aap] {10.1051/0004-6361/201629173},
  \href {https://ui.adsabs.harvard.edu/#abs/2017A&A...597A..82N} {597, A82}

\bibitem[\protect\citeauthoryear{{Noterdaeme}, {Ledoux}, {Zou}, {Petitjean},
  {Srianand}, {Balashev}  \& {L{\'o}pez}}{{Noterdaeme}
  et~al.}{2018}]{Noterdaeme2018}
{Noterdaeme} P.,  {Ledoux} C.,  {Zou} S.,  {Petitjean} P.,  {Srianand} R.,
  {Balashev} S.,   {L{\'o}pez} S.,  2018, \mn@doi [\aap]
  {10.1051/0004-6361/201732266}, \href
  {https://ui.adsabs.harvard.edu/#abs/2018A&A...612A..58N} {612, A58}

\bibitem[\protect\citeauthoryear{{Noterdaeme}, {Balashev}, {Krogager},
  {Srianand }, {Fathivavsari}, {Petitjean}  \& {Ledoux}}{{Noterdaeme}
  et~al.}{2019}]{Noterdaeme2019}
{Noterdaeme} P.,  {Balashev} S.,  {Krogager} J.~K.,  {Srianand } R.,
  {Fathivavsari} H.,  {Petitjean} P.,   {Ledoux} C.,  2019, \mn@doi [\aap]
  {10.1051/0004-6361/201935371}, 627, A32

\bibitem[\protect\citeauthoryear{{Petitjean}, {Ledoux}, {Noterdaeme}  \&
  {Srianand}}{{Petitjean} et~al.}{2006}]{Petitjean2006}
{Petitjean} P.,  {Ledoux} C.,  {Noterdaeme} P.,   {Srianand} R.,  2006, \mn@doi
  [\aap] {10.1051/0004-6361:20065769}, \href
  {https://ui.adsabs.harvard.edu/\#abs/2006A&A...456L...9P} {456, L9}

\bibitem[\protect\citeauthoryear{{Rachford} et~al.,}{{Rachford}
  et~al.}{2009}]{Rachford2009}
{Rachford} B.~L.,  et~al., 2009, \mn@doi [The Astrophysical Journal Supplement
  Series] {10.1088/0067-0049/180/1/125}, \href
  {https://ui.adsabs.harvard.edu/\#abs/2009ApJS..180..125R} {180, 125}

\bibitem[\protect\citeauthoryear{{Rafelski}, {Wolfe}, {Prochaska}, {Neeleman}
  \& {Mendez}}{{Rafelski} et~al.}{2012}]{Rafelski2012}
{Rafelski} M.,  {Wolfe} A.~M.,  {Prochaska} J.~X.,  {Neeleman} M.,   {Mendez}
  A.~J.,  2012, \mn@doi [\apj] {10.1088/0004-637X/755/2/89}, \href
  {https://ui.adsabs.harvard.edu/\#abs/2012ApJ...755...89R} {755, 89}

\bibitem[\protect\citeauthoryear{{Ranjan} et~al.,}{{Ranjan}
  et~al.}{2018}]{Ranjan2018}
{Ranjan} A.,  et~al., 2018, \mn@doi [\aap] {10.1051/0004-6361/201833446}, \href
  {https://ui.adsabs.harvard.edu/abs/2018A&A...618A.184R} {618, A184}

\bibitem[\protect\citeauthoryear{{Ranjan}, {Noterdaeme}, {Krogager},
  {Petitjean}, {Srianand}, {Balashev}, {Gupta}  \& {Ledoux}}{{Ranjan}
  et~al.}{2019}]{Ranjan2019}
{Ranjan} A.,  {Noterdaeme} P.,  {Krogager} J.-K.,  {Petitjean} P.,  {Srianand}
  R.,  {Balashev} S.,  {Gupta} N.,   {Ledoux} C.,  2019, submitted to \aap

\bibitem[\protect\citeauthoryear{{Rawlins}, {Srianand}, {Shaw}, {Rahmani},
  {Dutta}  \& {Chacko}}{{Rawlins} et~al.}{2018}]{Rawlins2018}
{Rawlins} K.,  {Srianand} R.,  {Shaw} G.,  {Rahmani} H.,  {Dutta} R.,
  {Chacko} S.,  2018, \mn@doi [\mnras] {10.1093/mnras/sty2321}, \href
  {https://ui.adsabs.harvard.edu/abs/2018MNRAS.481.2083R} {481, 2083}

\bibitem[\protect\citeauthoryear{{Roy}, {Chengalur}  \& {Srianand}}{{Roy}
  et~al.}{2006}]{Roy2006}
{Roy} N.,  {Chengalur} J.~N.,   {Srianand} R.,  2006, \mn@doi [\mnras]
  {10.1111/j.1745-3933.2005.00114.x}, \href
  {https://ui.adsabs.harvard.edu/abs/2006MNRAS.365L...1R} {365, L1}

\bibitem[\protect\citeauthoryear{{Savage}, {Bohlin}, {Drake}  \&
  {Budich}}{{Savage} et~al.}{1977}]{Savage1977}
{Savage} B.~D.,  {Bohlin} R.~C.,  {Drake} J.~F.,   {Budich} W.,  1977, \mn@doi
  [\apj] {10.1086/155471}, \href
  {https://ui.adsabs.harvard.edu/\#abs/1977ApJ...216..291S} {216, 291}

\bibitem[\protect\citeauthoryear{{Schruba} et~al.,}{{Schruba}
  et~al.}{2011}]{Schruba11}
{Schruba} A.,  et~al., 2011, \mn@doi [\aj] {10.1088/0004-6256/142/2/37}, \href
  {http://adsabs.harvard.edu/abs/2011AJ....142...37S} {142, 37}

\bibitem[\protect\citeauthoryear{{Selsing}, {Fynbo}, {Christensen}  \&
  {Krogager}}{{Selsing} et~al.}{2016}]{Selsing2016}
{Selsing} J.,  {Fynbo} J.~P.~U.,  {Christensen} L.,   {Krogager} J.~K.,  2016,
  \mn@doi [\aap] {10.1051/0004-6361/201527096}, \href
  {https://ui.adsabs.harvard.edu/abs/2016A&A...585A..87S} {585, A87}

\bibitem[\protect\citeauthoryear{{Shaw}, {Rawlins}  \& {Srianand}}{{Shaw}
  et~al.}{2016}]{Shaw2016}
{Shaw} G.,  {Rawlins} K.,   {Srianand} R.,  2016, \mn@doi [\mnras]
  {10.1093/mnras/stw788}, \href
  {https://ui.adsabs.harvard.edu/abs/2016MNRAS.459.3234S} {459, 3234}

\bibitem[\protect\citeauthoryear{{Silva} \& {Viegas}}{{Silva} \&
  {Viegas}}{2002}]{Silva2002}
{Silva} A.~I.,  {Viegas} S.~M.,  2002, \mn@doi [\mnras]
  {10.1046/j.1365-8711.2002.04956.x}, \href
  {https://ui.adsabs.harvard.edu/abs/2002MNRAS.329..135S} {329, 135}

\bibitem[\protect\citeauthoryear{{Srianand}, {Petitjean}, {Ledoux}, {Ferland}
  \& {Shaw}}{{Srianand} et~al.}{2005}]{Srianand2005}
{Srianand} R.,  {Petitjean} P.,  {Ledoux} C.,  {Ferland} G.,   {Shaw} G.,
  2005, \mn@doi [\mnras] {10.1111/j.1365-2966.2005.09324.x}, \href
  {https://ui.adsabs.harvard.edu/abs/2005MNRAS.362..549S} {362, 549}

\bibitem[\protect\citeauthoryear{{Srianand}, {Noterdaeme}, {Ledoux}  \&
  {Petitjean}}{{Srianand} et~al.}{2008}]{Srianand2008}
{Srianand} R.,  {Noterdaeme} P.,  {Ledoux} C.,   {Petitjean} P.,  2008, \mn@doi
  [\aap] {10.1051/0004-6361:200809727}, \href
  {https://ui.adsabs.harvard.edu/#abs/2008A&A...482L..39S} {482, L39}

\bibitem[\protect\citeauthoryear{{Srianand}, {Gupta}, {Petitjean},
  {Noterdaeme}, {Ledoux}, {Salter}  \& {Saikia}}{{Srianand}
  et~al.}{2012}]{Srianand2012}
{Srianand} R.,  {Gupta} N.,  {Petitjean} P.,  {Noterdaeme} P.,  {Ledoux} C.,
  {Salter} C.~J.,   {Saikia} D.~J.,  2012, \mn@doi [\mnras]
  {10.1111/j.1365-2966.2011.20342.x}, \href
  {https://ui.adsabs.harvard.edu/\#abs/2012MNRAS.421..651S} {421, 651}

\bibitem[\protect\citeauthoryear{{Sternberg}, {Le Petit}, {Roueff}  \& {Le
  Bourlot}}{{Sternberg} et~al.}{2014}]{Sternberg2014}
{Sternberg} A.,  {Le Petit} F.,  {Roueff} E.,   {Le Bourlot} J.,  2014, \mn@doi
  [\apj] {10.1088/0004-637X/790/1/10}, \href
  {https://ui.adsabs.harvard.edu/#abs/2014ApJ...790...10S} {790, 10}

\bibitem[\protect\citeauthoryear{{Vernet} et~al.,}{{Vernet}
  et~al.}{2011}]{Vernet2011}
{Vernet} J.,  et~al., 2011, \mn@doi [\aap] {10.1051/0004-6361/201117752}, \href
  {https://ui.adsabs.harvard.edu/#abs/2011A&A...536A.105V} {536, A105}

\bibitem[\protect\citeauthoryear{{Vladilo}, {Centuri{\'o}n}, {Levshakov},
  {P{\'e}roux}, {Khare}, {Kulkarni}  \& {York}}{{Vladilo}
  et~al.}{2006}]{Vladilo2006}
{Vladilo} G.,  {Centuri{\'o}n} M.,  {Levshakov} S.~A.,  {P{\'e}roux} C.,
  {Khare} P.,  {Kulkarni} V.~P.,   {York} D.~G.,  2006, \mn@doi [\aap]
  {10.1051/0004-6361:20054742}, \href
  {https://ui.adsabs.harvard.edu/abs/2006A&A...454..151V} {454, 151}

\bibitem[\protect\citeauthoryear{{Vladilo}, {Abate}, {Yin}, {Cescutti}  \&
  {Matteucci}}{{Vladilo} et~al.}{2011}]{Vladilo2011}
{Vladilo} G.,  {Abate} C.,  {Yin} J.,  {Cescutti} G.,   {Matteucci} F.,  2011,
  \mn@doi [\aap] {10.1051/0004-6361/201016330}, \href
  {https://ui.adsabs.harvard.edu/abs/2011A&A...530A..33V} {530, A33}

\bibitem[\protect\citeauthoryear{{Welty}, {Xue}  \& {Wong}}{{Welty}
  et~al.}{2012}]{Welty2012}
{Welty} D.~E.,  {Xue} R.,   {Wong} T.,  2012, \mn@doi [\apj]
  {10.1088/0004-637X/745/2/173}, \href
  {https://ui.adsabs.harvard.edu/\#abs/2012ApJ...745..173W} {745, 173}

\bibitem[\protect\citeauthoryear{{Welty}, {Lauroesch}, {Wong}  \&
  {York}}{{Welty} et~al.}{2016}]{Welty2016}
{Welty} D.~E.,  {Lauroesch} J.~T.,  {Wong} T.,   {York} D.~G.,  2016, \mn@doi
  [\apj] {10.3847/0004-637X/821/2/118}, \href
  {https://ui.adsabs.harvard.edu/abs/2016ApJ...821..118W} {821, 118}

\bibitem[\protect\citeauthoryear{{Wolfe}, {Gawiser}  \& {Prochaska}}{{Wolfe}
  et~al.}{2005}]{Wolfe2005}
{Wolfe} A.~M.,  {Gawiser} E.,   {Prochaska} J.~X.,  2005, \mn@doi [Annual
  Review of Astronomy and Astrophysics]
  {10.1146/annurev.astro.42.053102.133950}, \href
  {https://ui.adsabs.harvard.edu/#abs/2005ARA&A..43..861W} {43, 861}

\bibitem[\protect\citeauthoryear{{Wolfire}, {McKee}, {Hollenbach}  \&
  {Tielens}}{{Wolfire} et~al.}{2003}]{Wolfire2003}
{Wolfire} M.~G.,  {McKee} C.~F.,  {Hollenbach} D.,   {Tielens} A.~G.~G.~M.,
  2003, \mn@doi [\apj] {10.1086/368016}, \href
  {https://ui.adsabs.harvard.edu/abs/2003ApJ...587..278W} {587, 278}

\bibitem[\protect\citeauthoryear{{York} et~al.,}{{York}
  et~al.}{2000}]{York2000}
{York} D.~G.,  et~al., 2000, \mn@doi [\aj] {10.1086/301513}, \href
  {https://ui.adsabs.harvard.edu/abs/2000AJ....120.1579Y} {120, 1579}

\bibitem[\protect\citeauthoryear{{Zou}, {Petitjean}, {Noterdaeme}, {Ledoux},
  {Krogager}, {Fathivavsari}, {Srianand}  \& {L{\'o}pez}}{{Zou}
  et~al.}{2018}]{Zou2018}
{Zou} S.,  {Petitjean} P.,  {Noterdaeme} P.,  {Ledoux} C.,  {Krogager} J.~K.,
  {Fathivavsari} H.,  {Srianand} R.,   {L{\'o}pez} S.,  2018, \mn@doi [\aap]
  {10.1051/0004-6361/201732033}, \href
  {https://ui.adsabs.harvard.edu/\#abs/2018A&A...616A.158Z} {616, A158}

\bibitem[\protect\citeauthoryear{{Zwaan} \& {Prochaska}}{{Zwaan} \&
  {Prochaska}}{2006}]{Zwaan2006}
{Zwaan} M.~A.,  {Prochaska} J.~X.,  2006, \mn@doi [\apj] {10.1086/503191},
  \href {https://ui.adsabs.harvard.edu/#abs/2006ApJ...643..675Z} {643, 675}

\makeatother
\end{thebibliography}

\appendix
\section{Supplementary material}


\label{sect:DetailedAnalisys}
In this section we present the detailed results of the analysis of each DLA individually.




\subsection{\Jone}

\begin{figure}
\centering
\includegraphics[trim={0.0cm 0.0cm 0.0cm 0.0cm},clip,width=\hsize]{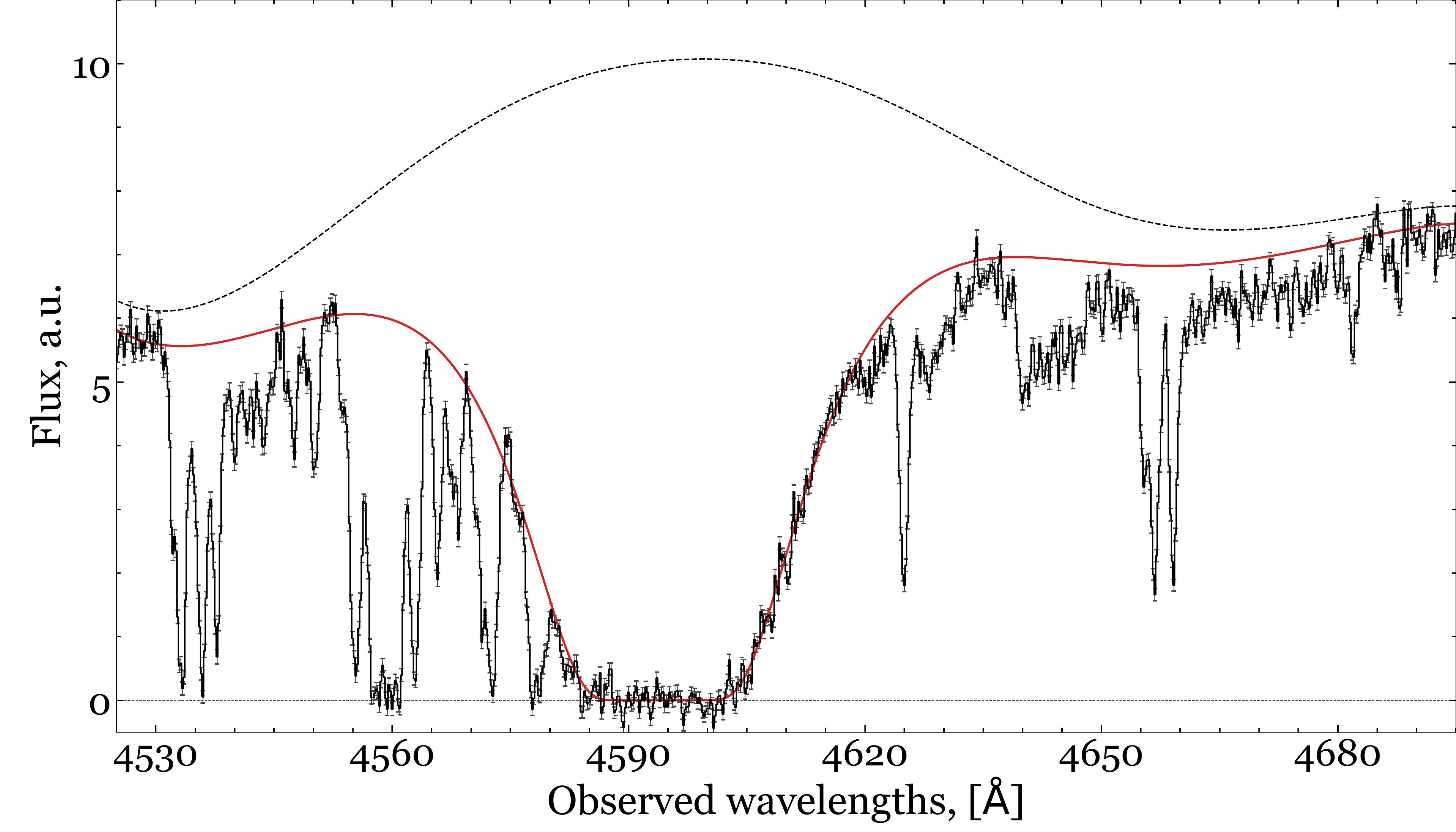}
		\caption{Fit to \HI\ Ly$\alpha$ absorption line in DLA at $z=2.779$ towards \Jones.
		\label{fig:J0136_HI}}
\end{figure}

We measured the \HI\ column density of the DLA at $z=2.779$ towards \Jones\ by fitting the Ly$\alpha$ absorption line. We fitted together the continuum using sixth order Chebyshev polynomial in the region of Ly$\alpha$ line with a one-component absorption model. We derived $\log N(\HI)=20.73\pm0.01$ at the best fit value $z=2.77916$. The Ly$\alpha$ profile, together with the reconstructed quasar continuum are shown in Fig.~\ref{fig:J0136_HI}. We note that the Ly$\alpha$ emission line almost coincides with the damped Ly$\alpha$ absorption and is partially reconstructed from continuum fitting but that its exact shape remains uncertain at the line center where the DLA completely absorb the corresponding photons.

\begin{figure*}
\centering
\includegraphics[trim={0.0cm 0.0cm 0.0cm 0.0cm},clip,width=\hsize]{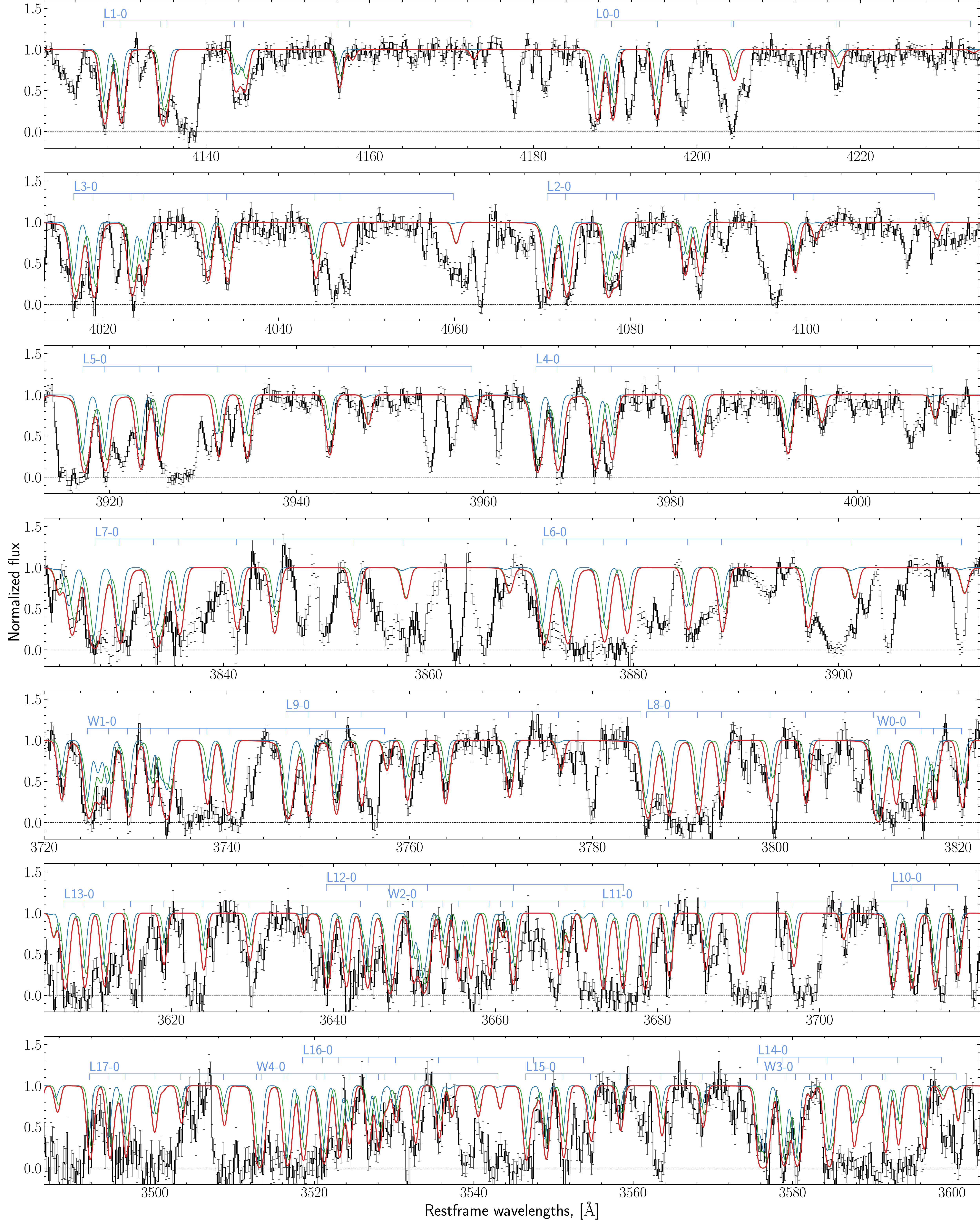}
		\caption{Fit of H$_2$ absorption lines in the DLA at $z=2.77924$ towards \Jones.
		\label{fig:J0136_H2}}
\end{figure*}

\begin{figure}
\centering
\includegraphics[trim={0.0cm 0.0cm 0.0cm 0.0cm},clip,width=\hsize]{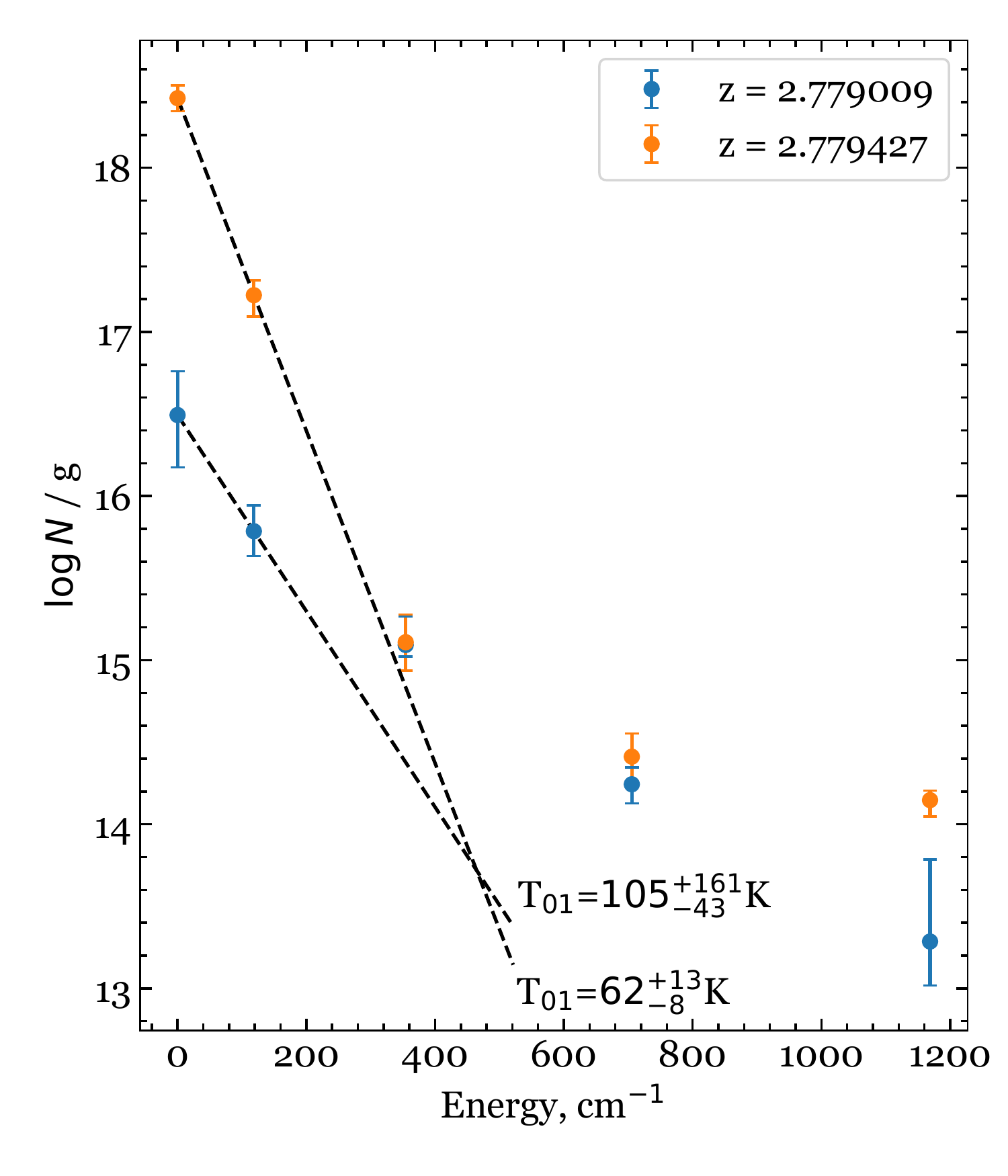}
		\caption{Excitation diagram for the two H$_2$ components at $z=2.779$ towards \Jones. 
		\label{fig:J0136_H2_exc}}
\end{figure}

The fit of H$_2$ absorption lines was done using a two components model. The result of the fit is given in Table~\ref{table:J0136_H2} and H$_2$ line profiles are shown in Fig.~\ref{fig:J0136_H2}. The excitation diagrams of H$_2$ rotation levels for both components are shown in Fig.~\ref{fig:J0136_H2_exc}, with derived $T_{01}$ temperatures   
of $102^{+161}_{-43}$\,K and $62^{+13}_{-8}$\,K at $z=2.77943$ and 2.77901, respectively.  

\begin{figure}
    \centering
    \includegraphics[trim={0.0cm 0.0cm 0.0cm 0.0cm},clip,width=\hsize]{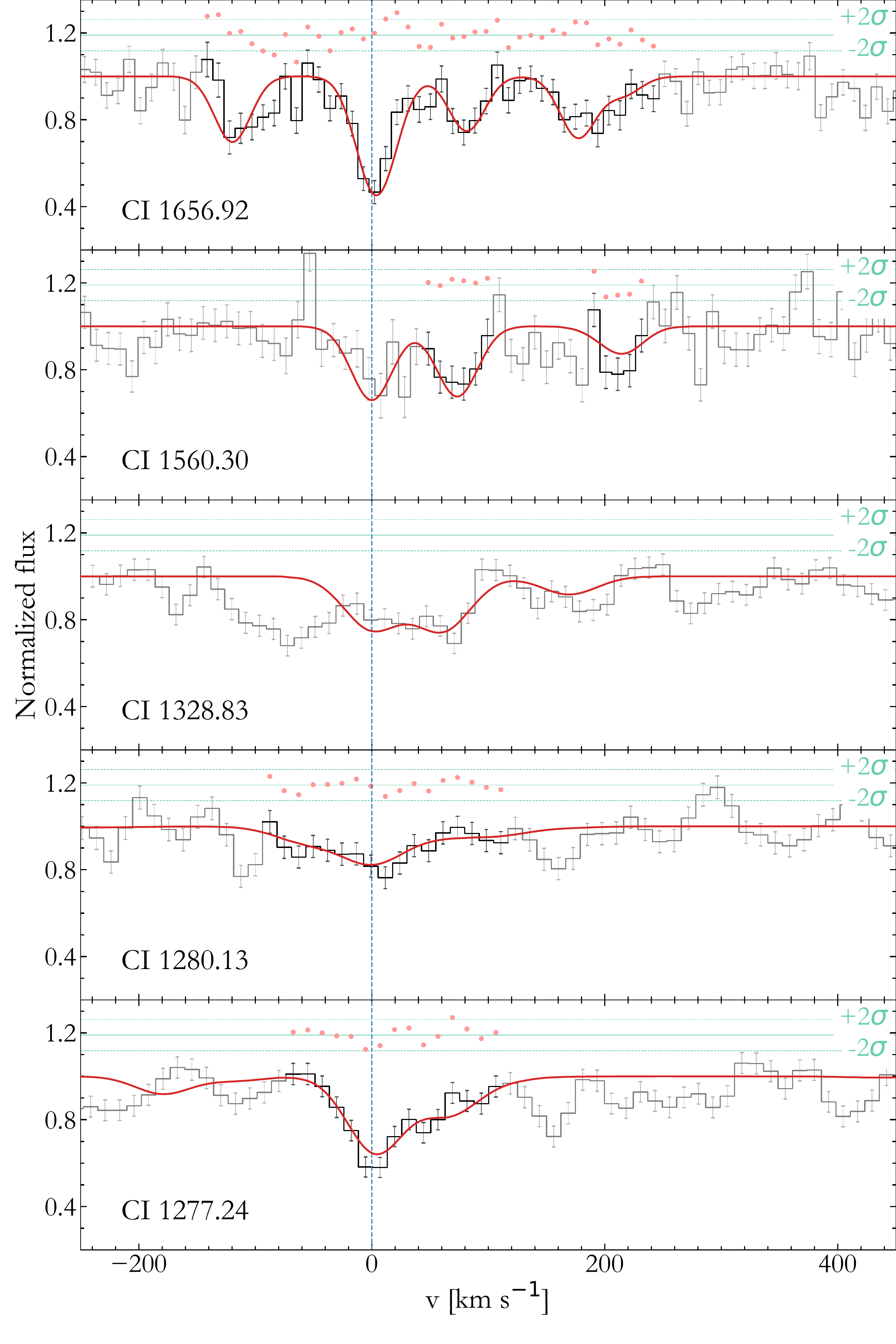}
	\caption{Fit to \CI\ absorption lines at $z=2.779$ towards \Jones.}
	\label{fig:J0136_CI}
\end{figure}

We detected \CI\ absorption lines in a single component arising from three fine structure levels, coinciding with the main H$_2$ component at $z=2.77943$. 
The result of the fit is given in Table~\ref{table:J0136_CI} and the \CI\ absorption profiles are shown in Fig.~\ref{fig:J0136_CI}.

\begin{figure}
\centering
\includegraphics[trim={0.0cm 0.0cm 0.0cm 0.0cm},clip,width=\hsize]{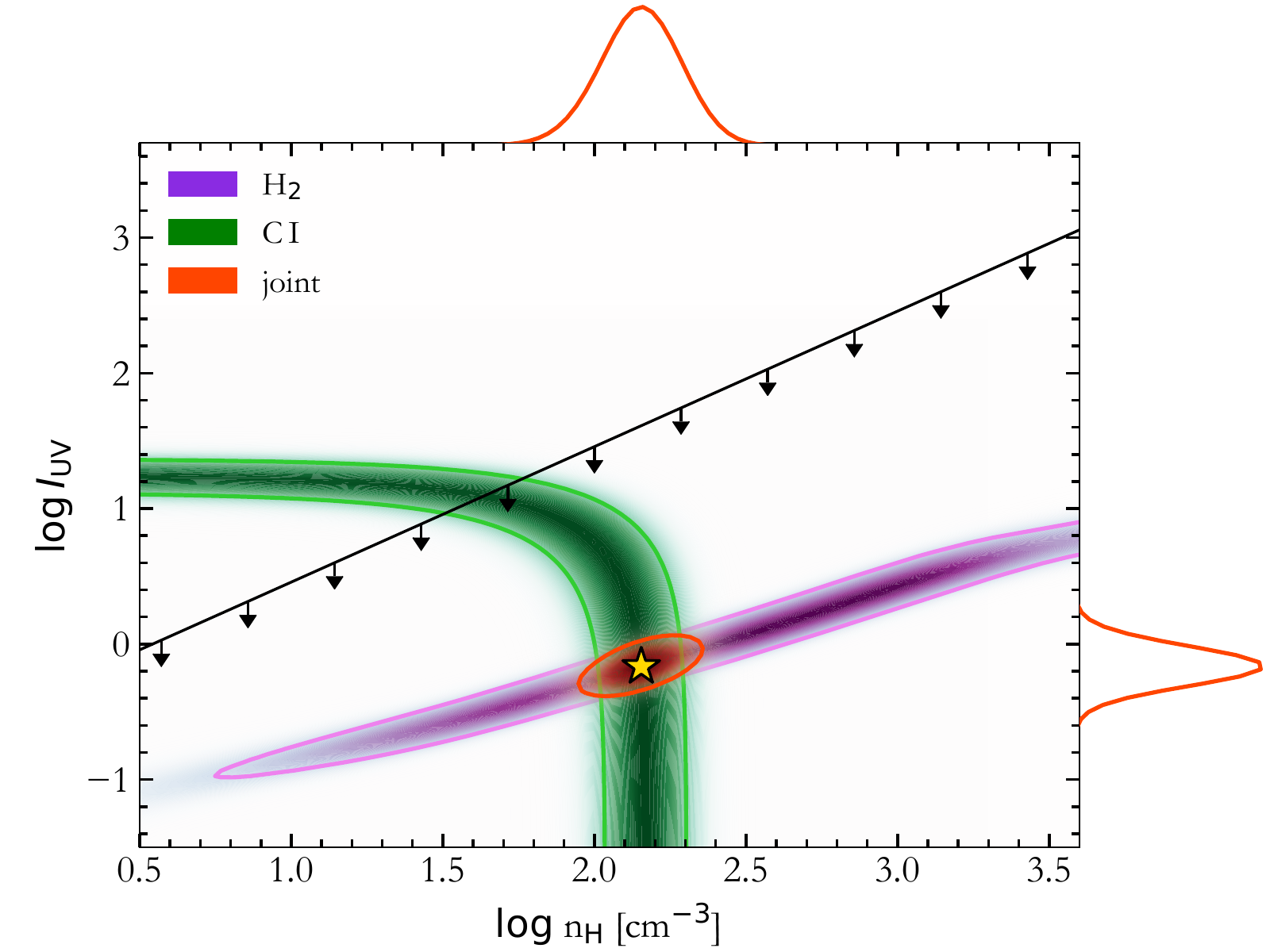}
		\caption{Constraints on the number density and UV field in the H$_2$-bearing medium at $z=2.77943$ 
		towards \Jones\ from the excitation of \CI\ fine-structure and H$_2$ rotational levels. The colors and lines are the same as in Fig.~2 }
\label{fig:J0136_phys}
\end{figure}

The constrained values of the hydrogen number density and UV field in the H$_2$/\CI-bearing medium, based on the excitation of H$_2$ and \CI\ levels are shown in Fig.~\ref{fig:J0136_phys}.  The comparison between observed and model column densities at the best fit value of the parameters (shown by yellow star in Fig.~\ref{fig:J0136_phys}) is provided in Table~\ref{table:J0136_comp}.

\begin{figure*}
\centering
\includegraphics[trim={0.0cm 0.0cm 0.0cm 0.0cm},clip,width=\hsize]{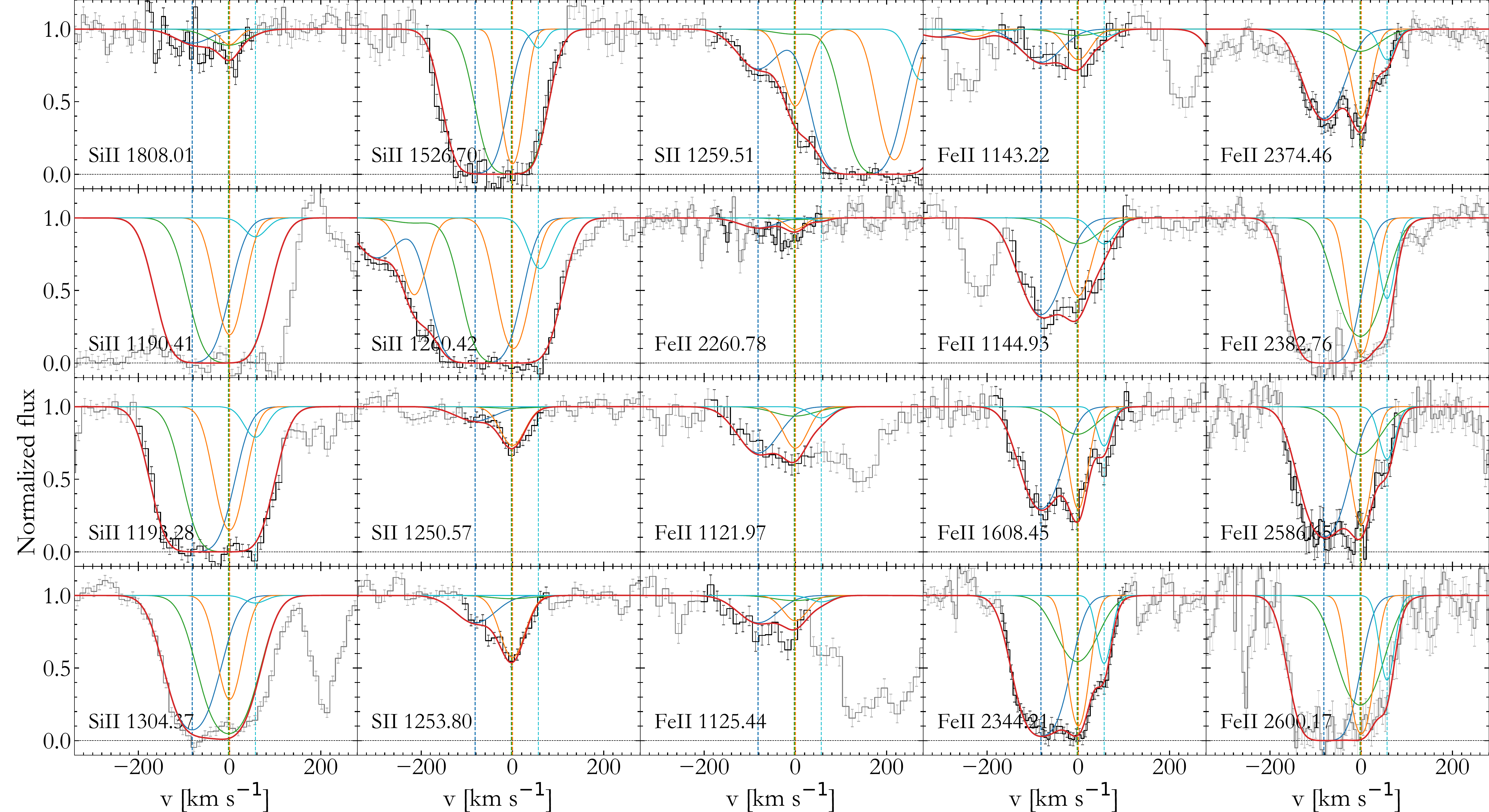}
		\caption{Fit to the metal absorption lines at $z=2.779$ towards \Jones.}
		\label{fig:J0136_metals}
\end{figure*}

Metal lines are fitted using a four components model. 
We estimated the metallicity in this system using \ion{S}{ii}. The results of the fit are given in Table~\ref{table:J0136_metals} and fit profiles are shown in Fig.~\ref{fig:J0136_metals}.

\begin{table}
\centering
\caption{Fitting results of H$_2$ towards \Jones.
\label{table:J0136_H2}
}
\begin{tabular}{cccc}
\hline
comp. & 1 & 2 & $\log N_{\rm tot}$ \\
\hline
z & $2.77901(^{+7}_{-4})$ & $2.77943(^{+4}_{-7})$ &  \\
$\Delta$v, km\,s$^{-1}$ & 0.0 & 29.9 &  \\
b, km\,s$^{-1}$ & $13.2^{+1.7}_{-3.2}$ & $7.7^{+2.4}_{-1.9}$ &  \\
$\log N$(H$_2$ J0) & $16.49^{+0.32}_{-0.27}$ & $18.42^{+0.08}_{-0.08}$ & $18.43^{+0.08}_{-0.07}$ \\
$\log N$(H$_2$ J1) & $16.74^{+0.15}_{-0.16}$ & $18.18^{+0.13}_{-0.09}$ & $18.19^{+0.13}_{-0.08}$ \\
$\log N$(H$_2$ J2) & $15.79^{+0.07}_{-0.17}$ & $15.81^{+0.17}_{-0.17}$ & $16.04^{+0.10}_{-0.05}$ \\
$\log N$(H$_2$ J3) & $15.57^{+0.12}_{-0.10}$ & $15.73^{+0.19}_{-0.14}$ & $15.94^{+0.09}_{-0.07}$ \\
$\log N$(H$_2$ J4) & $14.24^{+0.27}_{-0.50}$ & $15.10^{+0.10}_{-0.06}$ & $15.13^{+0.10}_{-0.06}$ \\
$\log N_{\rm tot}$ & $16.97^{+0.21}_{-0.14}$ & $18.64^{+0.06}_{-0.08}$ & $18.65^{+0.06}_{-0.07}$ \\
\hline
\end{tabular}

\end{table}

\begin{table}
\centering
\caption{Fit results of \CI\ towards \Jones.
\label{table:J0136_CI}}
\begin{tabular}{cc}
\hline
z & $2.779311(^{+7}_{-9})$ \\
b, km\,s$^{-1}$ & $4.4^{+0.2}_{-0.5}$ \\
$\log N$(CI) & $14.44^{+0.10}_{-0.12}$ \\
$\log N$(CI*) & $14.08^{+0.05}_{-0.05}$ \\
$\log N$(CI**) & $13.31^{+0.06}_{-0.06}$ \\
$\log N_{\rm tot}$ & $14.61^{+0.08}_{-0.09}$ \\
\hline
\end{tabular}

\end{table}

\begin{table}
\centering
\caption{Comparison of observed column densities with the 
in the DLA towards \Jones\ with the 
modelled values calculated at the best fit value of number density and UV field from Fig.~\ref{fig:J0136_phys}. \label{table:J0136_comp}
}
\begin{tabular}{ccc}
\hline
species & \multicolumn{2}{c}{$\log N$}\\ 
        & observed & model \\
\hline
CI & $14.44^{+0.10}_{-0.12}$ & 14.35$^a$ \\
CI* & $14.08^{+0.05}_{-0.05}$ & 14.20$^a$ \\
CI** & $13.31^{+0.06}_{-0.06}$ & 13.52$^a$ \\
H$_2$ J=0 & $18.42^{+0.08}_{-0.08}$ & 18.43 \\
H$_2$ J=1 & $18.18^{+0.13}_{-0.09}$ & 18.15 \\
H$_2$ J=2 & $15.81^{+0.17}_{-0.17}$ & 15.54 \\
\hline
H$_2$ J=3$^b$ & & 14.38 \\
H$_2$ J=4$^b$ & & 13.28 \\
\hline
\end{tabular}
\begin{tablenotes}
    \item (a) These values were obtained under then the additional assumption that the total \CI\ column density equates the observed one.     
    \item (b) We did not use these levels to compare with observed values, since the column densities measurements for these levels are subject to a large systematic uncertainty (see text).
\end{tablenotes}
\end{table}

\begin{table*}
\centering
\caption{Fit results of metals towards \Jones. 
\label{table:J0136_metals}}
\begin{tabular}{cccccccc}
\hline
comp & 1 & 2 & 3 & 4 & $\log N_{\rm tot}$ & $\rm [X/H]$ & $\rm [X/SII]$ \\
\hline
z & $2.778217(^{+31}_{-43})$ & $2.77922(^{+9}_{-14})$ & $2.779240(^{+15}_{-12})$ & $2.779960(^{+15}_{-23})$ &  &  &  \\
$\Delta$v, km\,s$^{-1}$ & -81.1 & -1.7 & 0.0 & 57.1 &  &  &  \\
b, km\,s$^{-1}$ & $51.0^{+1.8}_{-2.1}$ & $53.4^{+5.2}_{-5.2}$ & $19.1^{+1.6}_{-4.6}$ & $7.4^{+2.3}_{-2.9}$ &  &  &  \\
$\log N$(SiII) & $14.99^{+0.06}_{-0.09}$ & $15.19^{+0.11}_{-0.16}$ & $14.84^{+0.19}_{-0.36}$ & $<14.0$ & $15.45^{+0.04}_{-0.03}$ & $-0.80^{+0.04}_{-0.03}$ & $-0.22^{+0.04}_{-0.05}$ \\
$\log N$(SII) & $14.76^{+0.05}_{-0.05}$ & $<14.8$ & $15.11^{+0.05}_{-0.06}$ &  & $15.28^{+0.03}_{-0.02}$ & $-0.58^{+0.03}_{-0.03}$ & $0.00^{+0.04}_{-0.04}$ \\
$\log N$(FeII) & $14.68^{+0.03}_{-0.03}$ & $14.18^{+0.26}_{-0.53}$ & $14.49^{+0.05}_{-0.11}$ & $13.82^{+0.07}_{-0.09}$ & $14.96^{+0.01}_{-0.01}$ & $-1.28^{+0.02}_{-0.02}$ & $-0.70^{+0.03}_{-0.04}$ \\
\hline
\end{tabular}

\end{table*}

\clearpage

\subsection{\Jtwo}

\begin{figure}
\centering
\includegraphics[trim={0.0cm 0.0cm 0.0cm 0.0cm},clip,width=\hsize]{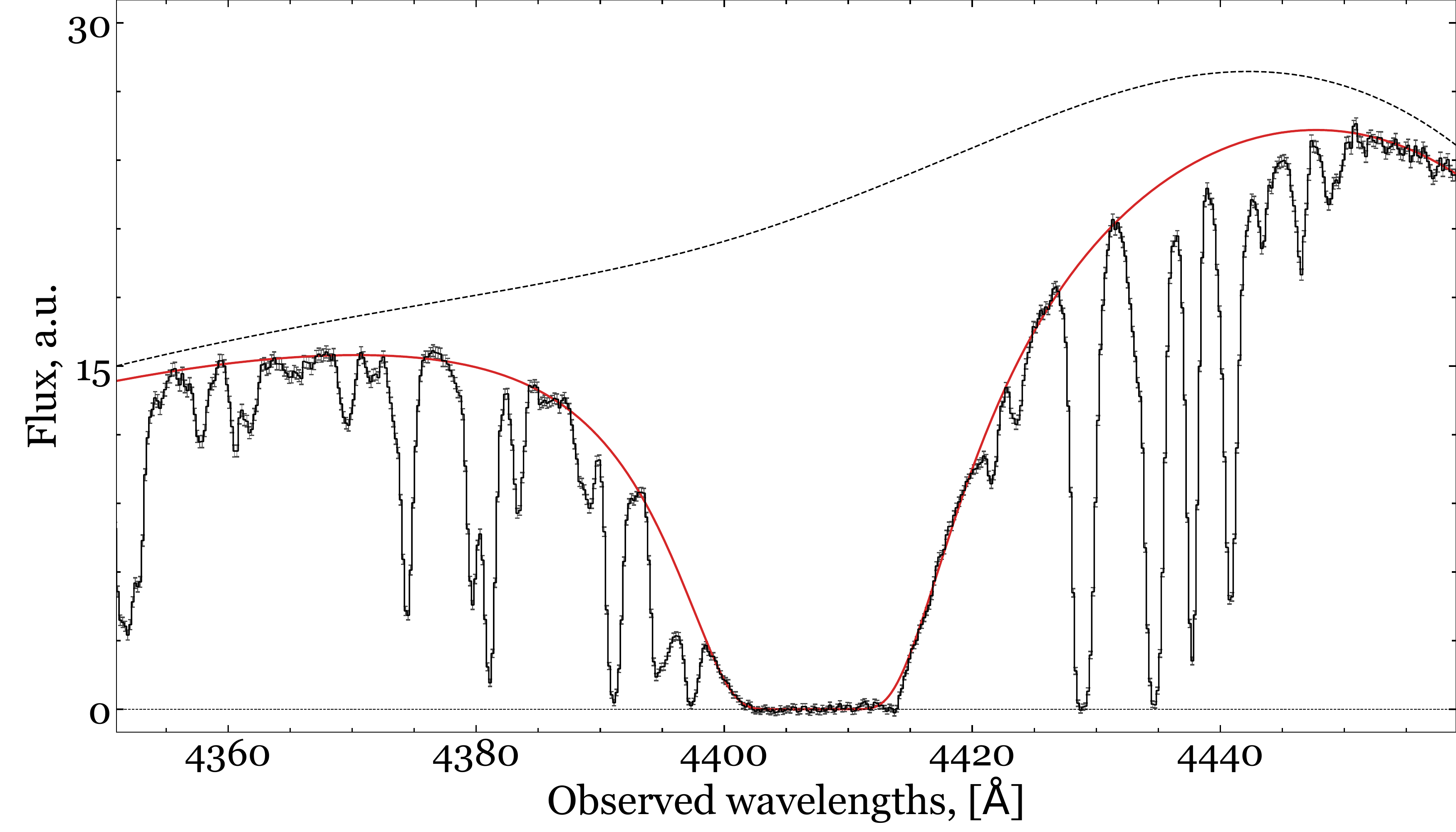}
		\caption{Fit to the damped \HI\ Ly$\alpha$ absorption line at $z=2.625$ towards \Jtwos.
		\label{fig:J0858_HI}}
\end{figure}

We measured the \HI\ column density of the DLA at $z=2.625$ towards \Jtwos\ by fitting the Ly$\alpha$ absorption line with a single component together with the continuum modelled using a sixth order Chebyshev polynomial. We derived $\log N(\HI)=20.40\pm0.01$ at the best fit value of $z=2.62526$. The Ly$\alpha$ profile, together with the reconstructed quasar continuum are shown in Fig.~\ref{fig:J0858_HI}.

\begin{figure*}
\centering
\includegraphics[trim={0.0cm 0.0cm 0.0cm 0.0cm},clip,width=\hsize]{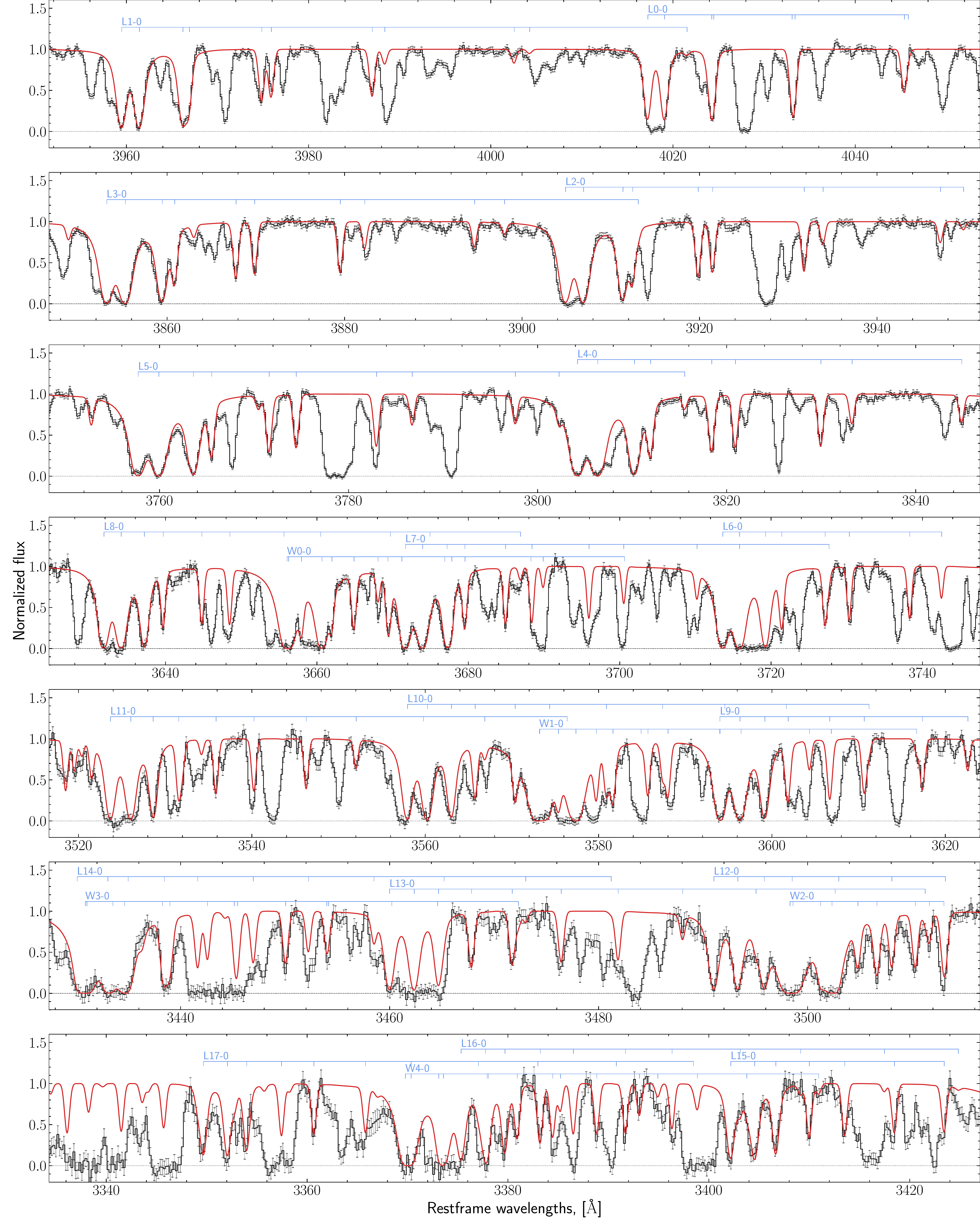}
		\caption{Fit to the H$_2$ absorption lines at $z=2.625$ towards \Jtwos.
		\label{fig:J0858_H2}}
\end{figure*}

\begin{figure}
\centering
\includegraphics[trim={0.0cm 0.0cm 0.0cm 0.0cm},clip,width=\hsize]{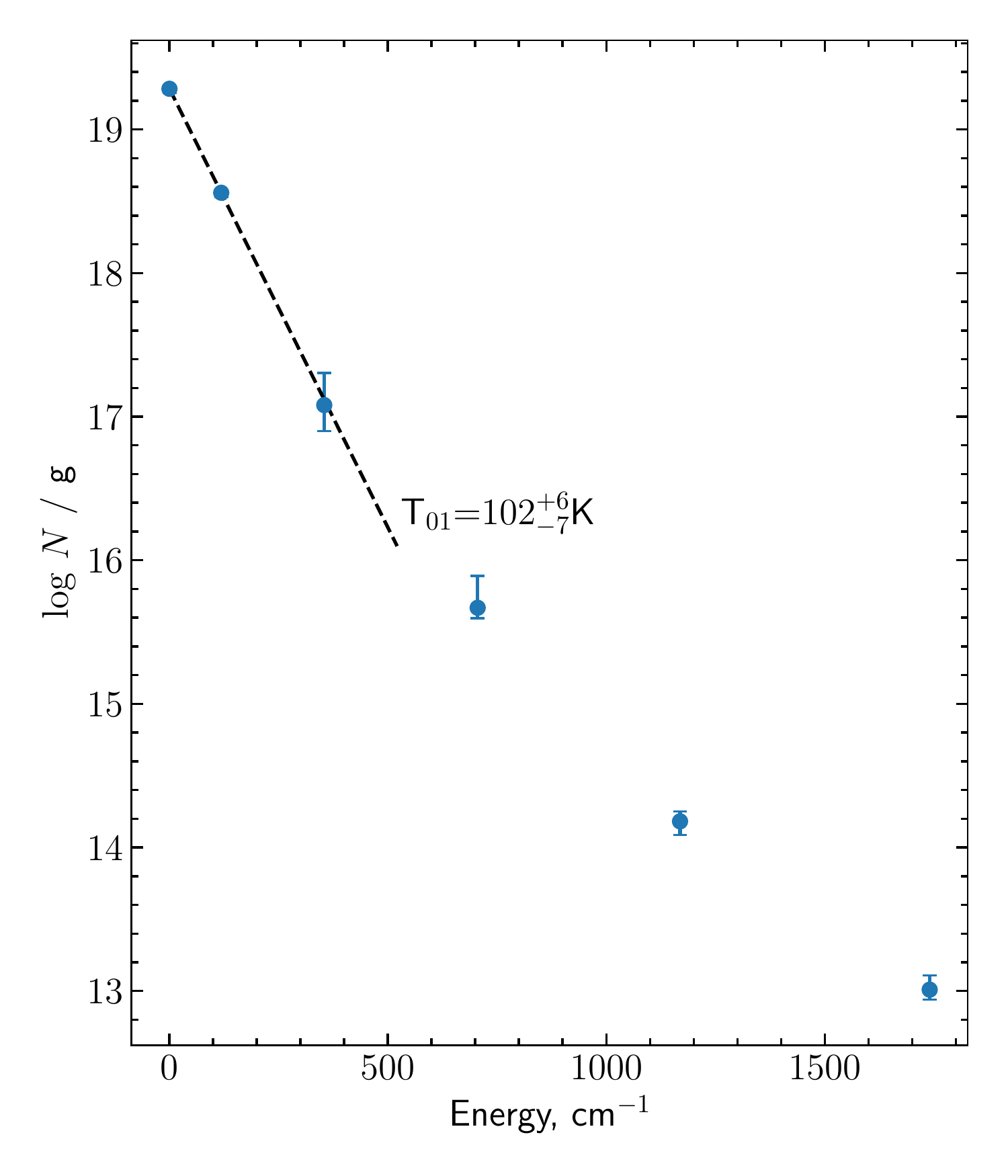}
		\caption{Excitation diagram of H$_2$ at $z=2.625$ towards \Jtwos.
		\label{fig:J0858_H2_exc}}
\end{figure}

We fitted the H$_2$ absorption lines towards \Jtwo\ using a one component model. 
The result of the fit is given in Table~\ref{table:J0858_H2} and H$_2$ line profiles are shown in Fig.~\ref{fig:J0858_H2}. The excitation diagram of H$_2$ rotation levels is shown in Fig.~\ref{fig:J0858_H2_exc} and we derived $T_{01} = 102^{+6}_{-7}$~K. 

\begin{figure}
    \centering
    \includegraphics[trim={0.0cm 0.0cm 0.0cm 0.0cm},clip,width=\hsize]{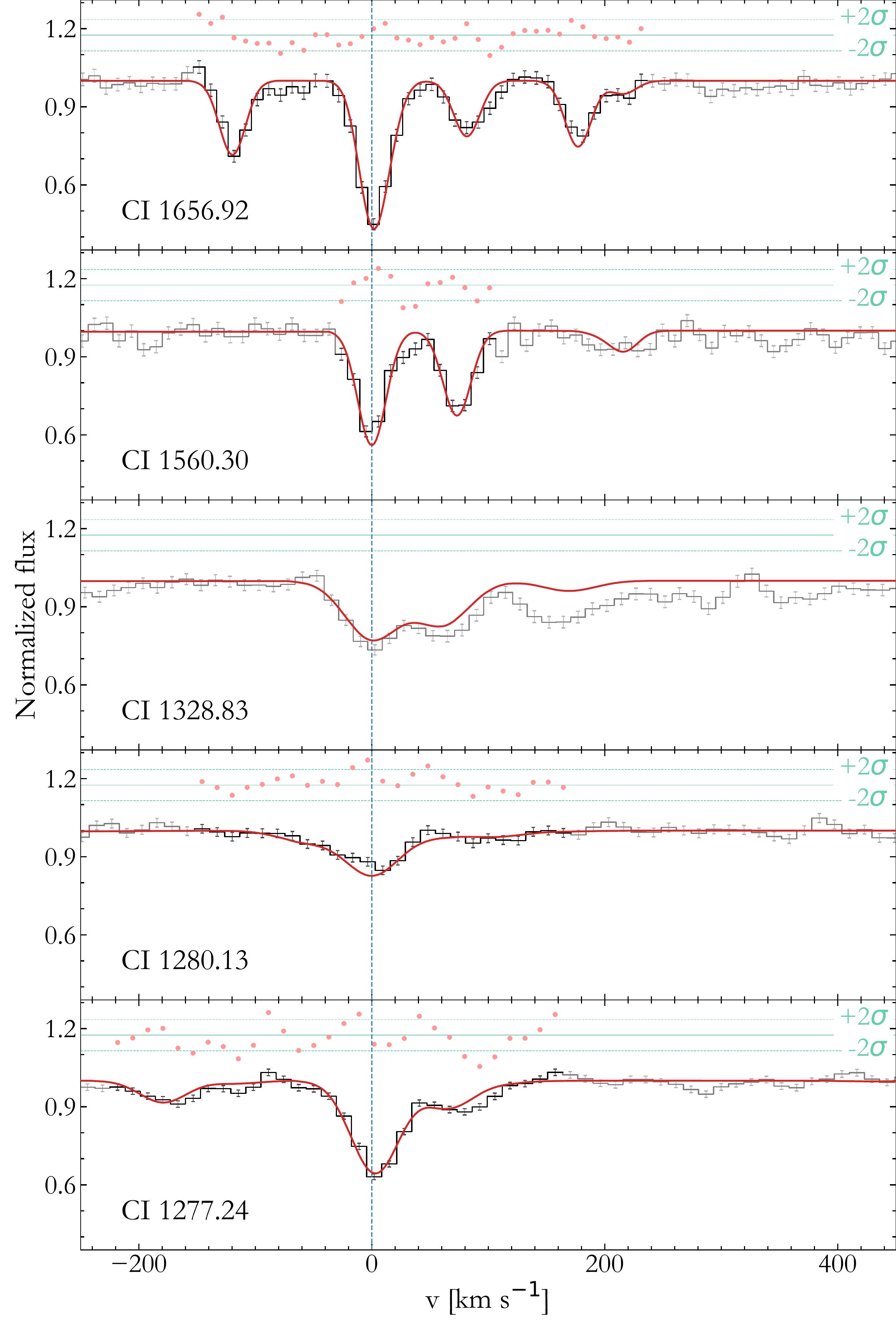}
	\caption{Fit to the \CI\ absorption lines at $z=2.625$ towards \Jtwos.}
	\label{fig:J0858_CI}
\end{figure}

We detected \CI\ absorption lines in three fine structure levels in a single component corresponding to H$_2$. 
The result from fitting these lines is given in Table~\ref{table:J0858_CI} and the corresponding profiles are shown in Fig.~\ref{fig:J0858_CI}.

The constrained values of the hydrogen number density and UV field in the H$_2$/\CI-bearing medium are shown in Fig.~2 and the comparison between observed and model column densities at the best fit value (yellow star in Fig.~2) is provided in Table~\ref{table:J0858_phys}.

\begin{figure*}
\centering
\includegraphics[trim={0.0cm 0.0cm 0.0cm 0.0cm},clip,width=\hsize]{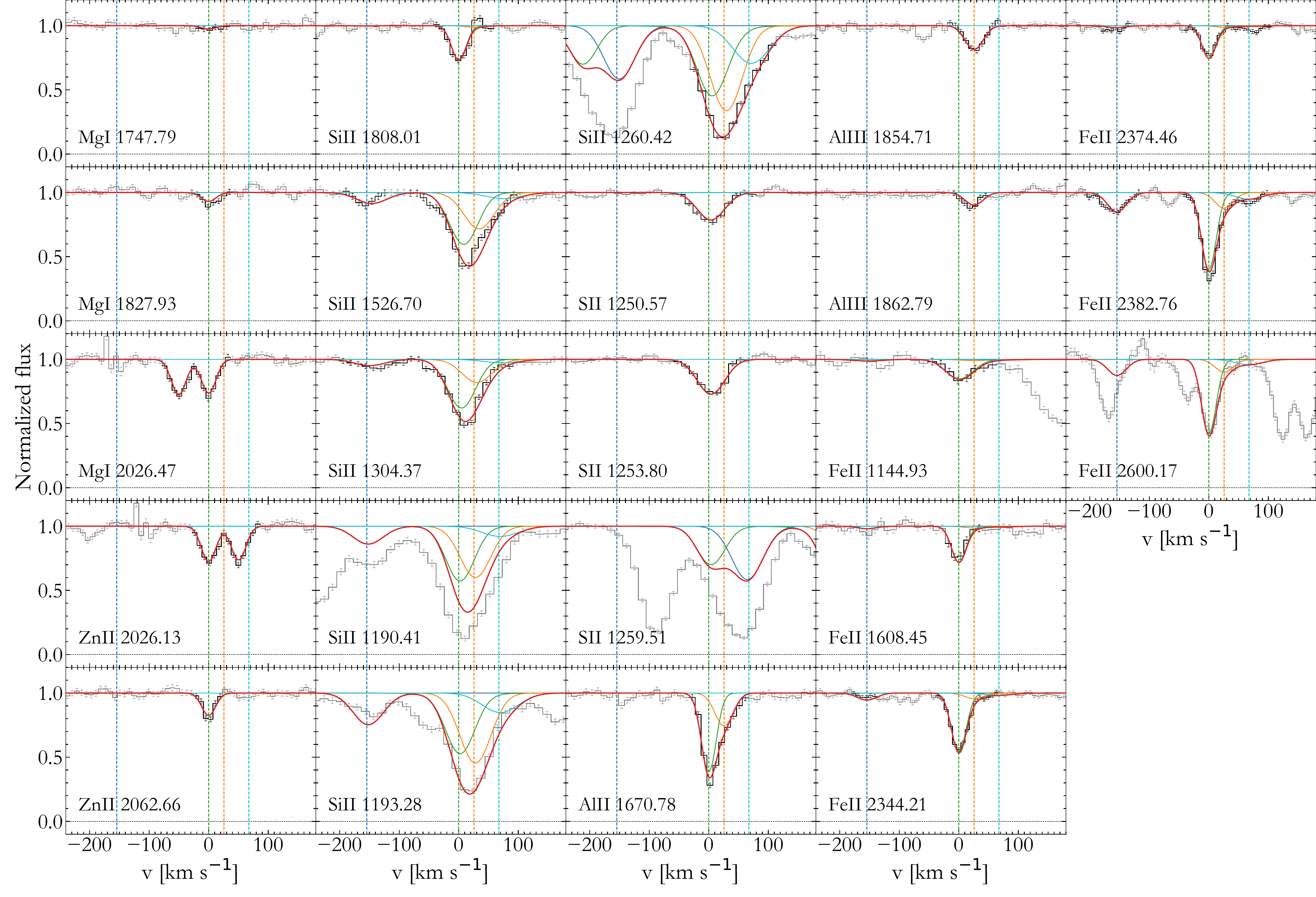}
		\caption{Fit to the metal absorption lines at $z=2.625$ towards \Jtwos.}
		\label{fig:J0858_metals}
\end{figure*}

We fitted the metal lines with four components. At the spectral resolution achieved with X-Shooter, the low-ionisation metal profile appears to be very simple with the second component dominating the profiles and coinciding with the position of H$_2$ and \CI. 
Since the S/N in this spectrum is high, we were able to fit weak \ion{Zn}{ii}, \ion{Mg}{i} and \ion{Al}{iii} transitions. We also added \ion{Al}{ii} to the fit. However, we caution that only a single \ion{Al}{ii} line is available and in an intrinsically saturated regime. Hence, the derived column density should be considered with care.
The result of the fit is given in Table~\ref{table:J0858_metals} and fit profiles are shown in Fig.~\ref{fig:J0858_metals}.
We estimate the metallicity in this system using \ion{S}{ii}, which we found to be in agreement with the metallicity measured from \ion{Zn}{ii}. We noticed a relatively high depletion of iron for the measured metallicity of $\sim-0.5$ (see Fig.~5). 
This does not appear to be due to a fitting failure since the column density of \ion{Fe}{ii} is well constrain by fitting five transitions out of which three are weak.

\begin{table}
\centering
\caption{Fit results of H$_2$ towards \Jtwos.
\label{table:J0858_H2}
}
\begin{tabular}{cc}
\hline
z & $2.625248(^{+5}_{-4})$ \\
b (km\,s$^{-1}$) & $7.9^{+0.4}_{-0.4}$ \\
$\log N$(H$_2$ J0) & $19.28^{+0.02}_{-0.03}$ \\
$\log N$(H$_2$ J1) & $19.51^{+0.01}_{-0.03}$ \\
$\log N$(H$_2$ J2) & $17.78^{+0.22}_{-0.18}$ \\
$\log N$(H$_2$ J3) & $16.99^{+0.22}_{-0.07}$ \\
$\log N$(H$_2$ J4) & $15.14^{+0.07}_{-0.09}$ \\
$\log N$(H$_2$ J5) & $14.53^{+0.10}_{-0.07}$ \\
$\log N_{\rm tot}$ & $19.72^{+0.01}_{-0.02}$ \\
\hline
\end{tabular}
\end{table}

\begin{table}
\centering
\caption{Fit results of \CI\ towards \Jtwos.
\label{table:J0858_CI}
}
\begin{tabular}{cc}
\hline
z & $2.6252859(^{+24}_{-16})$ \\
b, km\,s$^{-1}$ & $3.8^{+0.1}_{-0.1}$ \\
$\log N$(CI) & $14.26^{+0.02}_{-0.04}$ \\
$\log N$(CI*) & $13.74^{+0.02}_{-0.01}$ \\
$\log N$(CI**) & $12.96^{+0.02}_{-0.03}$ \\
$\log N_{\rm tot}$ & $14.39^{+0.02}_{-0.03}$ \\
\hline
\end{tabular}
\end{table}

\begin{table}
\caption{Comparison of observed column densities in the DLA towards \Jtwos\ with the model ones, calculated at the best fit value of number density and UV field from Fig.~2. 
\label{table:J0858_phys}
}
\begin{tabular}{ccc}
\hline
species & \multicolumn{2}{c}{$\log N$}\\ 
& observed & model \\
\hline
CI & $14.26^{+0.02}_{-0.04}$ & 14.23$^a$ \\
CI* & $13.74^{+0.02}_{-0.02}$ & 13.83$^a$ \\
CI** & $12.96^{+0.02}_{-0.03}$ & 12.88$^a$ \\
H$_2$ J0 & $19.28^{+0.02}_{-0.03}$ & 19.26 \\
H$_2$ J1 & $19.51^{+0.01}_{-0.03}$ & 19.44 \\
H$_2$ J2 & $17.78^{+0.22}_{-0.18}$ & 17.41 \\
\hline
H$_2$ J3$^b$ &  & 16.11 \\
H$_2$ J4$^b$ &  & 14.23 \\
H$_2$ J5$^b$ &  & 13.76 \\
\hline
\end{tabular}
\begin{tablenotes}
     \item (a) These values were obtained under then the additional assumption that the total \CI\ column density equates the observed one.   
     \item (b) We did not use these levels to compare with observed values, since the column densities measurements for these levels are subject to a large systematic uncertainty (see text).
\end{tablenotes}
\end{table}



\begin{table*}
\centering
\caption{Fit results of metals towards \Jtwos.
\label{table:J0858_metals}
}
\begin{tabular}{cccccccc}
\hline
comp & 1 & 2 & 3 & 4 & $\log N_{\rm tot}$ & $\rm [X/H]$ & $\rm [X/SII]$ \\
\hline
z & $2.623389(^{+9}_{-15})$ & $2.6252594(^{+25}_{-30})$ & $2.625565(^{+28}_{-12})$ & $2.62596(^{+4}_{-13})$ &  &  &  \\
$\Delta$v, km\,s$^{-1}$ & -155.5 & 0.0 & 25.0 & 70.6 &  &  &  \\
b, km\,s$^{-1}$ & $15.9^{+2.9}_{-1.8}$ & $5.7^{+0.2}_{-0.2}$ & $14.8^{+1.1}_{-4.5}$ & $42.5^{+5.0}_{-8.4}$ &  &  &  \\
$\log N$(SiII) & $13.02^{+0.06}_{-0.05}$ & $15.08^{+0.03}_{-0.03}$ & $13.60^{+0.06}_{-0.04}$ & $13.06^{+0.08}_{-0.13}$ & $15.10^{+0.02}_{-0.03}$ & $-0.96^{+0.02}_{-0.04}$ & $-0.41^{+0.05}_{-0.04}$ \\
$\log N$(FeII) & $12.55^{+0.02}_{-0.05}$ & $13.70^{+0.02}_{-0.02}$ & $12.23^{+0.17}_{-0.13}$ & $12.26^{+0.16}_{-0.13}$ & $13.75^{+0.02}_{-0.01}$ & $-2.30^{+0.02}_{-0.02}$ & $-1.75^{+0.05}_{-0.03}$ \\
$\log N$(SII) &  & $15.12^{+0.03}_{-0.04}$ &  &  & $15.12^{+0.03}_{-0.04}$ & $-0.55^{+0.03}_{-0.04}$ & $0.00^{+0.05}_{-0.05}$ \\
$\log N$(ZnII) &  & $12.64^{+0.04}_{-0.02}$ &  &  & $12.64^{+0.04}_{-0.02}$ & $-0.47^{+0.04}_{-0.02}$ & $0.08^{+0.06}_{-0.03}$ \\
$\log N$(MgI) &  & $13.30^{+0.03}_{-0.03}$ &  &  & $13.30^{+0.03}_{-0.03}$ & $-2.85^{+0.03}_{-0.03}$ & $-2.30^{+0.05}_{-0.04}$ \\
$\log N$(AlII) &  & $13.21^{+0.10}_{-0.09}$ & $12.18^{+0.05}_{-0.06}$ &  & $13.32^{+0.07}_{-0.08}$ & $-1.68^{+0.07}_{-0.08}$ & $-1.13^{+0.08}_{-0.09}$ \\
$\log N$(AlIII) &  &  & $12.46^{+0.02}_{-0.04}$ &  & $12.46^{+0.02}_{-0.04}$ & $-2.54^{+0.02}_{-0.04}$ & $-1.99^{+0.04}_{-0.05}$ \\
\hline
\end{tabular}
\end{table*}

\clearpage

\subsection{\Jthree}
\begin{figure}
\centering
\includegraphics[trim={0.0cm 0.0cm 0.0cm 0.0cm},clip,width=\hsize]{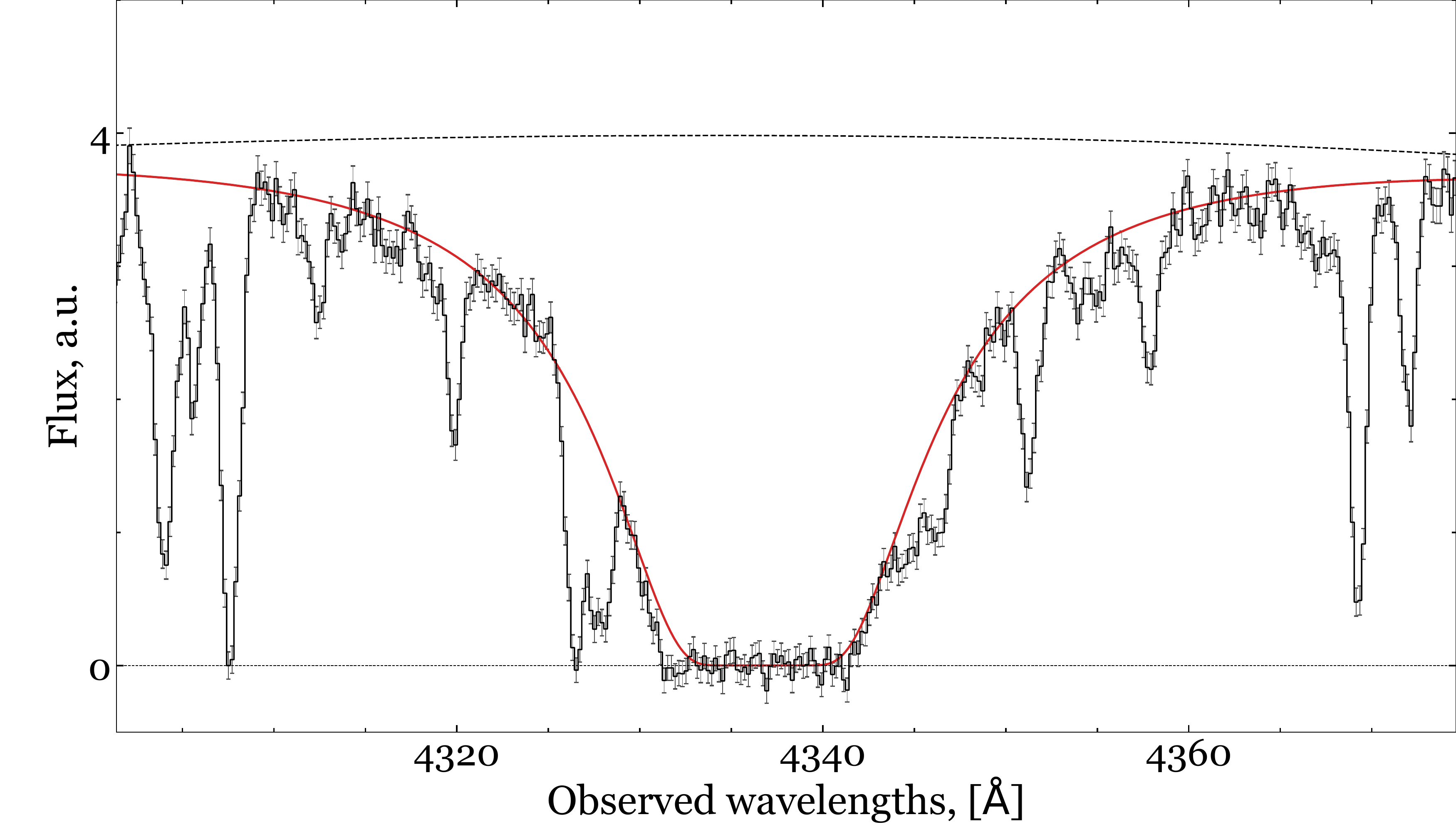}
		\caption{Fit to the \HI\ Ly$\alpha$ absorption line at $z=2.567$ towards \Jthrees.
		\label{fig:J0906_HI}}
\end{figure}

We measured the \HI\ column density of the DLA at $z=2.567$ towards \Jthrees\ by fitting the Ly$\alpha$ absorption line with a single component together with the continuum modelled using a sixth order Chebyshev polynomial. 
The derived column density, $\log N(\HI)=20.13\pm0.01$ at the best fit value $z=2.56742$, indicates that this system formally belong to the class of sub-DLAs. The Ly$\alpha$ profile, together with the estimated quasar continuum are shown in Fig.~\ref{fig:J0906_HI}.

\begin{figure*}
\centering
\includegraphics[trim={0.0cm 0.0cm 0.0cm 0.0cm},clip,width=\hsize]{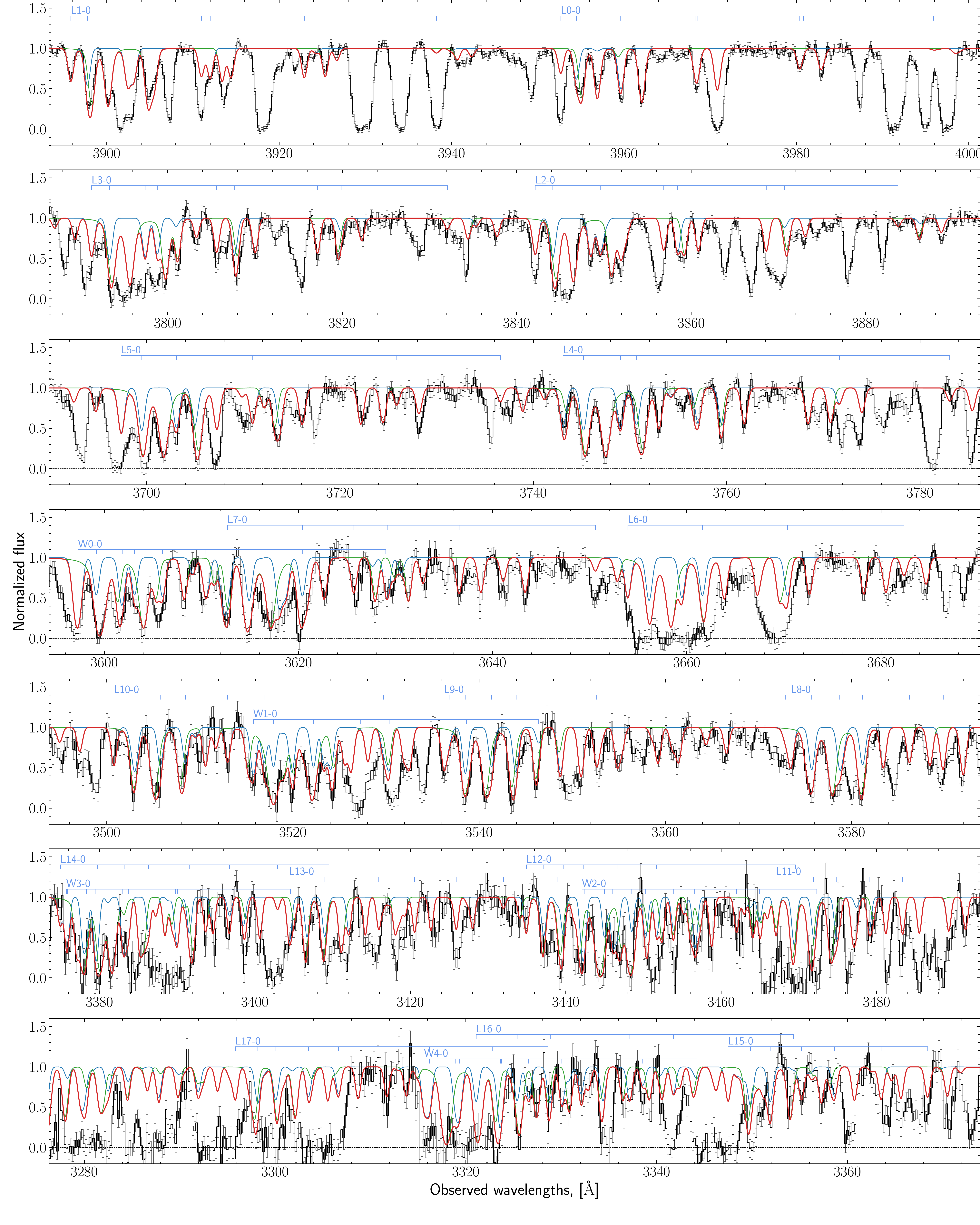}
		\caption{Fit to H$_2$ absorption lines at $z=2.567$ towards  \Jthrees.
		\label{fig:J0906_H2}}
\end{figure*}

\begin{figure}
\centering
\includegraphics[trim={0.0cm 0.0cm 0.0cm 0.0cm},clip,width=\hsize]{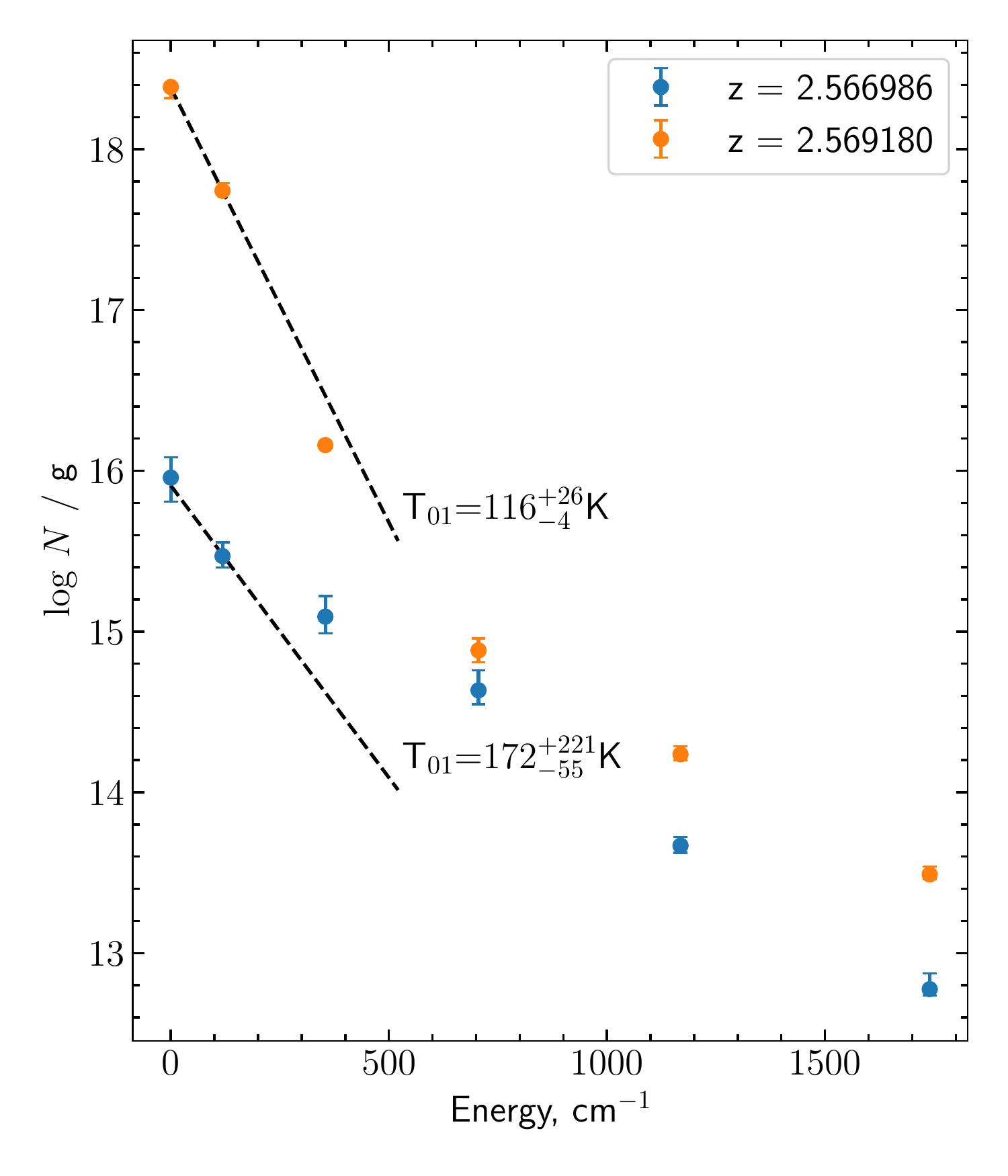}
		\caption{Excitation diagram for the two H$_2$ components at $z=2.567$ towards \Jthrees.
		}
		\label{fig:J0906_H2_exc}
\end{figure}

We fit H$_2$ absorption lines in \Jthrees\ using one component model. The result of the fit is given in Table~\ref{table:J0906_H2} and the profiles of H$_2$ lines are shown in Fig.~\ref{fig:J0906_H2}. The excitation diagrams of H$_2$ rotation levels for both components is shown in Fig.~\ref{fig:J0906_H2_exc}. We derived the $T_{01}$ temperature to be $172^{+221}_{-55}$\,K and $116^{+26}_{-4}$\,K for the H$_2$ components at $z=2.56698$ and 2.56918, respectively.  

\begin{figure}
    \centering
    \includegraphics[trim={0.0cm 15.0cm 0.0cm 0.0cm},clip,width=\hsize]{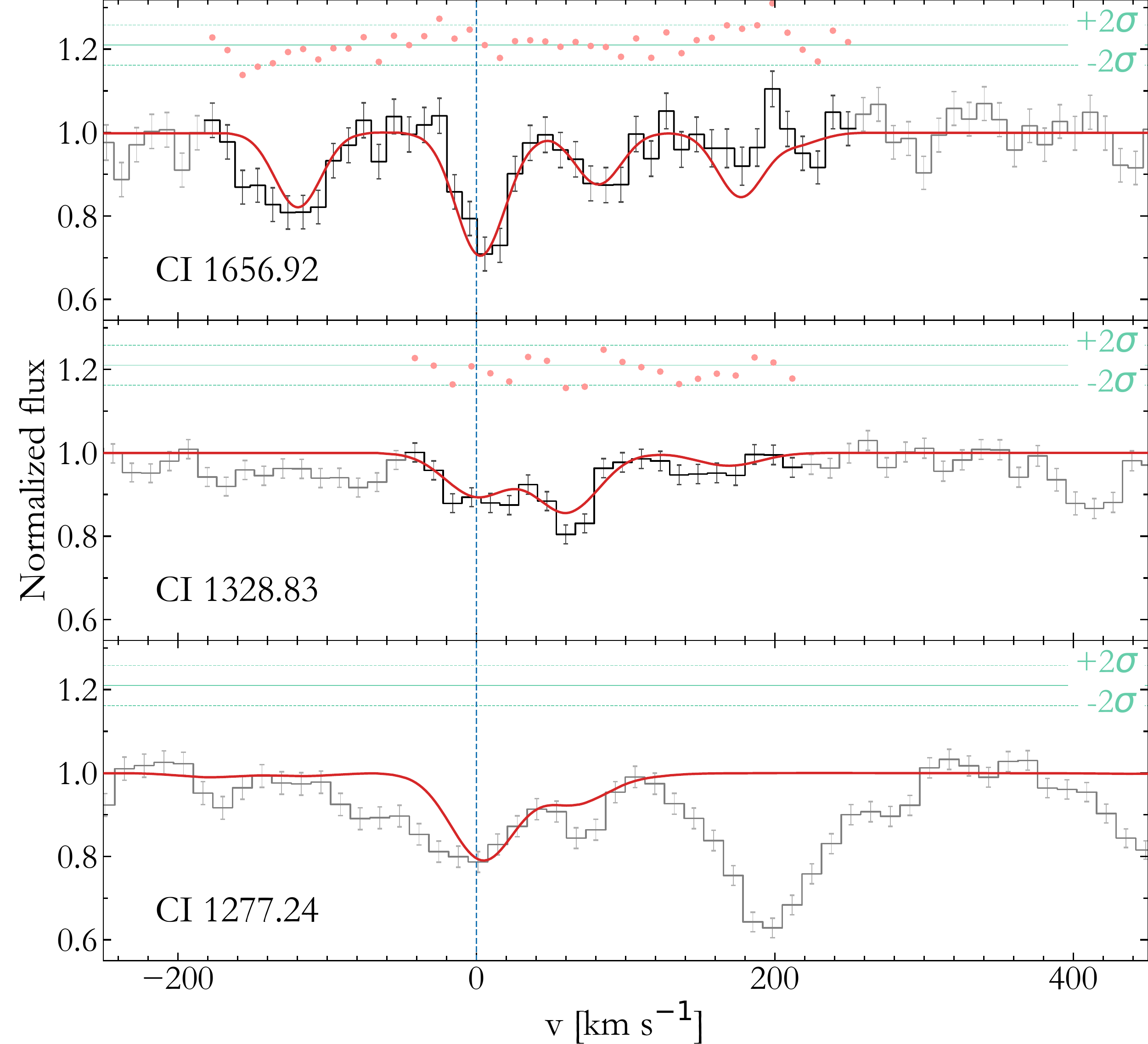}
	\caption{Fit to the \CI\ absorption lines at $z=2.567$ towards \Jthrees.}
	\label{fig:J0906_CI}
\end{figure}

We detected \CI\ absorption lines from its three fine structure levels only at the location of the reddest and strongest H$_2$ component that dominates the profile. 
We therefore fitted \CI\ lines using a one-component model. Unfortunately, \CI\ lines from the 1560\AA\ multiplet are located at the noisy blue side of the VIS exposure for both X-shooter exposures and were not considered to constrain the fit. 
The result of the fit is given in Table~\ref{table:J0906_CI} and the profiles of \CI\ absorption lines are shown in Fig.~\ref{fig:J0906_CI}. 

\begin{figure}
\centering
\includegraphics[trim={0.0cm 0.0cm 0.0cm 0.0cm},clip,width=\hsize]{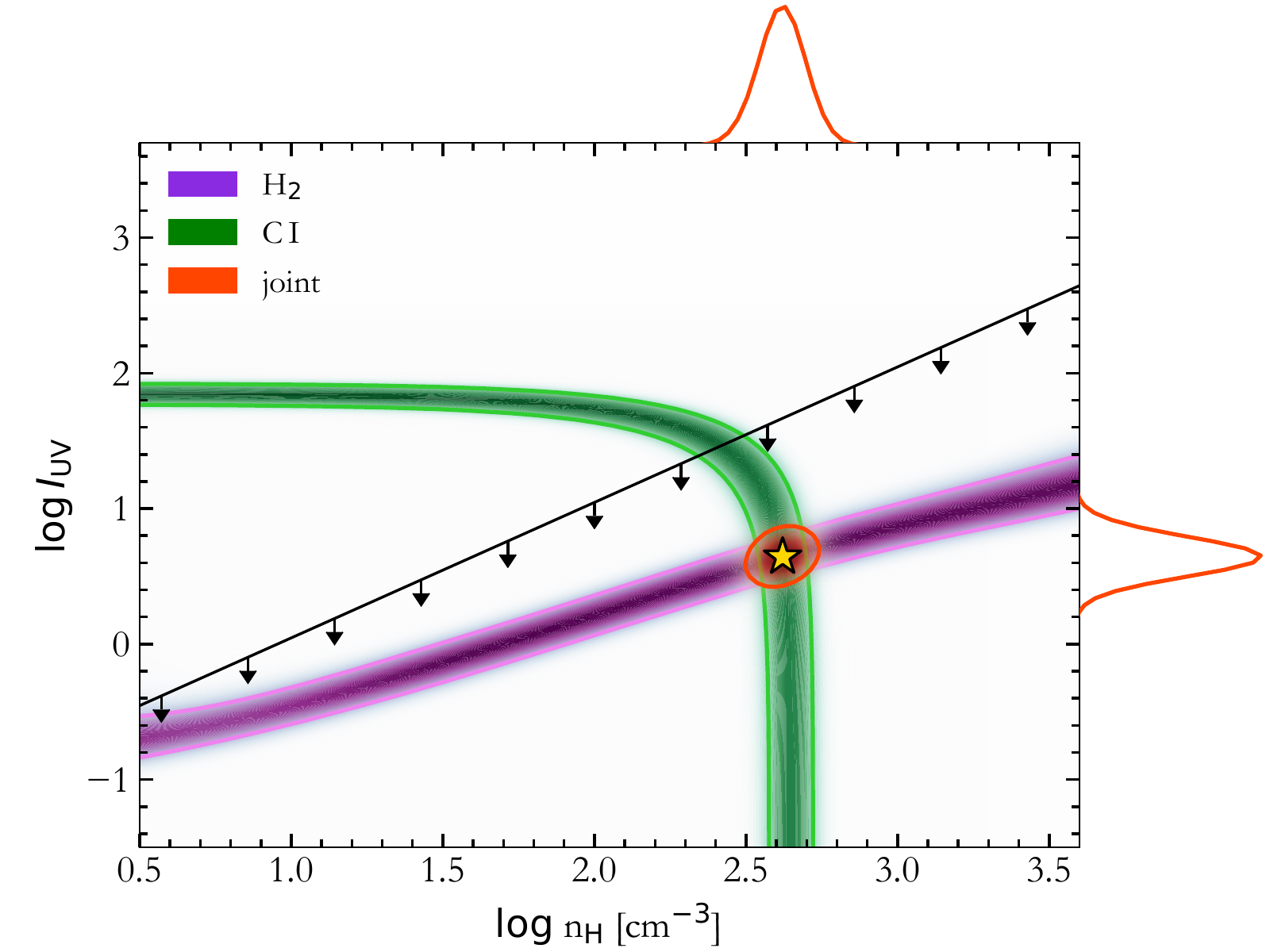}
			\caption{Constraints on the number density and UV field in the H$_2$-bearing medium at $z=2.569181$ 
		towards \Jthrees\ from the excitation of \CI\ fine-structure and H$_2$ rotational levels. The colors and lines are the same as in Fig.~2 }
		\label{fig:J0906_phys}
\end{figure}

The constrained values of the hydrogen number density and UV field in the H$_2$/\CI-bearing medium are shown in Fig.~\ref{fig:J0906_phys} and the comparison between observed and model column densities at the best fit value (yellow star in Fig.~\ref{fig:J0906_phys}) is provided in Table~\ref{table:J0906_phys}.

\begin{figure*}
\centering
\includegraphics[trim={0.0cm 0.0cm 0.0cm 0.0cm},clip,width=\hsize]{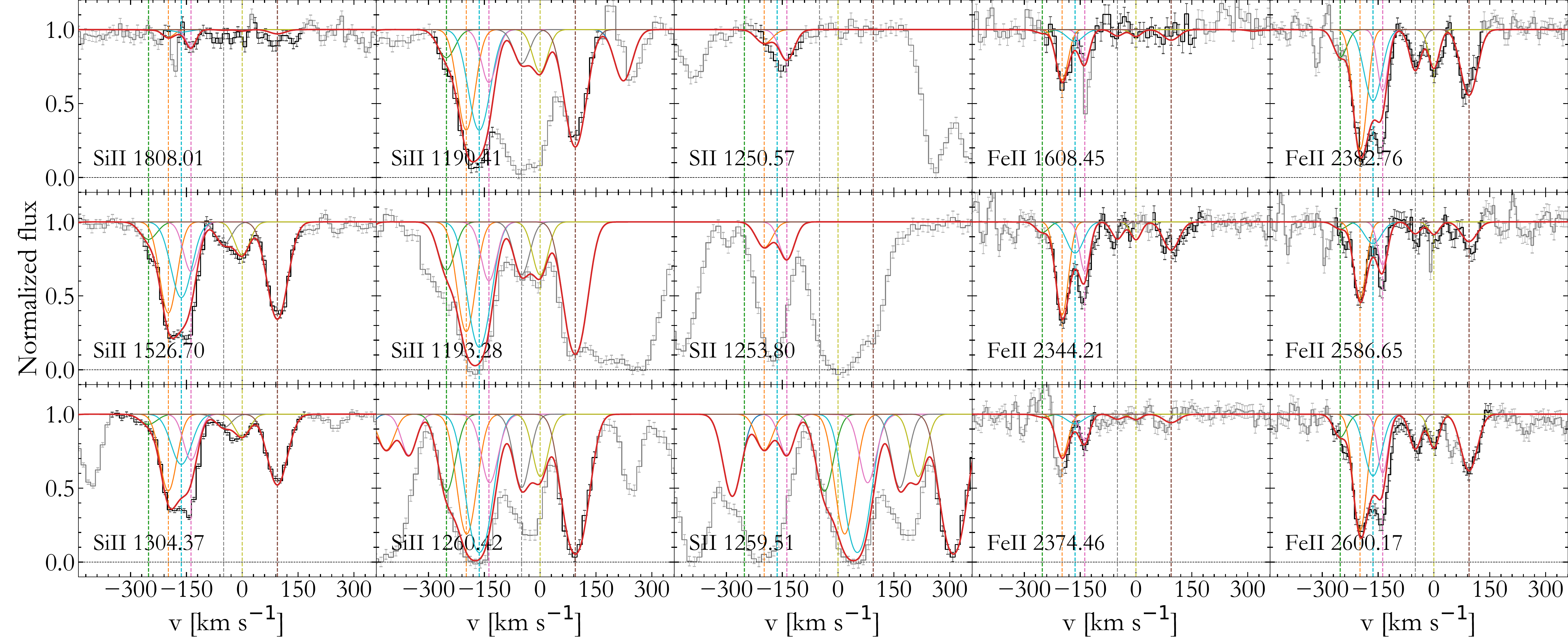}
		\caption{Fit to the metal absorption lines at $z=2.567$ towards \Jthrees.}
		\label{fig:J0906_metals}
\end{figure*}

Eight components were used to model the metal line profile. The fifth and seventh components correspond to the position of H$_2$ components. Since the \HI\ column density is relatively small we detected only a few low ionization metals and did not detect \ion{Zn}{ii}, \ion{Cr}{ii}, \ion{Ni}{ii}. We estimated the metallicity in this system using \ion{S}{ii}.
 The result of the fit is given in Table~\ref{table:J0906_metals} and fit profiles are shown in Fig.~\ref{fig:J0906_metals}. 

\begin{table}
\centering
\caption{Fit results of H$_2$ towards \Jthrees.
\label{table:J0906_H2}
}
\begin{tabular}{cccc}
\hline
comp & 1 & 2 & $\log N_{\rm tot}$ \\
\hline
z & $2.566980(^{+9}_{-4})$ & $2.5691809(^{+11}_{-73})$ &  \\
$\Delta$v, km\,s$^{-1}$ & 0.0 & 184.5 &  \\
b, km\,s$^{-1}$ & $7.5^{+0.3}_{-0.3}$ & $6.8^{+0.1}_{-0.1}$ &  \\
$\log N$(H$_2$ J0) & $15.91^{+0.13}_{-0.15}$ & $18.38^{+0.01}_{-0.07}$ & $18.38^{+0.01}_{-0.07}$ \\
$\log N$(H$_2$ J1) & $16.43^{+0.09}_{-0.07}$ & $18.70^{+0.05}_{-0.01}$ & $18.70^{+0.05}_{-0.01}$ \\
$\log N$(H$_2$ J2) & $15.69^{+0.13}_{-0.10}$ & $16.87^{+0.01}_{-0.01}$ & $16.89^{+0.02}_{-0.01}$ \\
$\log N$(H$_2$ J3) & $15.95^{+0.12}_{-0.09}$ & $16.15^{+0.07}_{-0.07}$ & $16.39^{+0.04}_{-0.08}$ \\
$\log N$(H$_2$ J4) & $14.74^{+0.05}_{-0.04}$ & $15.20^{+0.05}_{-0.04}$ & $15.32^{+0.05}_{-0.02}$ \\
$\log N$(H$_2$ J5) & $14.34^{+0.10}_{-0.04}$ & $15.00^{+0.05}_{-0.03}$ & $15.10^{+0.03}_{-0.03}$ \\
$\log N_{\rm tot}$ & $16.70^{+0.11}_{-0.06}$ & $18.87^{+0.02}_{-0.02}$ & $18.88^{+0.02}_{-0.02}$ \\
\hline
\end{tabular}

\end{table}

\begin{table}
\centering
\caption{Fit results of \CI\ towards \Jthrees.
\label{table:J0906_CI}
}
\begin{tabular}{cc}
\hline
z & $2.569225(^{+6}_{-7})$ \\
b, km\,s$^{-1}$ & $3.8^{+0.6}_{-0.5}$ \\
$\log N$(CI) & $13.50^{+0.08}_{-0.06}$ \\
$\log N$(CI*) & $13.61^{+0.03}_{-0.03}$ \\
$\log N$(CI**) & $12.89^{+0.07}_{-0.08}$ \\
$\log N_{\rm tot}$ & $13.90^{+0.04}_{-0.03}$ \\
\hline
\end{tabular}
\end{table}


\begin{table}
\centering
\caption{Comparison of observed column densities in the DLA towards \Jthrees\ with the model ones, calculated at the best fit value of number density and UV field from Fig.~\ref{fig:J0906_phys}.}
\begin{tabular}{ccc}
\hline
species & \multicolumn{2}{c}{$\log N$}\\ 
& observed & model \\
\hline
CI & $13.50^{+0.08}_{-0.06}$ & 13.55$^a$ \\
CI* & $13.61^{+0.03}_{-0.03}$ & 13.54$^a$ \\
CI** & $12.89^{+0.07}_{-0.08}$ & 13.00$^a$ \\
H$_2$ J0 & $18.38^{+0.01}_{-0.07}$ & 18.45 \\
H$_2$ J1 & $18.70^{+0.05}_{-0.01}$ & 18.68 \\
H$_2$ J2 & $16.87^{+0.01}_{-0.01}$ & 16.92 \\
\hline
H$_2$ J3$^b$ &  & 15.94 \\
H$_2$ J4$^b$ &  & 14.29 \\
H$_2$ J5$^b$ &  & 13.87 \\
\hline
\end{tabular}
\label{table:J0906_phys}
\begin{tablenotes}
    \item (a) These values were obtained under the additional assumption that the total \CI\ column density equates the  observed one. 
    \item (b) We did not use these levels to compare with observed values, since the column densities measurements for these levels are subject to a large systematic uncertainty (see text).
\end{tablenotes}
\end{table}

\begin{table*}
\caption{Fit results of metals towards \Jthrees.
\label{table:J0906_metals}
}
\begin{tabular}{ccccccc}
\hline
comp & z & $\Delta$v, km\,s$^{-1}$ & b, km\,s$^{-1}$ & $\log N$(SiII) & $\log N$(FeII) & $\log N$(SII) \\
\hline
1 & $2.563318(^{+5}_{-24})$ & -500.3 & $10.6^{+5.7}_{-4.7}$ & $13.20^{+0.05}_{-0.05}$ & $12.81^{+0.04}_{-0.04}$ &  \\
2 & $2.56625(^{+3}_{-3})$ & -254.2 & $12.7^{+6.2}_{-6.8}$ & $13.01^{+0.10}_{-0.06}$ & $12.64^{+0.08}_{-0.08}$ &  \\
3 & $2.566892(^{+16}_{-16})$ & -199.2 & $7.6^{+2.5}_{-2.7}$ & $14.21^{+0.16}_{-0.13}$ & $13.79^{+0.09}_{-0.08}$ & $14.40^{+0.30}_{-0.49}$ \\
4 & $2.56701(^{+8}_{-3})$ & -187.3 & $25.1^{+5.9}_{-3.7}$ & $14.22^{+0.06}_{-0.10}$ & $13.60^{+0.07}_{-0.13}$ &  \\
5 & $2.567609(^{+13}_{-10})$ & -138.2 & $10.6^{+1.0}_{-1.1}$ & $14.55^{+0.07}_{-0.05}$ & $13.64^{+0.03}_{-0.03}$ & $15.08^{+0.04}_{-0.09}$ \\
6 & $2.568677(^{+10}_{-11})$ & -48.8 & $7.8^{+1.4}_{-3.1}$ & $13.28^{+0.05}_{-0.03}$ & $12.84^{+0.04}_{-0.04}$ &  \\
7 & $2.569257(^{+14}_{-8})$ & 0.0 & $2.8^{+1.2}_{-0.2}$ & $13.80^{+0.26}_{-0.19}$ & $12.93^{+0.10}_{-0.08}$ &  \\
8 & $2.570366(^{+3}_{-5})$ & 92.9 & $23.4^{+0.6}_{-0.4}$ & $14.18^{+0.01}_{-0.01}$ & $13.18^{+0.02}_{-0.02}$ &  \\
$\log N_{\rm tot}$ &  &  &  & $14.99^{+0.04}_{-0.05}$ & $14.24^{+0.03}_{-0.02}$ & $15.14^{+0.05}_{-0.08}$ \\
$\rm [X/H]$ &  &  &  & $-0.70^{+0.04}_{-0.05}$ & $-1.44^{+0.03}_{-0.03}$ & $-0.16^{+0.05}_{-0.08}$ \\
$\rm [X/SII]$ &  &  &  & $-0.54^{+0.08}_{-0.07}$ & $-1.28^{+0.08}_{-0.06}$ & $0.00^{+0.09}_{-0.09}$ \\
\hline
\end{tabular}
\end{table*}

\clearpage

\subsection{\Jfour}

\begin{figure}
\centering
\includegraphics[trim={0.0cm 0.0cm 0.0cm 0.0cm},clip,width=\hsize]{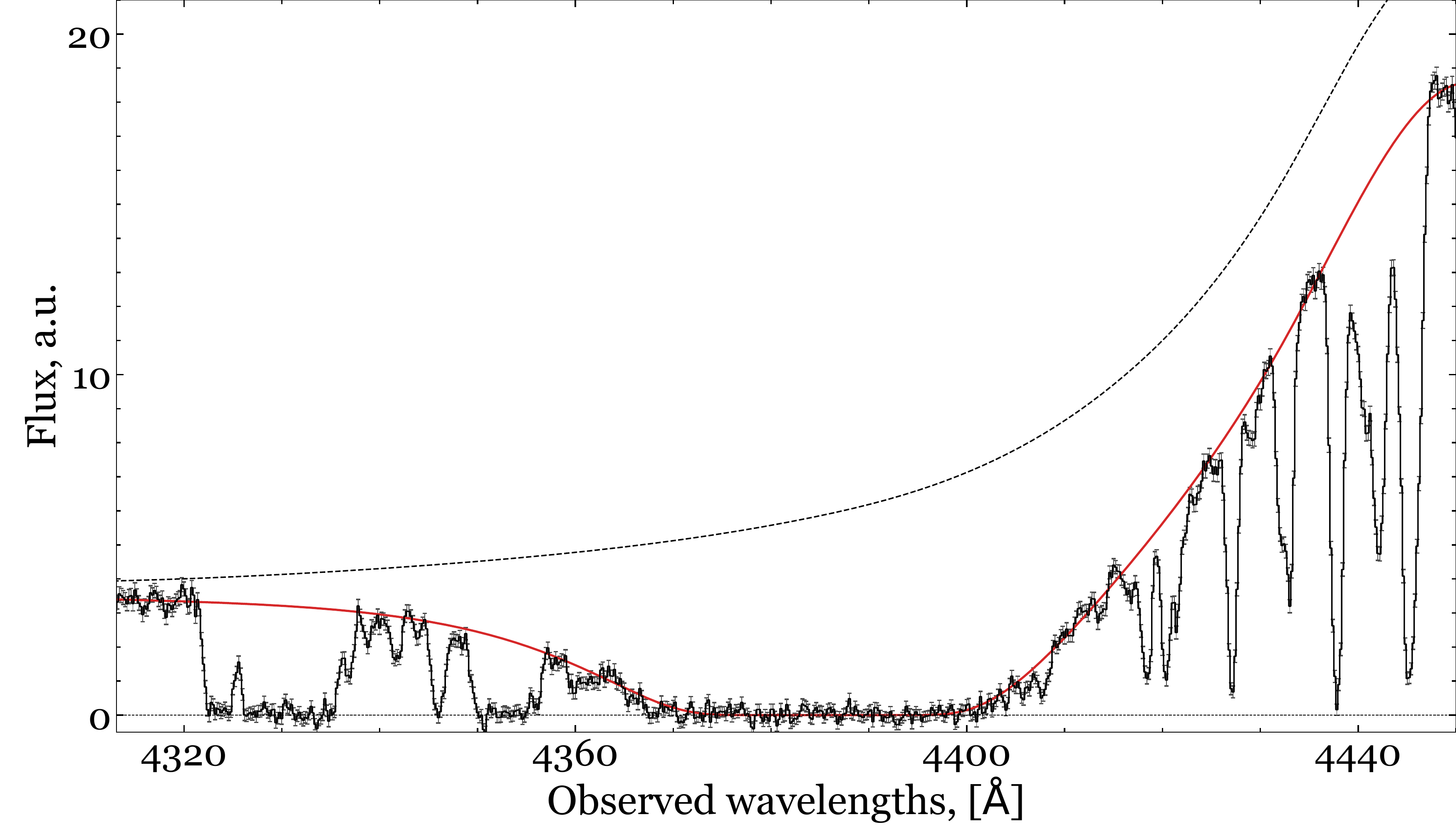}
		\caption{Fit to the damped \HI\ Ly$\alpha$ absorption line at $z=2.607$ towards \Jfours.
		\label{fig:J0946_HI}}
\end{figure}

We measured the \HI\ column density of the DLA at $z=2.607$ towards \Jfours\ by fitting the Ly$\alpha$ absorption line with a single component together with the continuum modelled using a seventh order Chebyshev polynomial. We derived $\log N(\HI)=21.15\pm0.02$ at the best fit value of $z=2.60771$. The Ly$\alpha$ profile, together with the reconstructed quasar continuum are shown in Fig.~\ref{fig:J0946_HI}.

\begin{figure*}
\centering
\includegraphics[trim={0.0cm 0.0cm 0.0cm 0.0cm},clip,width=\hsize]{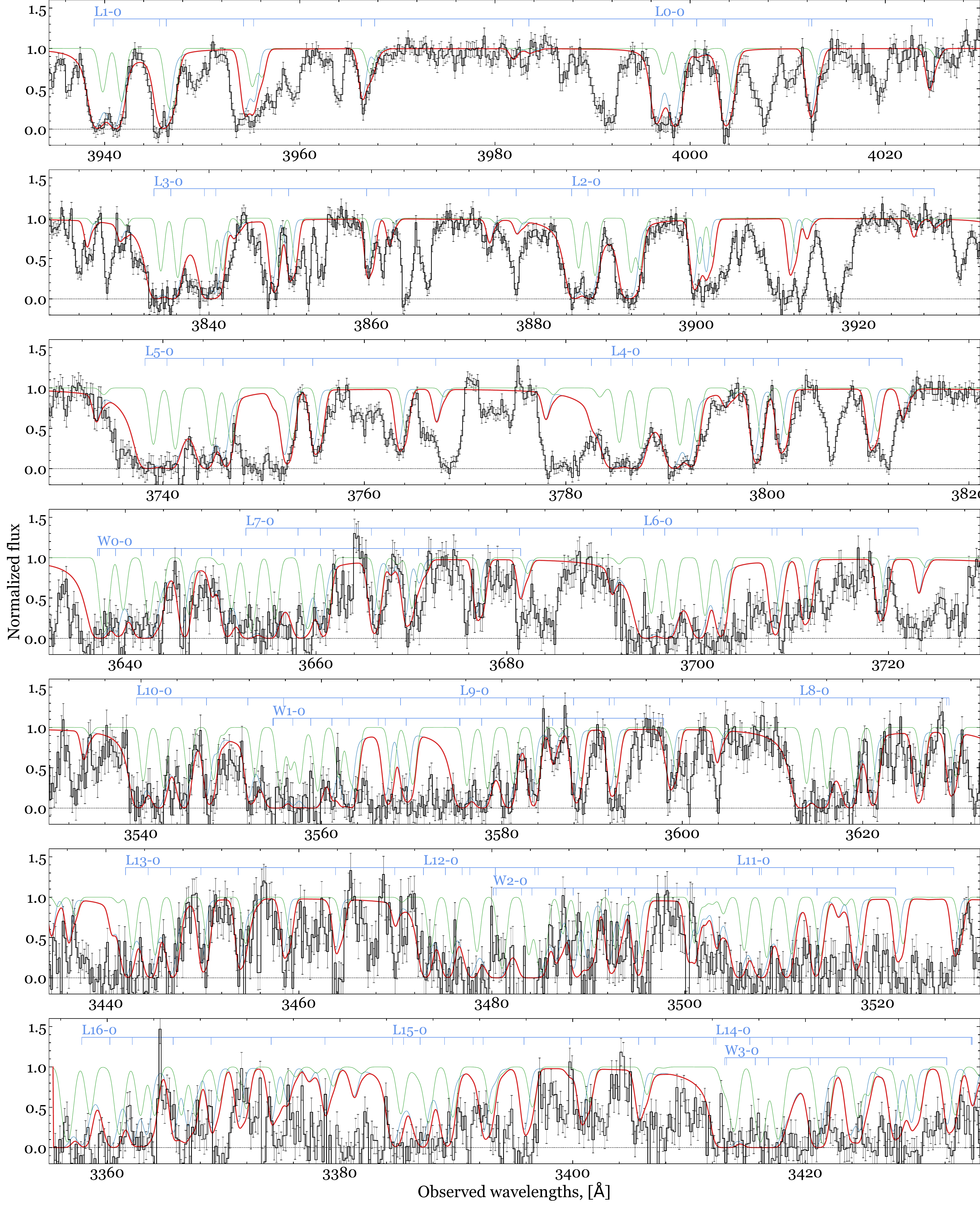}
		\caption{Fit to H$_2$ absorption lines at $z=2.607$ towards \Jfours.
		\label{fig:J0946_H2}}
\end{figure*}

\begin{figure}
\centering
\includegraphics[trim={0.0cm 0.0cm 0.0cm 0.0cm},clip,width=\hsize]{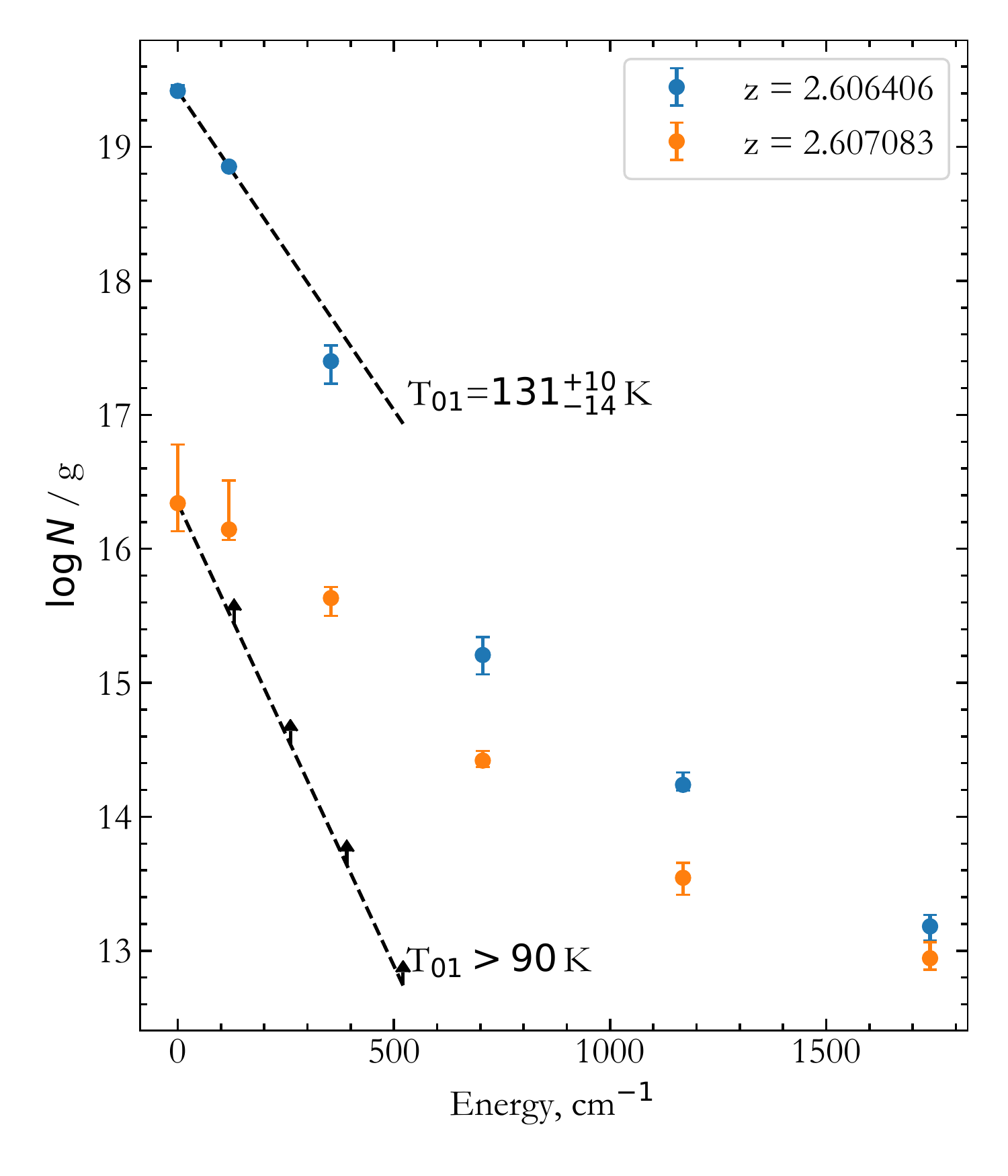}
		\caption{Excitation diagram for the two H$_2$ components at $z=2.607$ towards \Jfours. 
		\label{fig:J0946_H2_exc}}
\end{figure}

The fit of H$_2$ absorption lines was done using a two components model. The result of the fit is given in Table~\ref{table:J0946_H2} and H$_2$ line profiles are shown in Fig.~\ref{fig:J0946_H2}. The large reported Doppler parameters for the red component probably indicate to unresolved subcomponents in H$_2$ lines. The excitation diagrams of H$_2$ rotation levels for both components are shown in Fig.~\ref{fig:J0946_H2_exc}, with derived $T_{01}$ temperatures   
of $131^{+10}_{-14}$\,K and $>90$\,K (1$\sigma$ lower limit) for the component at $z=2.60641$ and 2.60708, respectively.  

\begin{figure}
    \centering
    \includegraphics[trim={0.0cm 0.0cm 0.0cm 0.0cm},clip,width=\hsize]{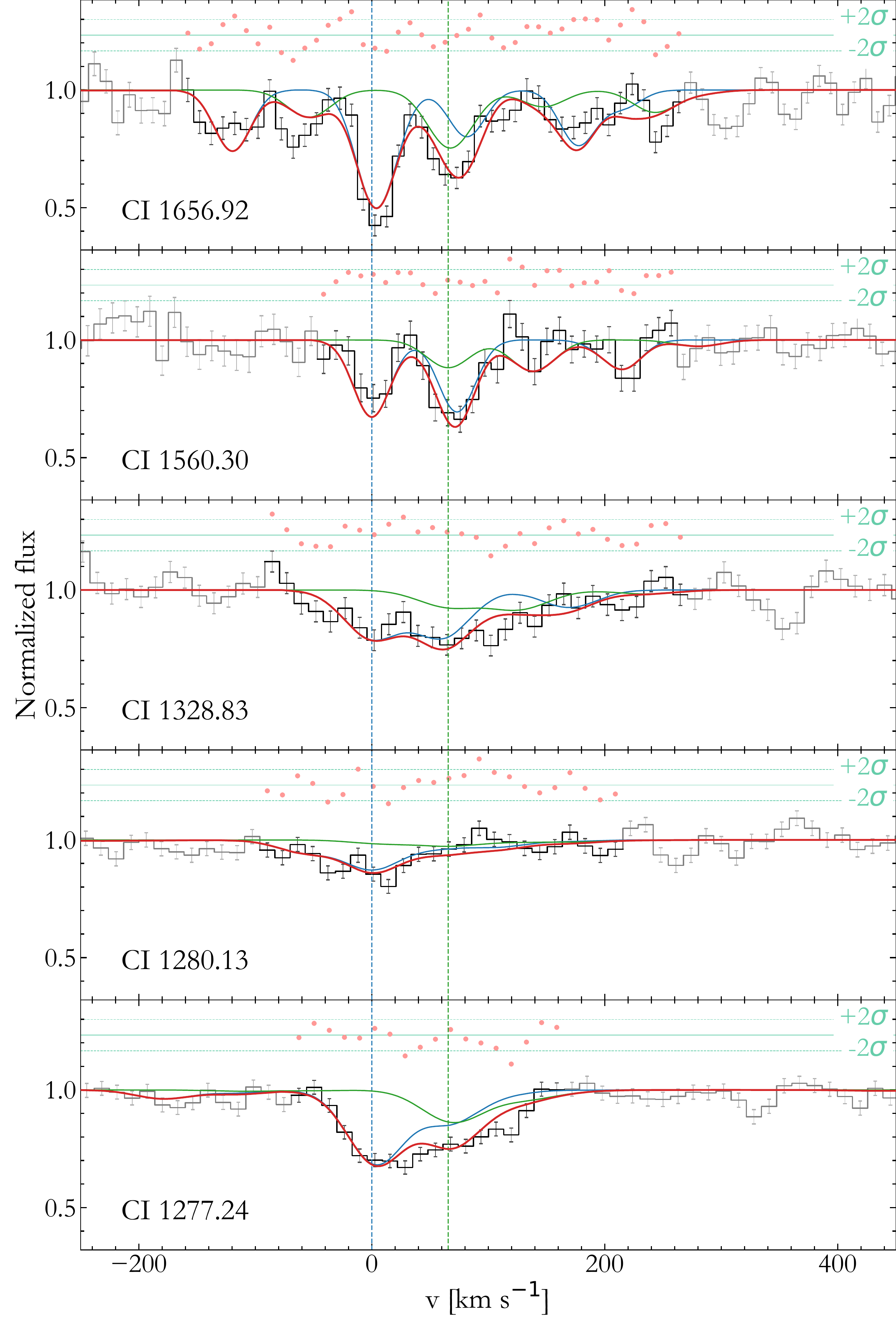}
	\caption{Fit to the \CI\ absorption lines at $z=2.607$ towards \Jfours.}
	\label{fig:J0946_CI}
\end{figure}

We detected \CI\ absorption lines in two components arising from three fine structure levels, coinciding with the H$_2$ components. 
The result of the fit is given in Table~\ref{table:J0946_CI} and the \CI\ absorption profiles are shown in Fig.~\ref{fig:J0946_CI}.

\begin{figure}
\centering
\includegraphics[trim={0.0cm 0.0cm 0.0cm 0.0cm},clip,width=\hsize]{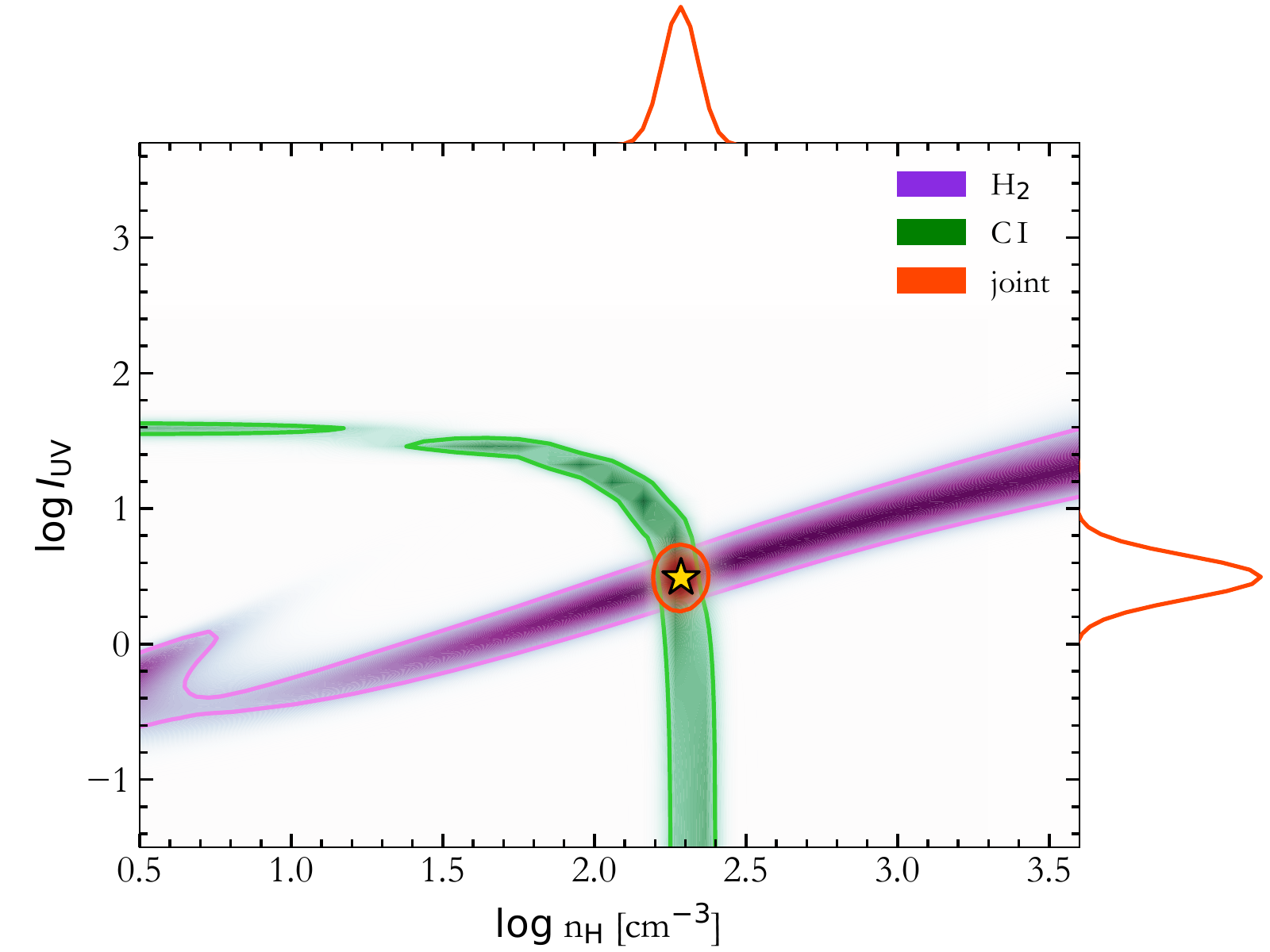}
		\caption{Constraints on number density and UV field in the H$_2$-bearing medium at $z=2.60641$ towards \Jfours\ from the excitation of \CI\ fine-structure and H$_2$ rotational levels. The colors and lines are the same as in Fig.~2.
		}
		\label{fig:J0946_phys}
\end{figure}

Since we measured only the lower limit on the temperature in the minor H$_2$/\CI-bearing component at $z=2.60708$, we constrained values of the hydrogen number density and UV field based on excitation of H$_2$ and \CI\ levels only in the main H$_2$/\CI-bearing component at $z=2.60641$ (shown in Fig.~\ref{fig:J0946_phys}). The comparison between observed and model column densities at the best fit value of the parameters (shown by yellow star in Fig.~\ref{fig:J0946_phys}) is provided in Table~\ref{table:J0946_phys}.

\begin{figure*}
\centering
\includegraphics[trim={0.0cm 0.0cm 0.0cm 0.0cm},clip,width=\hsize]{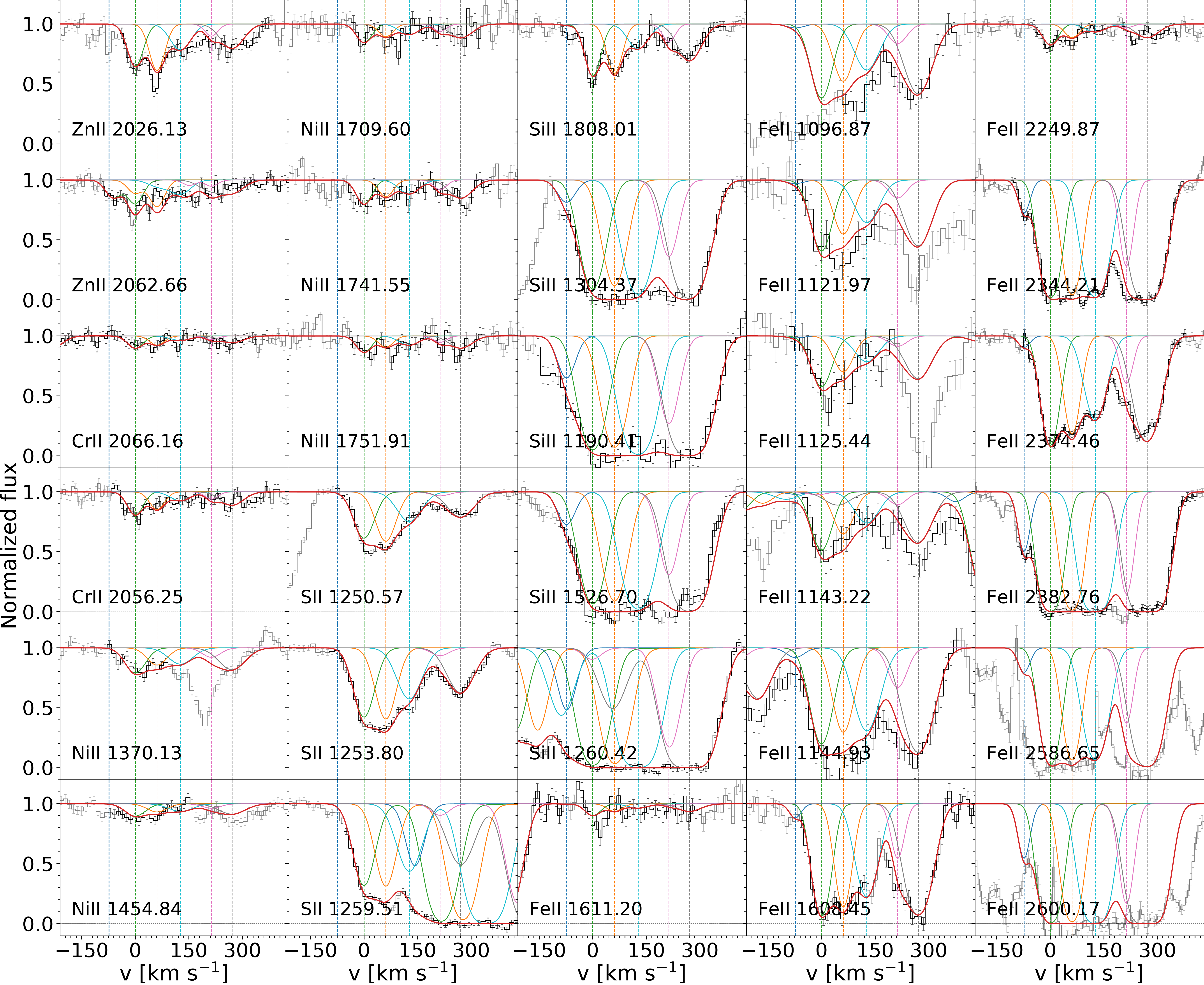}
		\caption{Fit to the metal absorption lines at $z=2.607$ towards \Jfours.}
		\label{fig:J0946_metals}
\end{figure*}

Metal lines are fitted using a six components model. 
We estimated the metallicity in this system using \ion{S}{ii}. The results of the fit are given in Table~\ref{table:J0946_metals} and fit profiles are shown in Fig.~\ref{fig:J0946_metals}.

\begin{table}
\caption{Fit results of H$_2$ towards \Jfours.
\label{table:J0946_H2}
}
\begin{tabular}{cccc}
\hline
comp & 1 & 2 & $\log N_{\rm tot}$ \\
\hline
z & $2.606406(^{+30}_{-8})$ & $2.607083(^{+16}_{-5})$ &  \\
$\Delta$v, km\,s$^{-1}$ & 0.0 & 57.2 &  \\
b, km\,s$^{-1}$ & $9.8^{+0.8}_{-0.3}$ & $17.6^{+1.6}_{-0.7}$ &  \\
$\log N$(H$_2$ J0) & $19.42^{+0.04}_{-0.02}$ & $16.34^{+0.44}_{-0.21}$ & $19.42^{+0.04}_{-0.02}$ \\
$\log N$(H$_2$ J1) & $19.81^{+0.02}_{-0.02}$ & $17.10^{+0.37}_{-0.08}$ & $19.81^{+0.02}_{-0.02}$ \\
$\log N$(H$_2$ J2) & $18.10^{+0.12}_{-0.17}$ & $16.33^{+0.08}_{-0.13}$ & $18.11^{+0.11}_{-0.17}$ \\
$\log N$(H$_2$ J3) & $16.53^{+0.13}_{-0.15}$ & $15.74^{+0.07}_{-0.05}$ & $16.58^{+0.12}_{-0.11}$ \\
$\log N$(H$_2$ J4) & $15.19^{+0.09}_{-0.04}$ & $14.50^{+0.11}_{-0.13}$ & $15.29^{+0.05}_{-0.05}$ \\
$\log N$(H$_2$ J5) & $14.70^{+0.09}_{-0.10}$ & $14.46^{+0.12}_{-0.08}$ & $14.92^{+0.04}_{-0.07}$ \\
$\log N_{\rm tot}$ & $19.96^{+0.01}_{-0.02}$ & $17.26^{+0.31}_{-0.08}$ & $19.97^{+0.01}_{-0.02}$ \\
\hline
\end{tabular}
\end{table}

\begin{table}
\caption{Fit results of \CI\ towards \Jfours.
\label{table:J0946_CI}
}
\begin{tabular}{cccc}
\hline
comp & 1 & 2 & $\log N_{\rm tot}$ \\
\hline
z & $2.606540(^{+5}_{-6})$ & $2.607271(^{+21}_{-15})$ &  \\
$\Delta$v, km\,s$^{-1}$ & 0.0 & 63.0 &  \\
b, km\,s$^{-1}$ & $4.6^{+0.3}_{-0.3}$ & $12.6^{+3.1}_{-1.8}$ &  \\
$\log N$(CI) & $14.10^{+0.05}_{-0.05}$ & $13.22^{+0.05}_{-0.06}$ & $14.15^{+0.04}_{-0.04}$ \\
$\log N$(CI*) & $13.89^{+0.03}_{-0.04}$ & $13.35^{+0.03}_{-0.03}$ & $14.10^{+0.02}_{-0.03}$ \\
$\log N$(CI**) & $13.28^{+0.05}_{-0.04}$ & $12.83^{+0.10}_{-0.18}$ & $13.41^{+0.05}_{-0.05}$ \\
$\log N_{\rm tot}$ & $14.33^{+0.05}_{-0.02}$ & $13.64^{+0.03}_{-0.02}$ & $14.42^{+0.04}_{-0.02}$ \\
\hline
\end{tabular}
\end{table}

\begin{table}
\centering
\caption{Comparison of observed column densities in the component at $z=2.60641$ towards \Jfours\ with the model ones, calculated at the best fit value of number density and UV field from Fig.~\ref{fig:J0946_phys}.
\label{table:J0946_phys}}
\begin{tabular}{ccc}
\hline
species & \multicolumn{2}{c}{$\log N$}\\ 
& observed & model \\
\hline
CI & $14.10^{+0.05}_{-0.05}$ & 14.09$^a$ \\
CI* & $13.89^{+0.03}_{-0.04}$ & 13.92$^a$ \\
CI** & $13.28^{+0.05}_{-0.04}$ & 13.25$^a$ \\
H$_2$ J0 & $19.42^{+0.04}_{-0.02}$ & 19.53 \\
H$_2$ J1 & $19.81^{+0.02}_{-0.02}$ & 19.72 \\
H$_2$ J2 & $18.10^{+0.12}_{-0.17}$ & 17.87 \\
\hline
H$_2$ J3$^b$ &  & 16.29 \\
H$_2$ J4$^b$ &  & 15.19 \\
H$_2$ J5$^b$ &  & 14.70 \\
\hline
\end{tabular}
\begin{tablenotes}
    \item (a) These values were obtained under the additional assumption that the total \CI\ column density equates the observed one.
    \item (b) We did not use these levels to compare with observed values, since the column densities measurements for these levels are subject to a large systematic uncertainty (see text).
\end{tablenotes}
\end{table}

\begin{table*}
\caption{Fit results of metals towards \Jfours.
\label{table:J0946_metals}
}
\begin{tabular}{cccccccccc}
\hline
comp & z & $\Delta$v, km\,s$^{-1}$ & b, km\,s$^{-1}$ & $\log N$(SiII) & $\log N$(FeII) & $\log N$(SII) & $\log N$(NiII) & $\log N$(CrII) & $\log N$(ZnII) \\
\hline
1 & $2.605717(^{+9}_{-8})$ & -74.5 & $8.7^{+0.8}_{-0.6}$ & $13.74^{+0.06}_{-0.06}$ & $13.36^{+0.03}_{-0.03}$ &  &  &  &  \\
2 & $2.606617(^{+9}_{-11})$ & 0.0 & $21.7^{+0.6}_{-0.7}$ & $15.53^{+0.03}_{-0.03}$ & $14.99^{+0.03}_{-0.02}$ & $15.35^{+0.02}_{-0.02}$ & $13.81^{+0.02}_{-0.03}$ & $13.33^{+0.03}_{-0.03}$ & $12.96^{+0.04}_{-0.02}$ \\
3 & $2.607342(^{+18}_{-15})$ & 58.2 & $17.8^{+3.3}_{-2.8}$ & $15.47^{+0.03}_{-0.05}$ & $14.76^{+0.03}_{-0.04}$ & $15.37^{+0.02}_{-0.02}$ & $13.58^{+0.06}_{-0.07}$ & $13.14^{+0.05}_{-0.07}$ & $12.97^{+0.03}_{-0.05}$ \\
4 & $2.60812(^{+4}_{-4})$ & 119.9 & $35.9^{+4.8}_{-3.8}$ & $15.26^{+0.06}_{-0.07}$ & $14.66^{+0.04}_{-0.04}$ & $15.14^{+0.05}_{-0.04}$ & $13.64^{+0.08}_{-0.05}$ & $12.95^{+0.09}_{-0.11}$ & $12.82^{+0.04}_{-0.06}$ \\
5 & $2.609050(^{+14}_{-20})$ & 212.0 & $11.5^{+1.4}_{-2.2}$ & $14.75^{+0.10}_{-0.10}$ & $14.10^{+0.04}_{-0.04}$ & $14.01^{+0.14}_{-0.16}$ & $13.21^{+0.14}_{-0.14}$ & $12.69^{+0.10}_{-0.14}$ & $12.31^{+0.09}_{-0.10}$ \\
6 & $2.609836(^{+9}_{-10})$ & 269.7 & $39.2^{+0.5}_{-0.6}$ & $15.44^{+0.02}_{-0.02}$ & $14.94^{+0.01}_{-0.01}$ & $15.09^{+0.01}_{-0.01}$ & $13.82^{+0.05}_{-0.04}$ & $13.21^{+0.04}_{-0.05}$ & $12.82^{+0.03}_{-0.04}$ \\
$\log N_{\rm tot}$ &  &  &  & $16.06^{+0.01}_{-0.01}$ & $15.48^{+0.01}_{-0.01}$ & $15.86^{+0.01}_{-0.01}$ & $14.36^{+0.02}_{-0.02}$ & $13.81^{+0.02}_{-0.02}$ & $13.52^{+0.02}_{-0.01}$ \\
$\rm [X/H]$ &  &  &  & $-0.65^{+0.02}_{-0.02}$ & $-1.22^{+0.02}_{-0.02}$ & $-0.46^{+0.02}_{-0.02}$ & $-1.06^{+0.03}_{-0.03}$ & $-1.03^{+0.03}_{-0.03}$ & $-0.24^{+0.03}_{-0.02}$ \\
$\rm [X/SII]$ &  &  &  & $-0.19^{+0.02}_{-0.02}$ & $-0.76^{+0.01}_{-0.02}$ & $0.00^{+0.01}_{-0.02}$ & $-0.60^{+0.02}_{-0.03}$ & $-0.57^{+0.02}_{-0.03}$ & $0.22^{+0.02}_{-0.02}$ \\
\hline
\end{tabular}
\end{table*}

\clearpage

\subsection{\Jfive}

\begin{figure}
\centering
\includegraphics[trim={0.0cm 0.0cm 0.0cm 0.0cm},clip,width=\hsize]{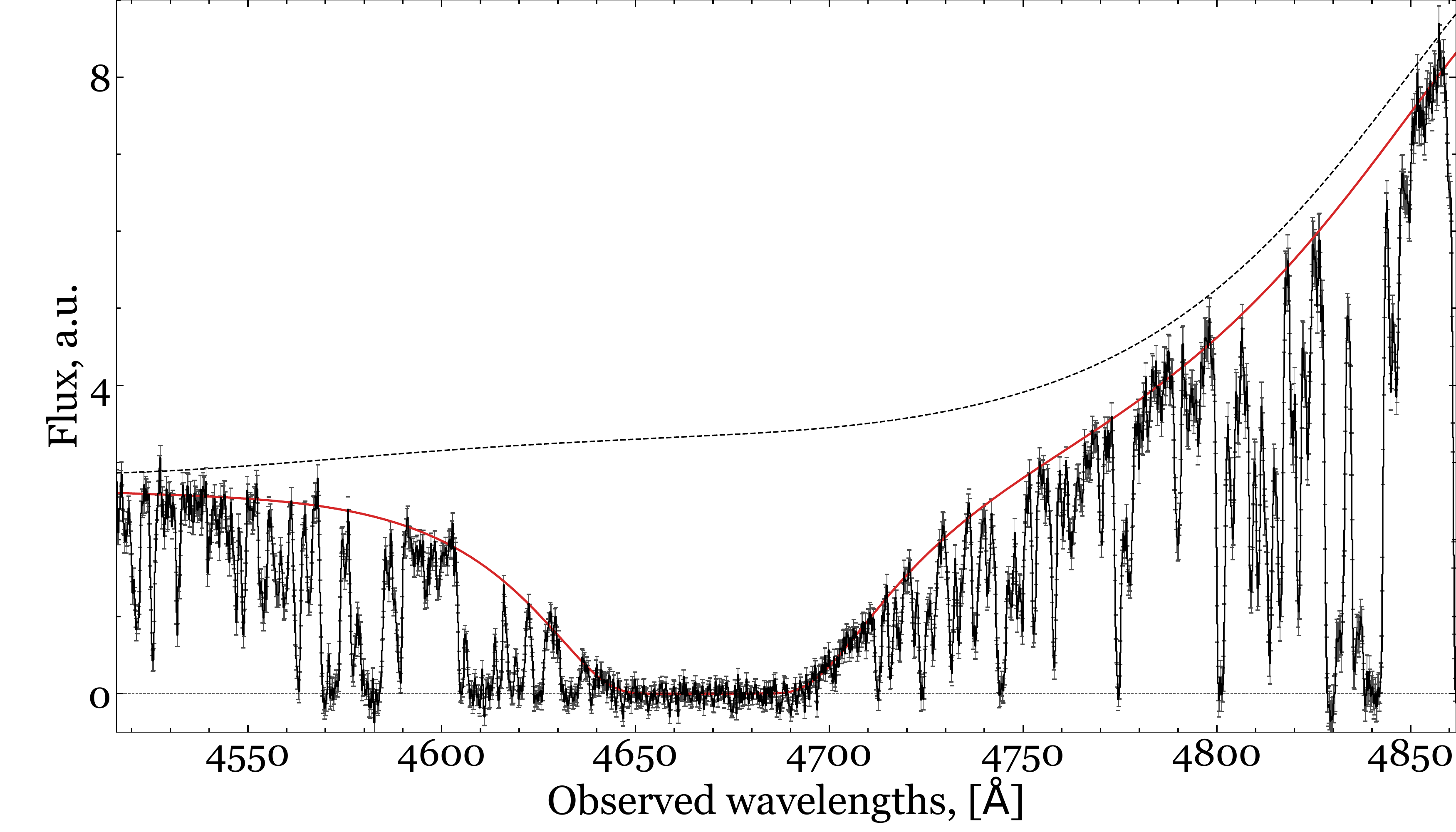}
		\caption{Fit to the damped \HI\ Ly$\alpha$ absorption line at $z=2.840$ towards \Jfives.
		\label{fig:J1146_HI}}
\end{figure}

We measured the \HI\ column density of the DLA at $z=2.840$ towards \Jfives\ by fitting the Ly$\alpha$ absorption line. We fitted together the continuum using seventh order Chebyshev polynomial in the region of Ly$\alpha$ line with a one-component absorption model. We derived $\log N(\HI)=21.54\pm0.01$ at the best fit value $z=2.84059$. The Ly$\alpha$ profile, together with the reconstructed quasar continuum are shown in Fig.~\ref{fig:J1146_HI}. 

\begin{figure*}
\centering
\includegraphics[trim={0.0cm 0.0cm 0.0cm 0.0cm},clip,width=\hsize]{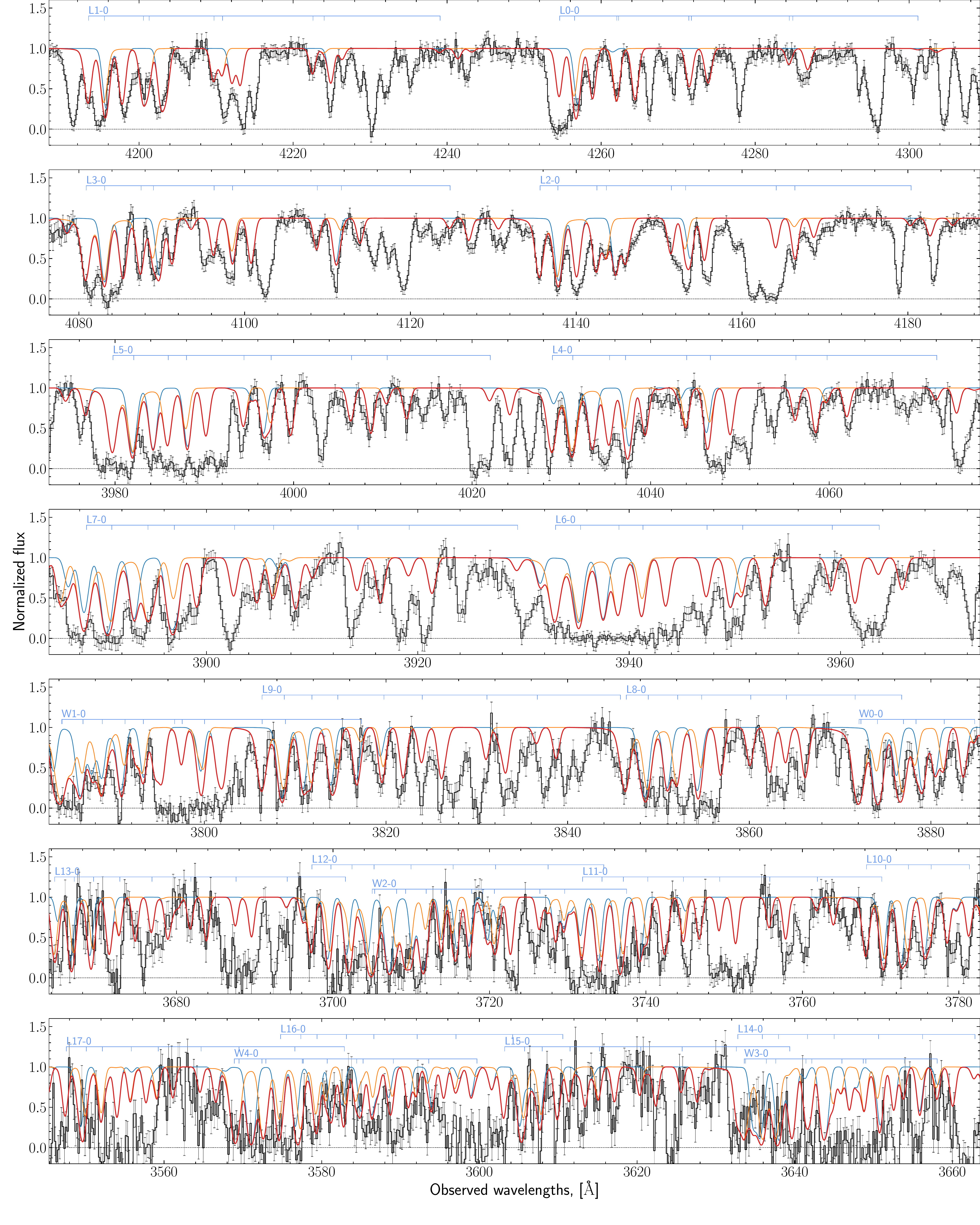}
		\caption{Fit to H$_2$ absorption lines at $z=2.840$ towards \Jfives.
		\label{fig:J1146_H2}}
\end{figure*}

\begin{figure}
\centering
\includegraphics[trim={0.0cm 0.0cm 0.0cm 0.0cm},clip,width=\hsize]{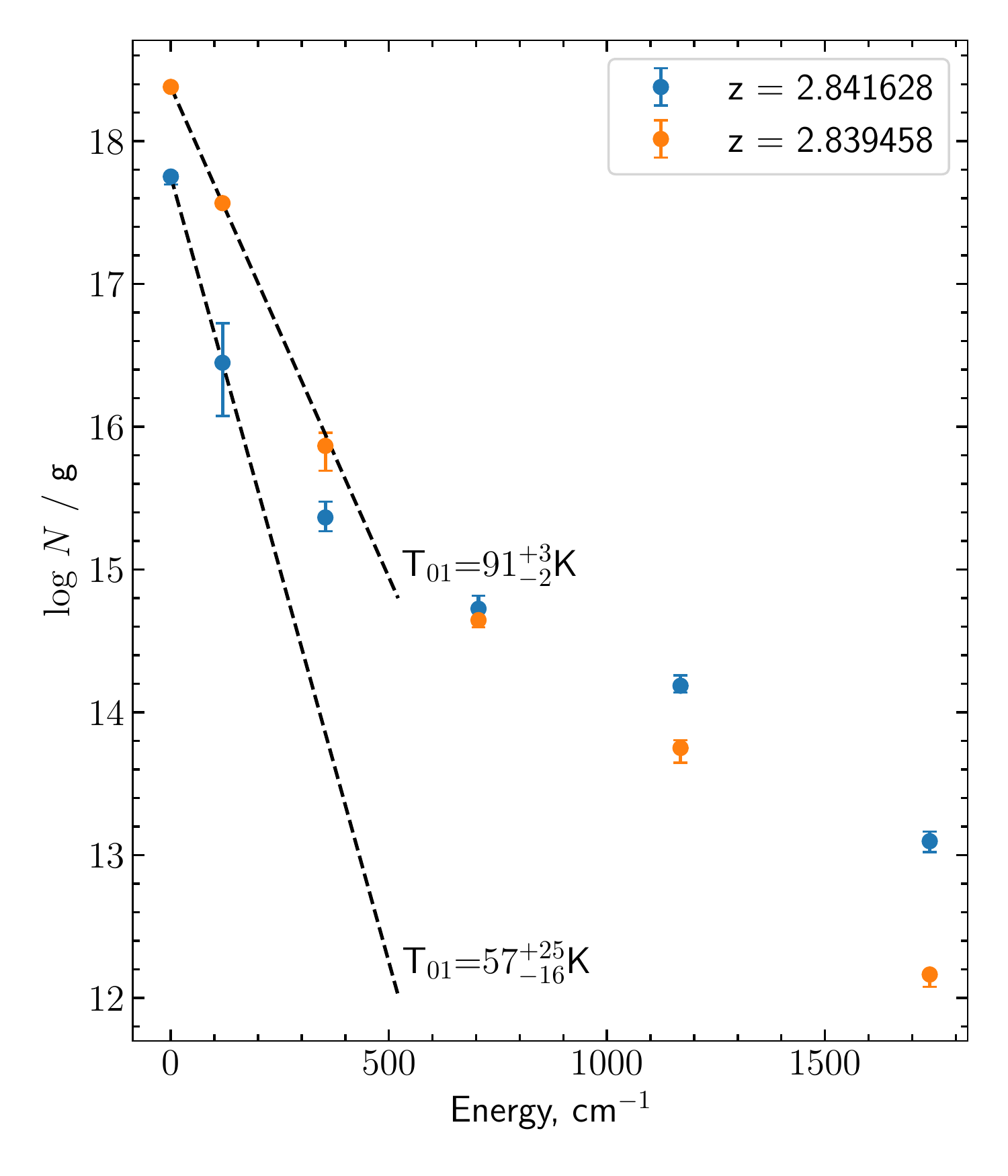}
		\caption{Excitation diagram for two H$_2$ components at $z=2.840$ towards \Jfives.
		\label{fig:J1146_H2_exc}}
\end{figure}

The fit of H$_2$ absorption lines was done using a two components model. The result of the fit is given in Table~\ref{table:J1146_H2} and H$_2$ line profiles are shown in Fig.~\ref{fig:J1146_H2}. The excitation diagrams of H$_2$ rotation levels for both components are shown in Fig.~\ref{fig:J1146_H2_exc}, with derived $T_{01}$ temperatures   
of $91^{+3}_{-2}$\,K and $57^{+25}_{-16}$\,K at $z=2.83946$ and 2.84163, respectively.  

\begin{figure}
    \centering
    \includegraphics[trim={0.0cm 0.0cm 0.0cm 0.0cm},clip,width=\hsize]{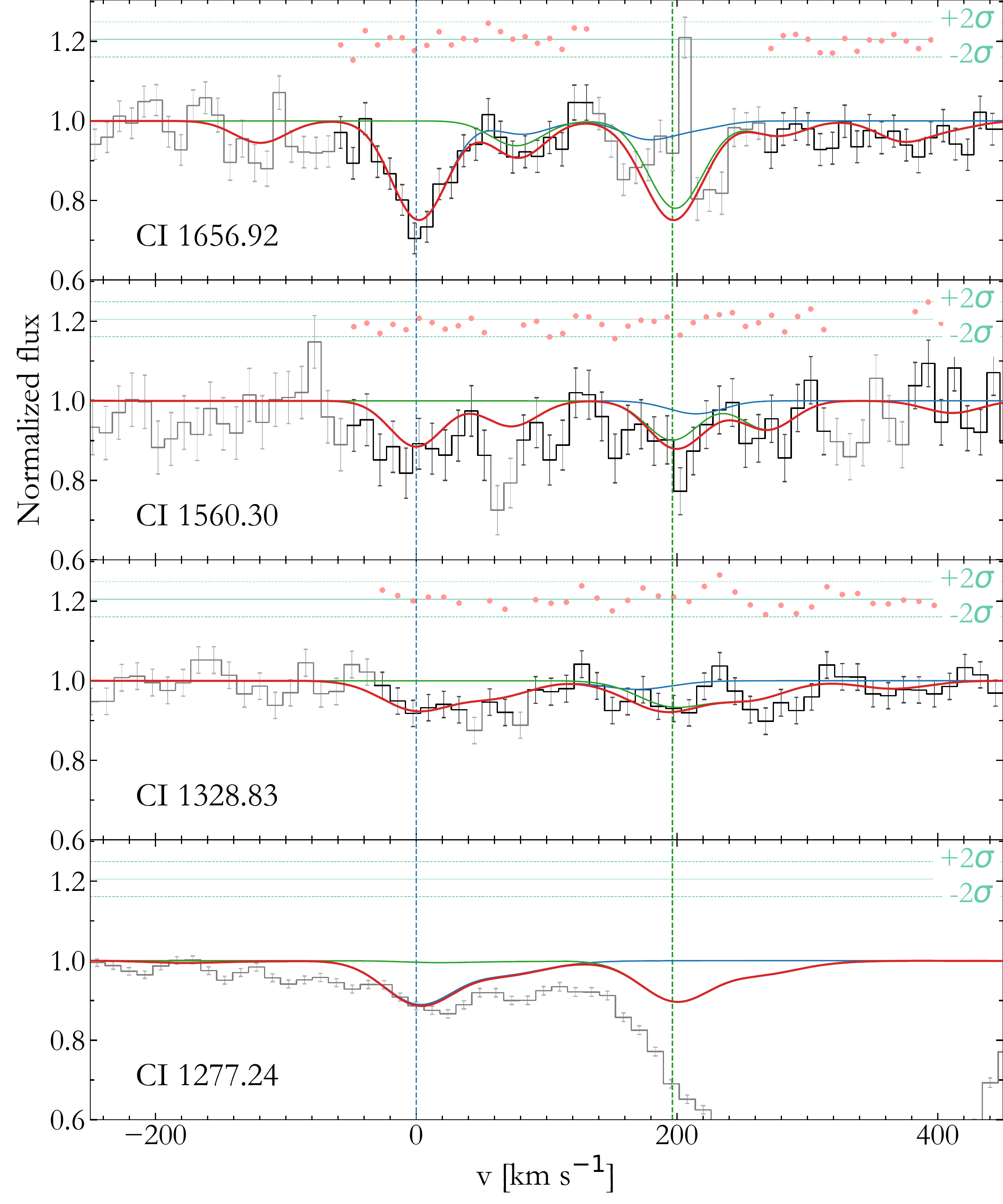}
	\caption{Fit to the \CI\ absorption lines at $z=2.840$ towards \Jfives. }
	\label{fig:J1146_CI}
\end{figure}

We detected \CI\ absorption lines in two components arising from three fine structure levels, coinciding with the H$_2$ components of DLA. 
The red component of \CI\,1656.92\AA\ line coincides with sky line, therefore we excluded corresponding pixels from the fit.
The result of the fit is given in Table~\ref{table:J1146_CI} and the \CI\ absorption profiles are shown in Fig.~\ref{fig:J1146_CI}.

\begin{figure}
\centering
\includegraphics[trim={0.0cm 0.0cm 0.0cm 0.0cm},clip,width=\hsize]{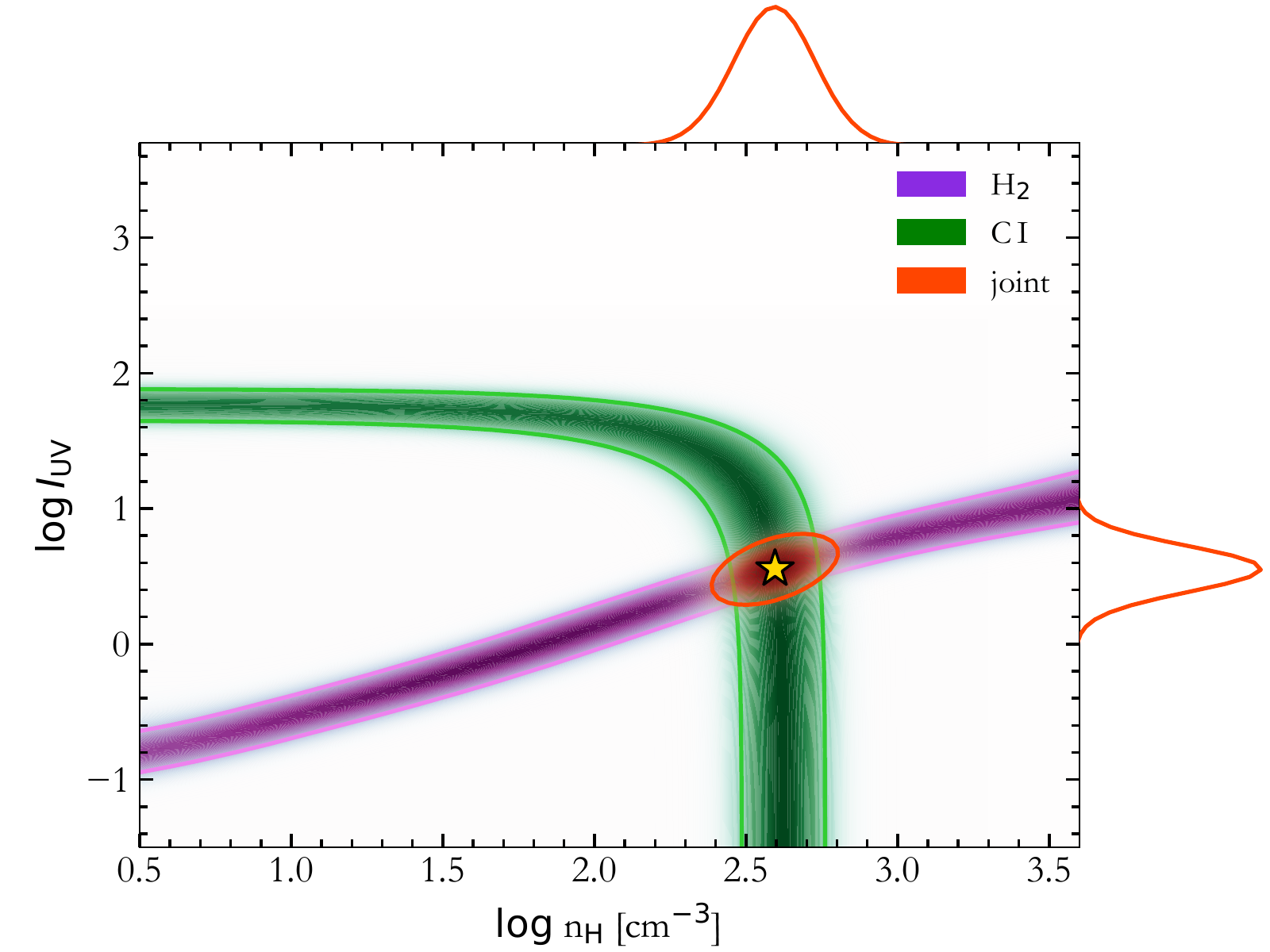}
		\caption{Constrains on the number density and UV field in H$_2$-bearing medium at $z=2.83946$ towards \Jfives\ from the excitation of \CI\ fine-structure and H$_2$ rotational levels. The colors and lines are the same as in Fig.~2.
		}
		\label{fig:J1146_phys_0}
\end{figure}

\begin{figure}
\centering
\includegraphics[trim={0.0cm 0.0cm 0.0cm 0.0cm},clip,width=\hsize]{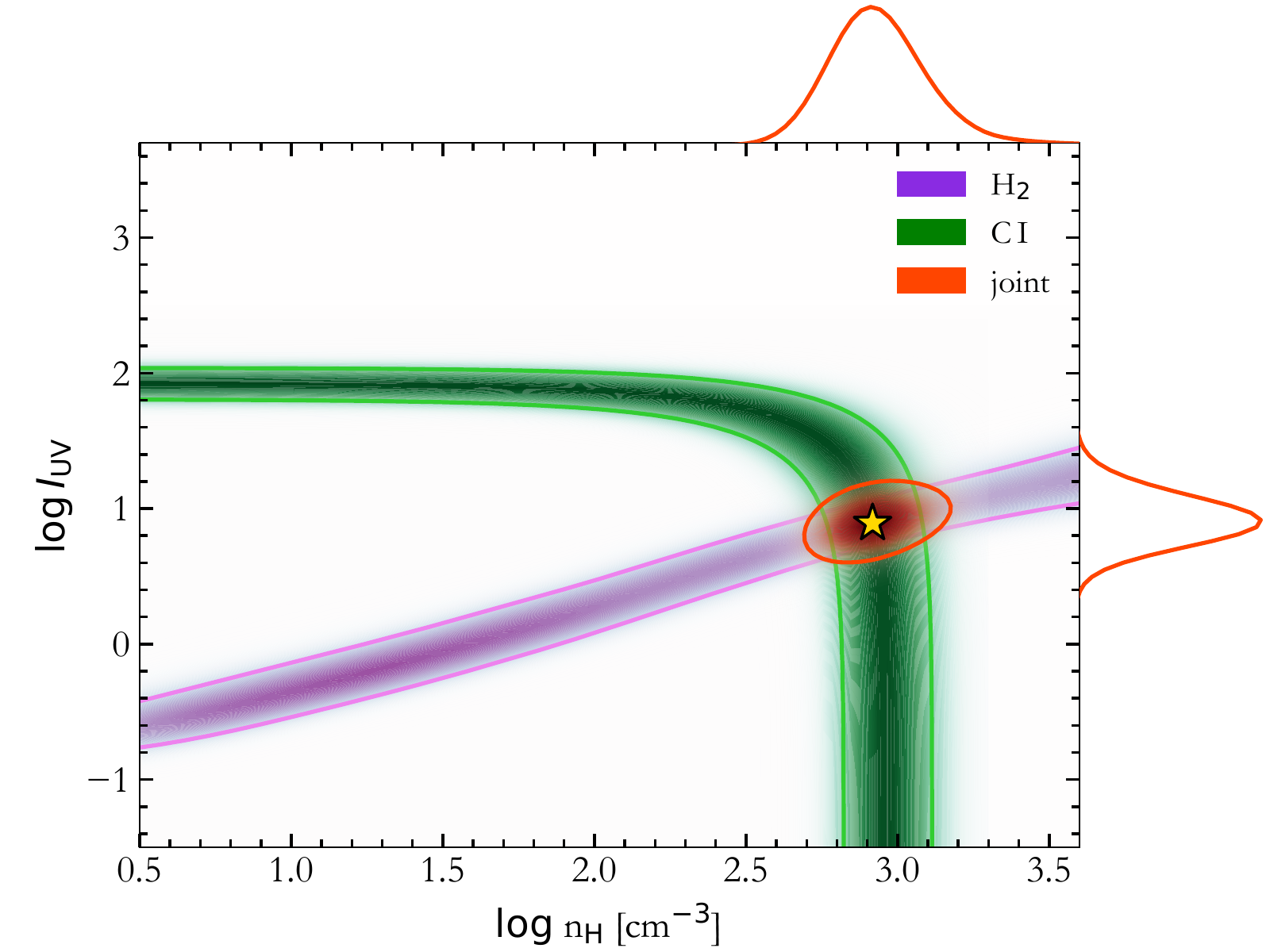}
		\caption{Constraints on the number density and UV field in H$_2$-bearing medium at $z=2.84163$ towards \Jfives\ from the excitation of \CI\ fine-structure and H$_2$ rotational levels. The colors and lines are the same as in Fig.~2.
		}
		\label{fig:J1146_phys_1}
\end{figure}

The constrained values of the hydrogen number density and UV field in the H$_2$/\CI-bearing medium, based on the excitation of H$_2$ and \CI\ levels are shown in Fig.~\ref{fig:J1146_phys_0} and \ref{fig:J1146_phys_1} for component at $z=2.83946$ and 2.84163, respectively.  The comparison between observed and model column densities at the best fit value of the parameters (shown by yellow star in Fig.~\ref{fig:J0136_phys}) is provided in Tables~\ref{table:J1146_phys_0} and \ref{table:J1146_phys_1}.

\begin{figure*}
\centering
\includegraphics[trim={0.0cm 0.0cm 0.0cm 0.0cm},clip,width=\hsize]{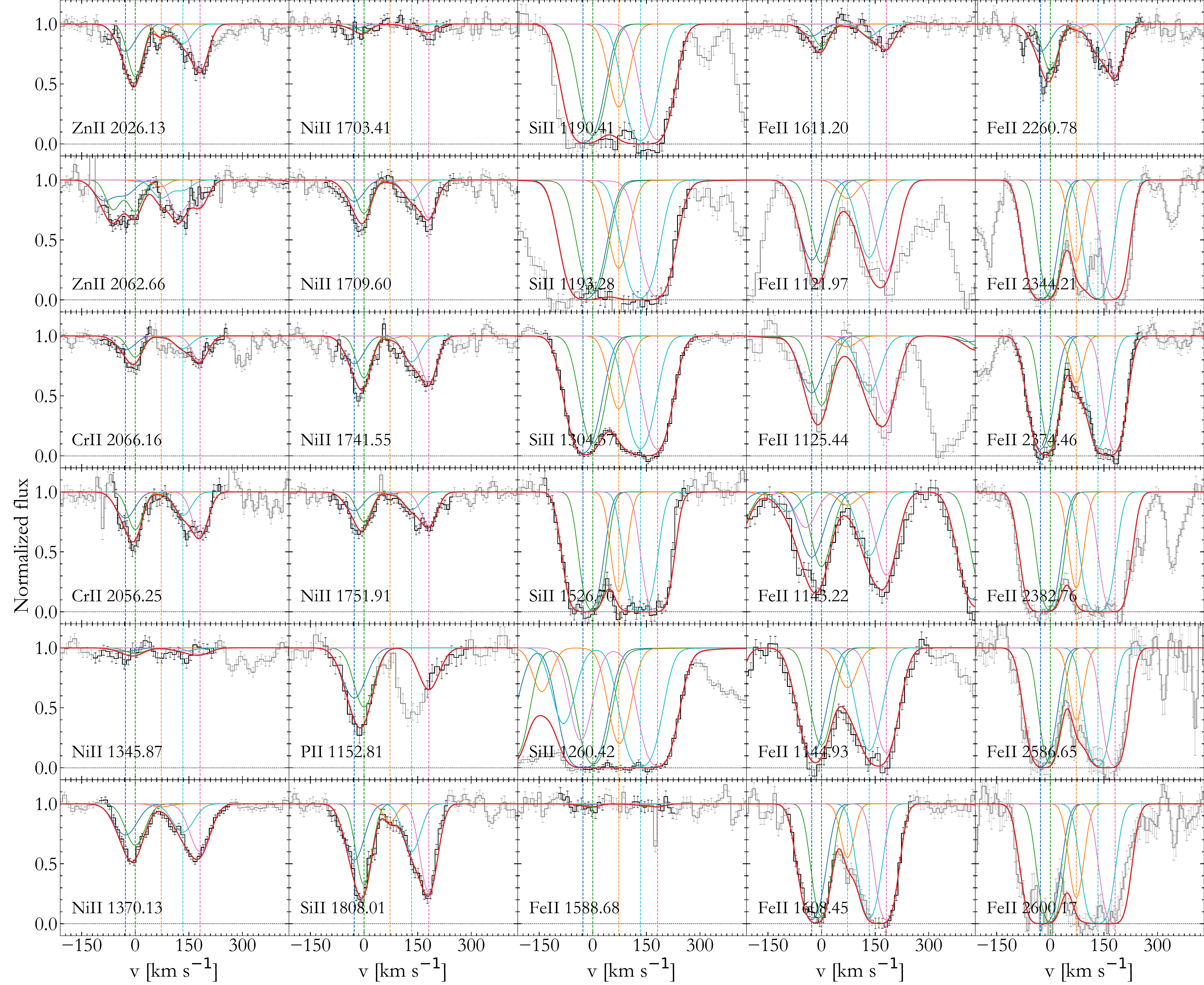}
		\caption{Fit to the metal absorption lines at $z=2.840$ towards \Jfives.}
		\label{fig:J1146_metals}
\end{figure*}

Metal lines are fitted using a five components model. The two strongest components seen in metal profiles coincide with the two components seen in H$_2$ absorption lines.
The results of the fit are given in Table~\ref{table:J1146_metals} and fit profiles are shown in Fig.~\ref{fig:J1146_metals}.

\begin{table}
\caption{Fit results of H$_2$ towards \Jfives.
\label{table:J1146_H2}
}
\begin{tabular}{cccc}
\hline
comp & 1 & 2 & $\log N_{\rm tot}$ \\
\hline
z & $2.839459(^{+19}_{-8})$ & $2.841629(^{+10}_{-24})$ &  \\
$\Delta$v, km\,s$^{-1}$ & 0 & 166.7 &  \\
b, km\,s$^{-1}$ & $7.6^{+0.1}_{-0.4}$ & $11.4^{+0.5}_{-0.7}$ &  \\
$\log N$(H$_2$ J0) & $18.38^{+0.01}_{-0.01}$ & $17.75^{+0.02}_{-0.05}$ & $18.47^{+0.01}_{-0.01}$ \\
$\log N$(H$_2$ J1) & $18.52^{+0.01}_{-0.01}$ & $17.40^{+0.28}_{-0.37}$ & $18.54^{+0.04}_{-0.01}$ \\
$\log N$(H$_2$ J2) & $16.57^{+0.09}_{-0.17}$ & $16.06^{+0.11}_{-0.10}$ & $16.67^{+0.11}_{-0.10}$ \\
$\log N$(H$_2$ J3) & $15.97^{+0.09}_{-0.05}$ & $16.05^{+0.09}_{-0.05}$ & $16.34^{+0.04}_{-0.06}$ \\
$\log N$(H$_2$ J4) & $14.70^{+0.05}_{-0.10}$ & $15.14^{+0.07}_{-0.05}$ & $15.28^{+0.05}_{-0.06}$ \\
$\log N$(H$_2$ J5) & $13.68^{+0.34}_{-2.18}$ & $14.62^{+0.07}_{-0.08}$ & $14.66^{+0.06}_{-0.10}$ \\
$\log N_{\rm tot}$ & $18.76^{+0.01}_{-0.01}$ & $17.94^{+0.11}_{-0.13}$ & $18.82^{+0.03}_{-0.02}$ \\
\hline
\end{tabular}
\end{table}

\begin{table}
\caption{Fit results of \CI\ towards \Jfives.
\label{table:J1146_CI}
}
\begin{tabular}{cccc}
\hline
comp & 1 & 2 & $\log N_{\rm tot}$ \\
\hline
z & $2.839015(^{+26}_{-18})$ & $2.841553(^{+3}_{-12})$ &  \\
$\Delta$v, km\,s$^{-1}$ & 0.0 & 196.4 &  \\
b, km\,s$^{-1}$ & $16.8^{+2.6}_{-4.0}$ & $2.7^{+14.9}_{-1.9}$ &  \\
$\log N$(CI) & $13.28^{+0.05}_{-0.04}$ & $13.17^{+0.10}_{-0.11}$ & $13.53^{+0.05}_{-0.05}$ \\
$\log N$(CI*) & $13.10^{+0.09}_{-0.10}$ & $13.07^{+0.05}_{-0.08}$ & $13.56^{+0.07}_{-0.04}$ \\
$\log N$(CI**) & $12.93^{+0.12}_{-0.18}$ & $12.85^{+0.12}_{-0.11}$ & $13.23^{+0.04}_{-0.16}$ \\
$\log N_{\rm tot}$ & $13.59^{+0.03}_{-0.04}$ & $13.51^{+0.05}_{-0.06}$ & $13.85^{+0.03}_{-0.03}$ \\
\hline
\end{tabular}
\end{table}

\begin{table}
\centering
\caption{Comparison of observed column densities in the component at $z=2.83946$ of DLA towards \Jfives\ with the model ones, calculated at the best fit value of number density and UV field from Fig.~\ref{fig:J1146_phys_0}.
\label{table:J1146_phys_0}
}
\begin{tabular}{ccc}
\hline
species & \multicolumn{2}{c}{$\log N$}\\ 
& observed & model \\
\hline
CI & $13.28^{+0.05}_{-0.04}$ & 13.28$^a$ \\
CI* & $13.10^{+0.09}_{-0.10}$ & 13.22$^a$ \\
CI** & $12.93^{+0.12}_{-0.18}$ & 12.64$^a$ \\
H$_2$ J0 & $18.38^{+0.01}_{-0.01}$ & 18.38 \\
H$_2$ J1 & $18.52^{+0.01}_{-0.01}$ & 18.47 \\
H$_2$ J2 & $16.57^{+0.09}_{-0.17}$ & 16.47 \\
\hline
H$_2$ J3$^{b}$ &  & 15.53 \\
H$_2$ J4$^{b}$ &  & 14.04 \\
H$_2$ J5$^{b}$ &  & 13.62 \\
\end{tabular}
\begin{tablenotes}
    \item (a) These values were obtained uder the additional assumption that the total \CI\ column density equates the observed one. 
    \item (b) We did not use these levels to compare with observed values, since the column densities measurements for these levels are subject to large systematic uncertainty (see text).
\end{tablenotes}
\end{table}

\begin{table}
\centering
\caption{Comparison of observed column densities in the component at z=2.84163 of DLA in \Jfives\ with the modelled ones, taken at the best fit values of number density and UV field from Fig.~\ref{fig:J1146_phys_1}.
\label{table:J1146_phys_1}
}
\begin{tabular}{ccc}
\hline
species & \multicolumn{2}{c}{$\log N$}\\ 
& observed & model \\
\hline
CI & $13.17^{+0.10}_{-0.11}$ & 13.15$^a$ \\
CI* & $13.07^{+0.05}_{-0.08}$ & 13.18$^a$ \\
CI** & $12.85^{+0.12}_{-0.11}$ & 12.67$^a$ \\
H$_2$ J0 & $17.75^{+0.02}_{-0.05}$ & 17.28 \\
H$_2$ J1 & $17.40^{+0.28}_{-0.37}$ & 17.51 \\
H$_2$ J2 & $16.06^{+0.11}_{-0.10}$ & 15.84 \\
\hline
H$_2$ J3$^b$ &  & 14.86 \\
H$_2$ J4$^b$ &  & 13.68 \\
H$_2$ J5$^b$ &  & 13.23 \\
\end{tabular}
\begin{tablenotes}
    \item (a) These values were obtained under the additional assumption that the total \CI\ column density equates the observed one. 
    \item (b) We did not use these levels to compare with observed values, since the column densities measurements for these levels are subject to a large systematic uncertainty (see text).
\end{tablenotes}
\end{table}

\begin{table*}
\caption{Fit results of metals towards \Jfives.
\label{table:J1146_metals}
}
\begin{tabular}{ccccccccc}
\hline
comp & 1 & 2 & 3 & 4 & 5 & $\log N_{\rm tot}$ & $\rm [X/H]$ & $\rm [X/ZnII]$ \\
\hline
z & $2.838898(^{+38}_{-25})$ & $2.839268(^{+16}_{-18})$ & $2.840190(^{+24}_{-26})$ & $2.84102(^{+4}_{-3})$ & $2.841599(^{+16}_{-13})$ &  &  &  \\
$\Delta$v, km\,s$^{-1}$ & -209.7 & -182.2 & -108.3 & -45.8 & 0.0 &  &  &  \\
b, km\,s$^{-1}$ & $29.0^{+1.0}_{-1.0}$ & $18.0^{+1.0}_{-1.0}$ & $11.8^{+1.7}_{-1.0}$ & $28.5^{+2.9}_{-4.4}$ & $21.0^{+0.6}_{-0.5}$ &  &  &  \\
$\log N$(SiII) & $15.65^{+0.04}_{-0.10}$ & $15.84^{+0.07}_{-0.06}$ & $14.91^{+0.08}_{-0.07}$ & $15.56^{+0.05}_{-0.07}$ & $15.89^{+0.04}_{-0.03}$ & $16.36^{+0.02}_{-0.01}$ & $-0.69^{+0.02}_{-0.02}$ & $-0.13^{+0.02}_{-0.02}$ \\
$\log N$(FeII) & $15.06^{+0.06}_{-0.08}$ & $15.27^{+0.04}_{-0.06}$ & $14.14^{+0.07}_{-0.06}$ & $15.09^{+0.06}_{-0.05}$ & $15.31^{+0.04}_{-0.03}$ & $15.81^{+0.01}_{-0.01}$ & $-1.23^{+0.01}_{-0.01}$ & $-0.68^{+0.01}_{-0.02}$ \\
$\log N$(ZnII) & $12.81^{+0.07}_{-0.09}$ & $13.07^{+0.05}_{-0.04}$ & $12.32^{+0.06}_{-0.07}$ & $12.63^{+0.09}_{-0.06}$ & $13.02^{+0.03}_{-0.03}$ & $13.54^{+0.01}_{-0.01}$ & $-0.56^{+0.02}_{-0.02}$ & $0.00^{+0.02}_{-0.02}$ \\
$\log N$(CrII) & $13.49^{+0.06}_{-0.08}$ & $13.56^{+0.06}_{-0.06}$ & $12.33^{+0.24}_{-0.61}$ & $13.41^{+0.05}_{-0.07}$ & $13.64^{+0.04}_{-0.03}$ & $14.13^{+0.01}_{-0.01}$ & $-1.05^{+0.02}_{-0.01}$ & $-0.49^{+0.02}_{-0.02}$ \\
$\log N$(NiII) & $13.94^{+0.07}_{-0.06}$ & $14.09^{+0.04}_{-0.05}$ & $12.87^{+0.16}_{-0.20}$ & $13.92^{+0.05}_{-0.06}$ & $14.12^{+0.03}_{-0.04}$ & $14.63^{+0.01}_{-0.01}$ & $-1.13^{+0.01}_{-0.01}$ & $-0.57^{+0.01}_{-0.01}$ \\
$\log N$(PII) & $13.79^{+0.07}_{-0.12}$ & $13.93^{+0.10}_{-0.12}$ &  &  & $13.53^{+0.10}_{-0.08}$ & $14.24^{+0.04}_{-0.05}$ & $-0.71^{+0.04}_{-0.05}$ & $-0.15^{+0.04}_{-0.05}$ \\
\hline
\end{tabular}
\end{table*}

\clearpage

\subsection{\Jsix}

\begin{figure}
\centering
\includegraphics[trim={0.0cm 0.0cm 0.0cm 0.0cm},clip,width=\hsize]{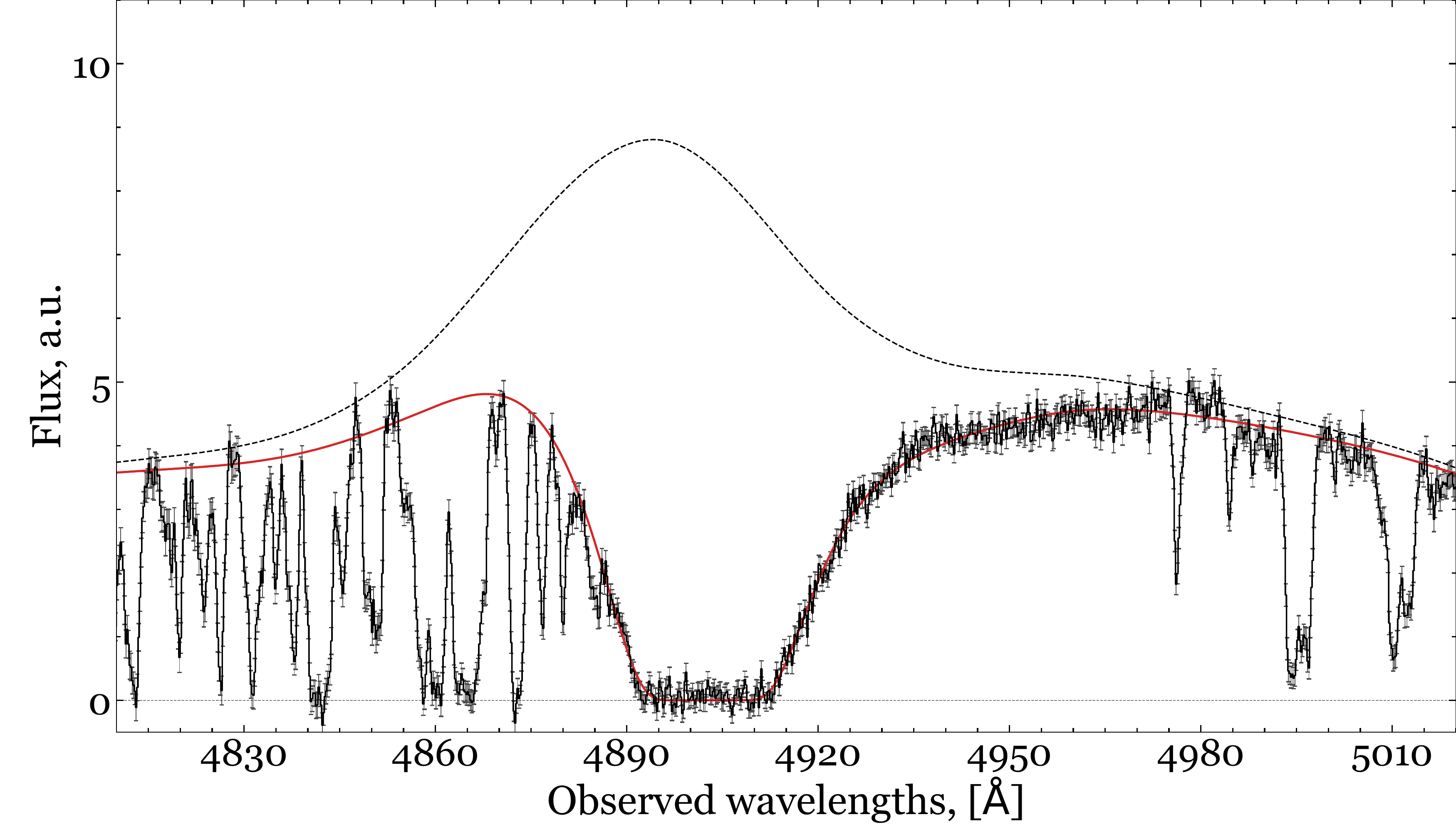}
		\caption{Fit to the damped \HI\ Ly$\alpha$ absorption line at $z=3.033$ towards \Jsixs.
		\label{fig:J1236_HI}}
\end{figure}

We measured the \HI\ column density of the DLA at $z=3.033$ towards \Jsixs\ by fitting the Ly$\alpha$ absorption line with a single component together with the continuum modelled using a seventh order Chebyshev polynomial. We derived $\log N(\HI)=20.74\pm0.01$ at the best fit value of $z=3.03284$. The Ly$\alpha$ profile, together with the reconstructed quasar continuum are shown in Fig.~\ref{fig:J1236_HI}. Interestingly, that reconstructed continuum indicates that DLA is redshifted regarding Ly$\alpha$ emission line of quasar by $\sim500$ km\,s$^{-1}$ (see Fig.~\ref{fig:J1236_HI}).

\begin{figure*}
\centering
\includegraphics[trim={0.0cm 0.0cm 0.0cm 0.0cm},clip,width=\hsize]{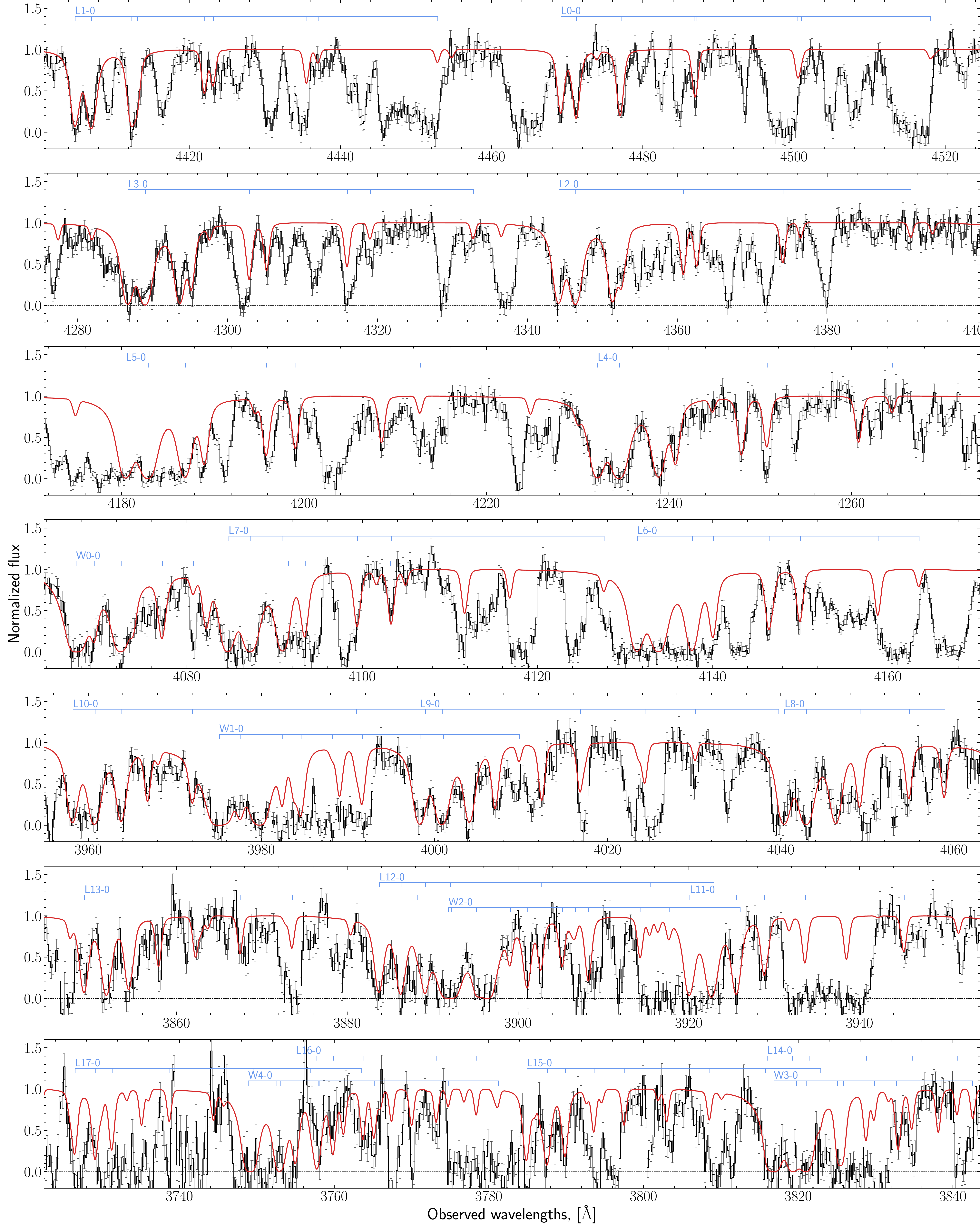}
		\caption{Fit to the H$_2$ absorption lines at $z=3.033$ towards \Jsixs.
		\label{fig:J1236_H2}}
\end{figure*}

\begin{figure}
\centering
\includegraphics[trim={0.0cm 0.0cm 0.0cm 0.0cm},clip,width=\hsize]{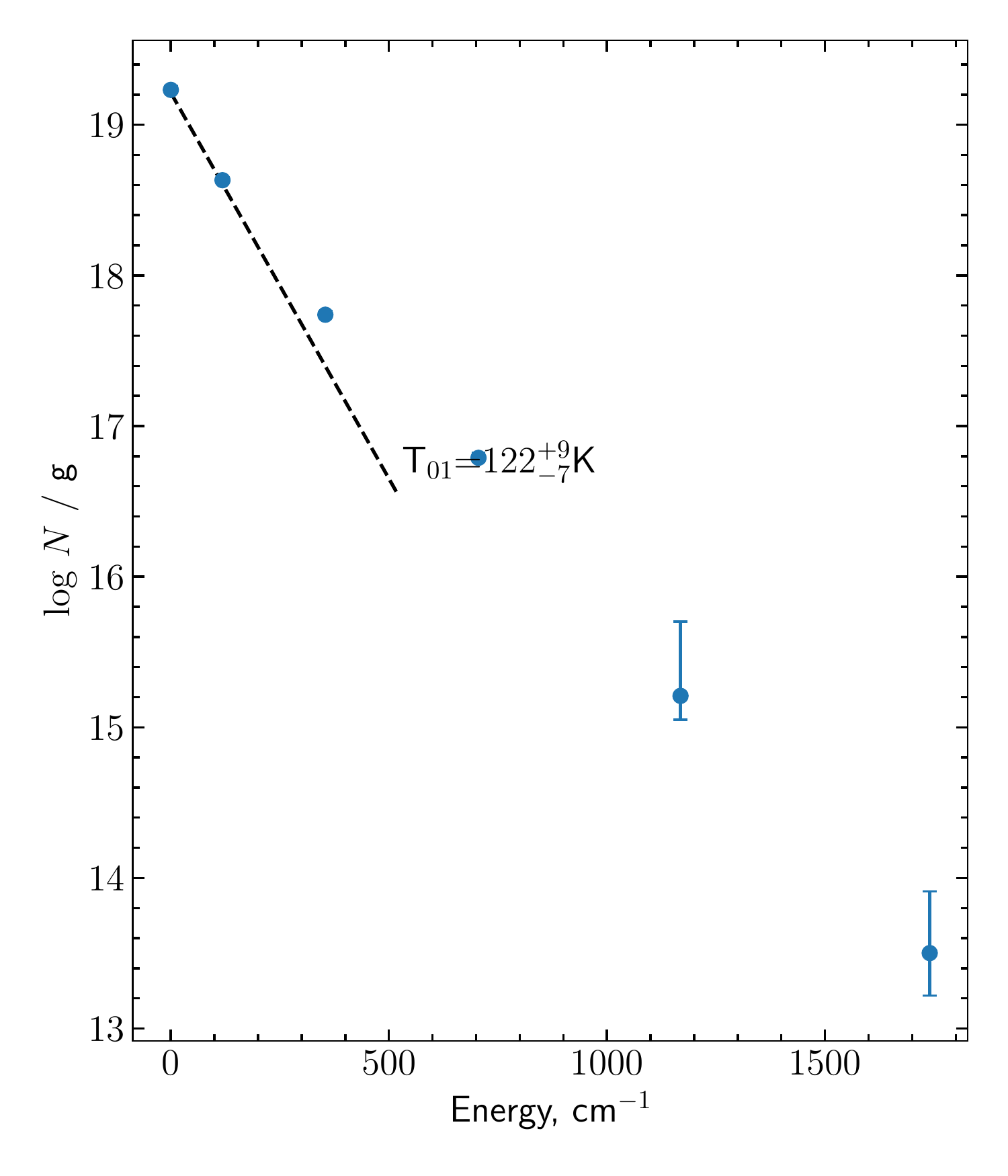}
		\caption{Excitation diagram of H$_2$ at $z=3.033$ towards \Jsixs.
		\label{fig:J1236_H2_exc}}
\end{figure}

We fitted the H$_2$ absorption lines towards \Jsixs\ using a one component model. The result of the fit is given in Table~\ref{table:J1236_H2} and H$_2$ line profiles are shown in Fig.~\ref{fig:J1236_H2}. The excitation diagram of H$_2$ rotation levels is shown in Fig.~\ref{fig:J1236_H2_exc} and we derived $T_{01} = 122^{+9}_{-7}$~K. 

\begin{figure}
    \centering
    \includegraphics[trim={0.0cm 0.0cm 0.0cm 0.0cm},clip,width=\hsize]{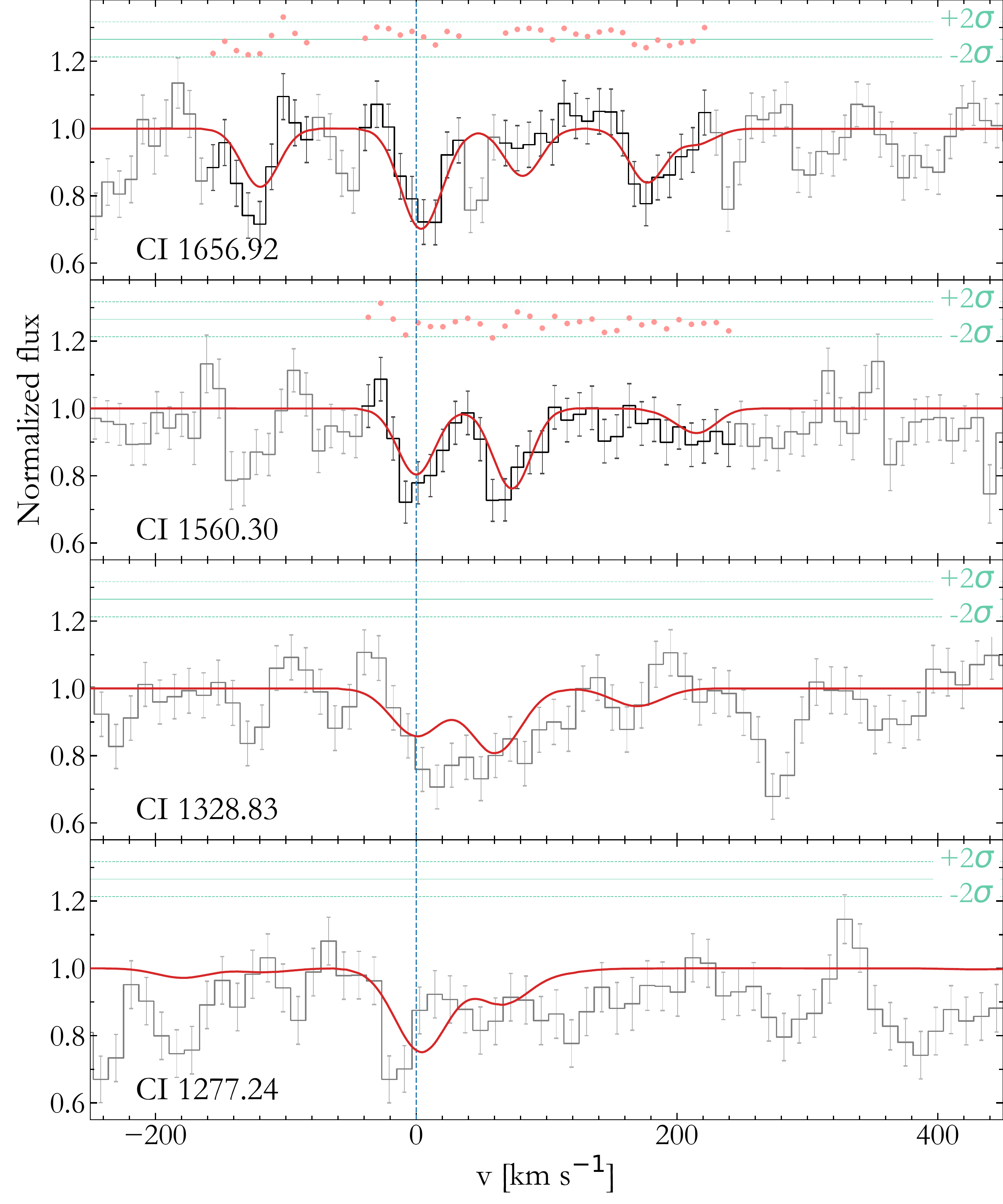}
	\caption{Fit to the \CI\ absorption lines at $z=3.033$ towards \Jsixs.}
	\label{fig:J1236_CI}
\end{figure}

We detected \CI\ absorption lines in three fine structure levels in a single component corresponding to H$_2$. 
The result from fitting these lines is given in Table~\ref{table:J1236_CI} and the corresponding profiles are shown in Fig.~\ref{fig:J1236_CI}. 

\begin{figure}
\centering
\includegraphics[trim={0.0cm 0.0cm 0.0cm 0.0cm},clip,width=\hsize]{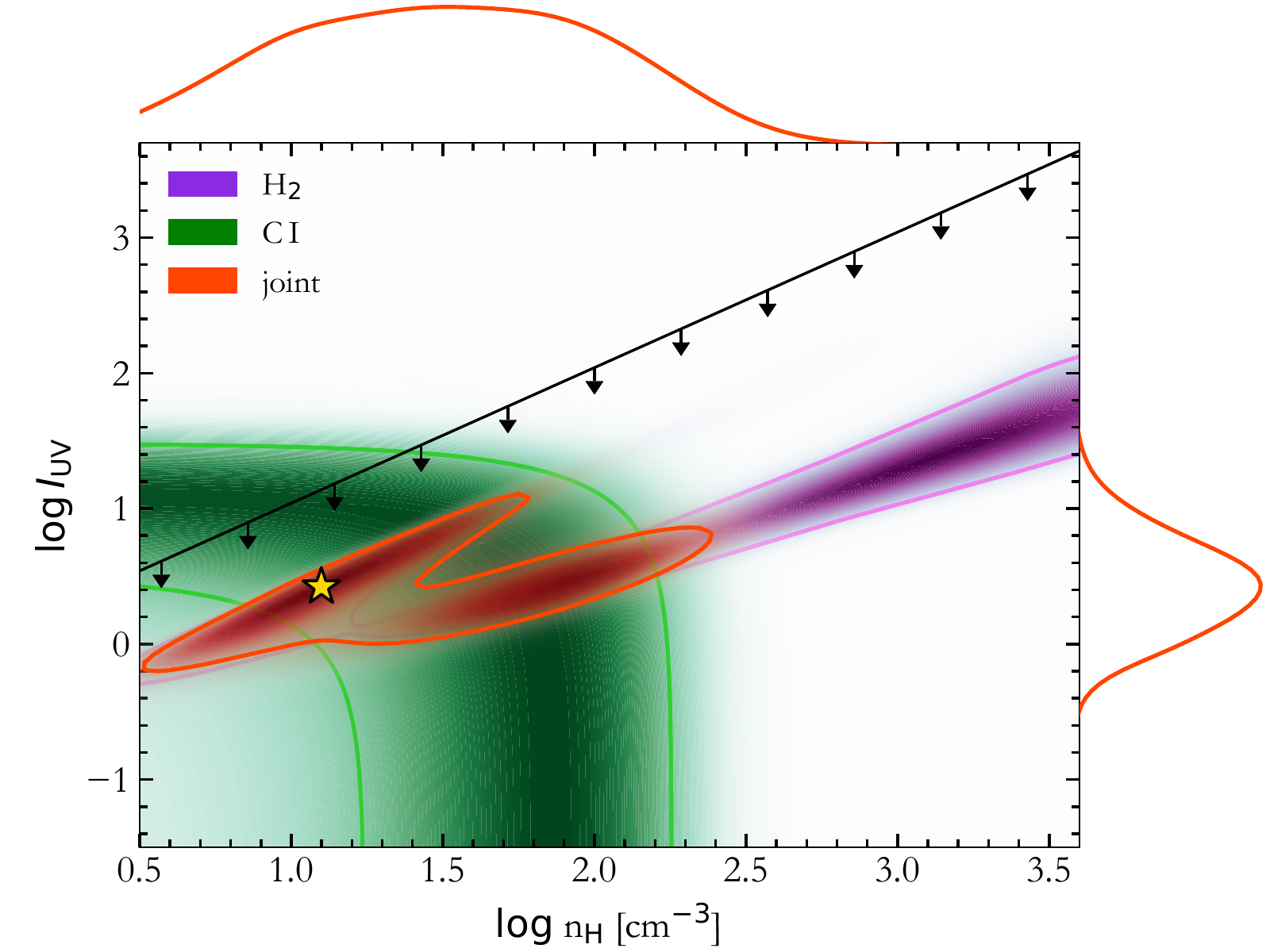}
		\caption{Constarints on the number density and UV field in H$_2$-bearing medium at $z=3.033$ towards \Jsixs\ from the excitation of \CI\ fine-structure and H$_2$ rotational levels. The colors and lines are the same as in Fig.~2.
		}
		\label{fig:J1236_phys}
\end{figure}

The constrained values of the hydrogen number density and UV field in the H$_2$/\CI-bearing medium are shown in Fig.~\ref{fig:J1236_phys} and the comparison between observed and model column densities at the best fit value (yellow star in Fig.~\ref{fig:J1236_phys}) is provided in Table~\ref{table:J1236_phys}. 

\begin{figure*}
\centering
\includegraphics[trim={0.0cm 0.0cm 0.0cm 0.0cm},clip,width=\hsize]{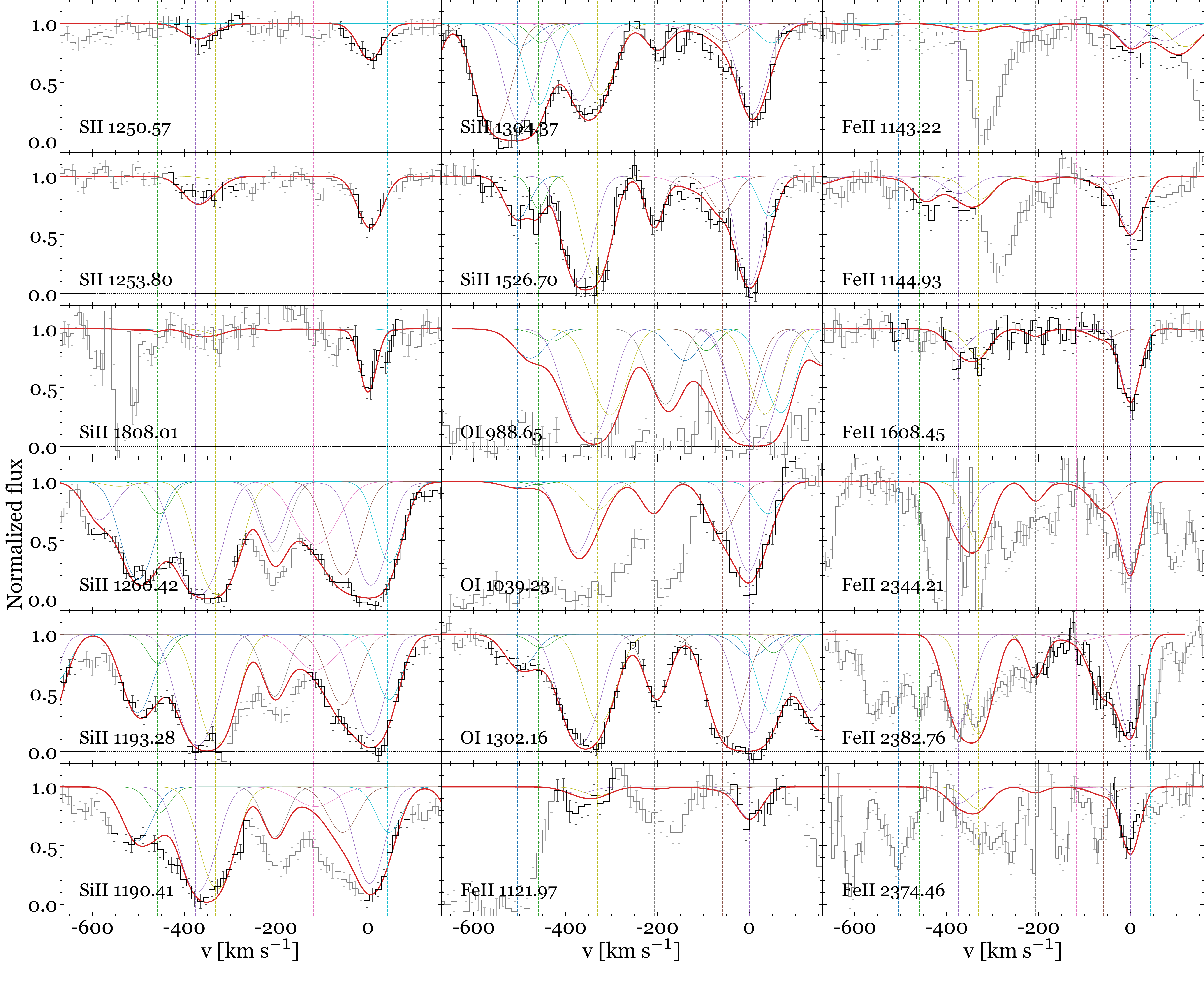}
		\caption{Fit to the metal absorption lines at $z=3.033$ towards \Jsixs.}
		\label{fig:J1236_metals}
\end{figure*}

Metal lines are fitted using a nine components model. The second component of metals from the red side coincides with position of H$_2$/\CI-bearing component of the DLA. 
The results of the fit are given in Table~\ref{table:J1236_metals} and the fit profiles are shown in Fig.~\ref{fig:J1236_metals}.


\begin{table}
\caption{Fit results of H$_2$ towards \Jsixs.
\label{table:J1236_H2}
}
\begin{tabular}{cc}
\hline
z & $3.03292(^{+27}_{-18})$ \\
b, km\,s$^{-1}$ & $2.3^{+0.2}_{-0.2}$ \\
$\log N$(H$_2$ J0) & $19.21^{+0.03}_{-0.02}$ \\
$\log N$(H$_2$ J1) & $19.56^{+0.02}_{-0.00}$ \\
$\log N$(H$_2$ J2) & $18.46^{+0.03}_{-0.02}$ \\
$\log N$(H$_2$ J3) & $18.10^{+0.04}_{-0.02}$ \\
$\log N$(H$_2$ J4) & $16.10^{+0.49}_{-0.16}$ \\
$\log N$(H$_2$ J5) & $15.11^{+0.41}_{-0.28}$ \\
$\log N_{\rm tot}$ & $19.76^{+0.01}_{-0.01}$ \\
\hline
\end{tabular}
\end{table}

\begin{table}
\caption{Fit results of \CI\ towards \Jsixs.
\label{table:J1236_CI}
}
\begin{tabular}{cc}
\hline
z & $3.032969(^{+17}_{-8})$ \\
b, km\,s$^{-1}$ & $2.0^{+0.4}_{-0.2}$ \\
$\log N$(CI) & $14.28^{+0.36}_{-0.37}$ \\
$\log N$(CI*) & $13.70^{+0.11}_{-0.08}$ \\
$\log N$(CI**) & $13.18^{+0.11}_{-0.16}$ \\
$\log N_{\rm tot}$ & $14.35^{+0.33}_{-0.25}$ \\
\hline
\end{tabular}
\end{table}

\begin{table}
\caption{Comparison of the observed column densities in the DLA towards \Jsixs\ with the model ones, calculated at the best fit value of the number density and UV field from Fig.~\ref{fig:J1236_phys}.
\label{table:J1236_phys}
}
\begin{tabular}{ccc}
\hline
species & \multicolumn{2}{c}{$\log N$}\\ 
& observed & model \\
\hline
CI & $14.28^{+0.36}_{-0.37}$ & 14.24$^a$ \\
CI* & $13.70^{+0.11}_{-0.08}$ & 13.88$^a$ \\
CI** & $13.18^{+0.11}_{-0.16}$ & 12.89$^a$ \\
H$_2$ J0 & $19.21^{+0.03}_{-0.02}$ & 19.14 \\
H$_2$ J1 & $19.56^{+0.02}_{-0.01}$ & 19.61 \\
H$_2$ J2 & $18.46^{+0.03}_{-0.02}$ & 18.26 \\
\hline
H$_2$ J3$^{b}$ &  & 17.42 \\
H$_2$ J4$^{b}$ &  & 15.29 \\
H$_2$ J5$^{b}$ &  & 14.59 \\
\hline
\end{tabular}
\begin{tablenotes}
    \item (a) These values were obtained under the additional assumption that the total \CI\ column density equates the observed one.  
    \item (b) We did not use these levels to compare with observed values, since the column densities measurements for these levels are subject to a large systematic uncertainty (see text).
\end{tablenotes}
\end{table}

\begin{table*}
\caption{Fit results of metals towards \Jsixs.
\label{table:J1236_metals}
}
\begin{tabular}{cccccccc}
\hline
comp & z & $\Delta$v, km\,s$^{-1}$ & b, km\,s$^{-1}$ & $\log N$(SiII) & $\log N$(OI) & $\log N$(SII) & $\log N$(FeII) \\
\hline
1 & $3.026187(^{+29}_{-34})$ & -512.9 & $14.7^{+2.0}_{-2.2}$ & $13.80^{+0.07}_{-0.04}$ & $14.18^{+0.05}_{-0.09}$ &  &  \\
2 & $3.02696(^{+4}_{-3})$ & -457.0 & $13.0^{+2.0}_{-1.7}$ & $13.71^{+0.07}_{-0.06}$ & $14.11^{+0.10}_{-0.08}$ &  &  \\
3 & $3.02797(^{+3}_{-6})$ & -380.4 & $16.7^{+3.0}_{-3.3}$ & $14.43^{+0.12}_{-0.11}$ & $15.47^{+0.30}_{-0.23}$ & $14.65^{+0.07}_{-0.07}$ & $13.63^{+0.10}_{-0.13}$ \\
4 & $3.028675(^{+28}_{-52})$ & -334.7 & $31.6^{+1.4}_{-2.9}$ & $14.58^{+0.05}_{-0.07}$ & $15.17^{+0.05}_{-0.10}$ & $14.52^{+0.12}_{-0.14}$ & $13.94^{+0.07}_{-0.06}$ \\
5 & $3.030418(^{+13}_{-14})$ & -202.7 & $8.4^{+2.9}_{-0.9}$ & $13.80^{+0.12}_{-0.04}$ & $16.21^{+0.15}_{-1.48}$ &  & $13.09^{+0.10}_{-0.06}$ \\
6 & $3.03142(^{+12}_{-8})$ & -123.5 & $49.1^{+0.9}_{-16.1}$ & $13.30^{+0.07}_{-0.10}$ &  &  & $12.50^{+0.20}_{-0.32}$ \\
7 & $3.03234(^{+4}_{-3})$ & -60.7 & $24.4^{+2.5}_{-2.5}$ & $13.66^{+0.08}_{-0.06}$ & $15.28^{+0.07}_{-0.08}$ &  & $13.42^{+0.04}_{-0.05}$ \\
8 & $3.033152(^{+13}_{-7})$ & 0.0 & $13.7^{+0.8}_{-0.7}$ & $15.58^{+0.08}_{-0.04}$ & $16.65^{+0.52}_{-0.33}$ & $15.21^{+0.06}_{-0.03}$ & $14.46^{+0.04}_{-0.04}$ \\
9 & $3.033699(^{+33}_{-28})$ & 39.5 & $21.1^{+1.6}_{-2.3}$ & $13.74^{+0.08}_{-0.06}$ & $14.97^{+0.12}_{-0.08}$ &  &  \\
$\log N_{\rm tot}$ &  &  &  & $15.69^{+0.06}_{-0.05}$ & $16.77^{+0.44}_{-0.23}$ & $15.39^{+0.03}_{-0.04}$ & $14.66^{+0.03}_{-0.03}$ \\
$\rm [X/H]$ &  &  &  & $-0.60^{+0.06}_{-0.05}$ & $-0.70^{+0.44}_{-0.23}$ & $-0.51^{+0.03}_{-0.04}$ & $-1.62^{+0.04}_{-0.03}$ \\
$\rm [X/SII]$ &  &  &  & $-0.09^{+0.07}_{-0.06}$ & $-0.19^{+0.44}_{-0.23}$ & $0.00^{+0.05}_{-0.05}$ & $-1.11^{+0.05}_{-0.04}$ \\
\hline
\end{tabular}
\end{table*}

\clearpage

\subsection{\Jseven}

\begin{figure}
\centering
\includegraphics[trim={0.0cm 0.0cm 0.0cm 0.0cm},clip,width=\hsize]{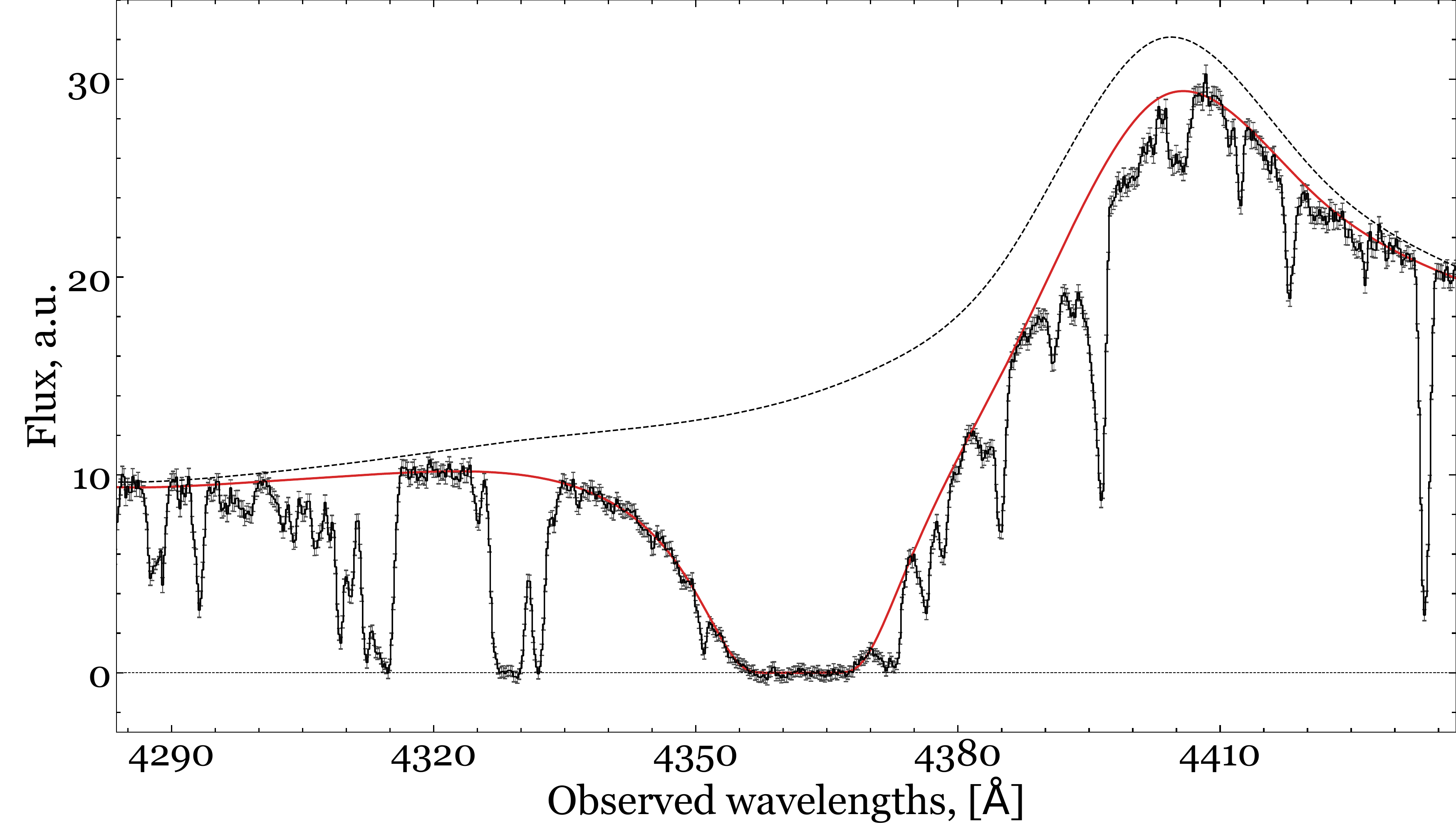}
		\caption{Fit to the damped \HI\ Ly$\alpha$ absorption line at $z=2.588$ towards \Jsevens.
		\label{fig:J2347_HI}}
\end{figure}

We measured the \HI\ column density of the DLA at $z=2.588$ towards \Jsevens\ by fitting the Ly$\alpha$ absorption line with a single component together with the continuum modelled using a seventh order Chebyshev polynomial. We derived $\log N(\HI)=20.47\pm0.01$ at the best fit value of $z=2.58813$. The Ly$\alpha$ profile, together with the reconstructed quasar continuum are shown in Fig.~\ref{fig:J2347_HI}. There is additional DLA at $z=1.9742$ in this spectrum with strong metal absorption lines.

\begin{figure*}
\centering
\includegraphics[trim={0.0cm 0.0cm 0.0cm 0.0cm},clip,width=\hsize]{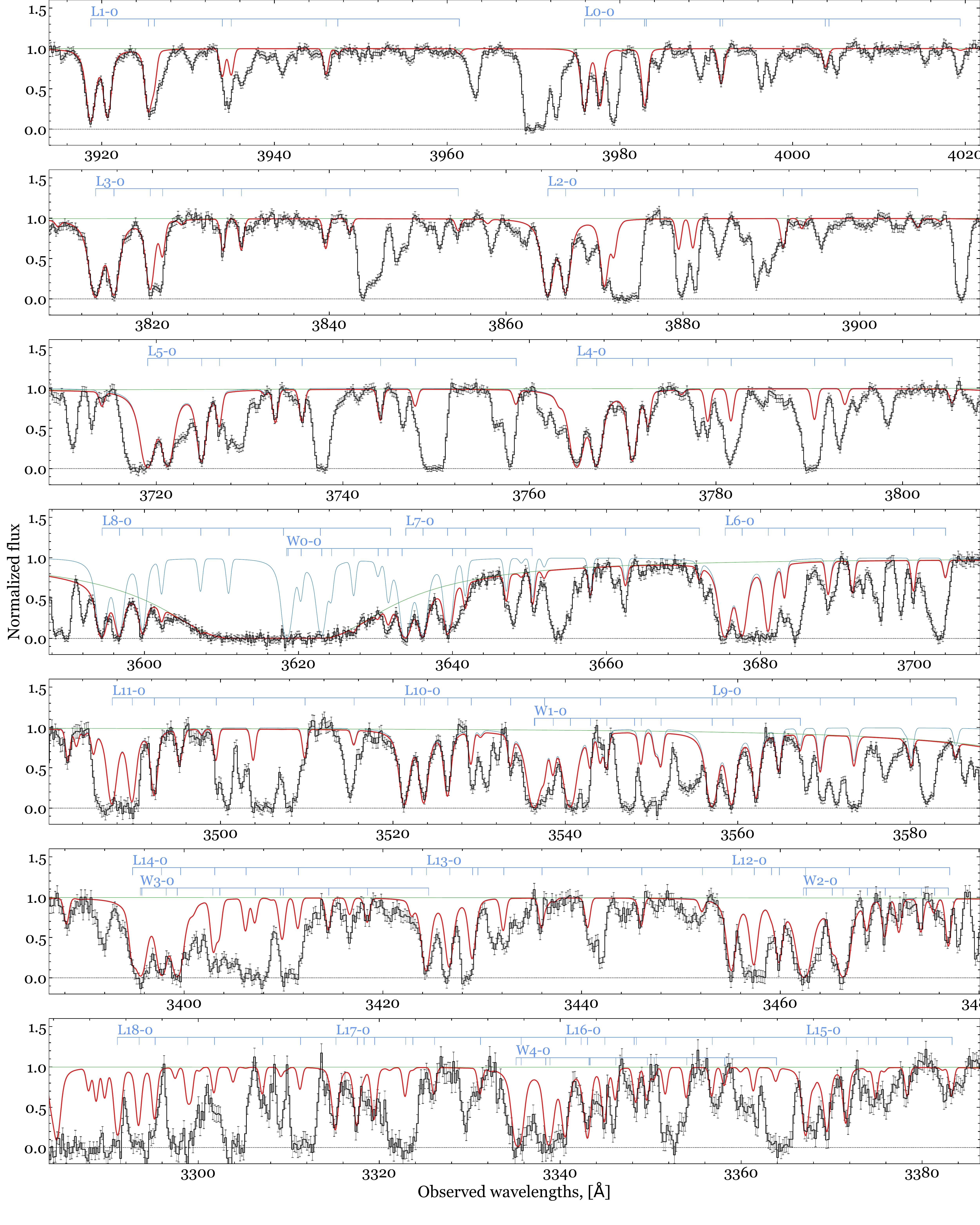}
		\caption{Fit to the H$_2$ absorption lines at $z=2.588$ towards \Jsevens.
		\label{fig:J2347_H2}}
\end{figure*}

\begin{figure}
\centering
\includegraphics[trim={0.0cm 0.0cm 0.0cm 0.0cm},clip,width=\hsize]{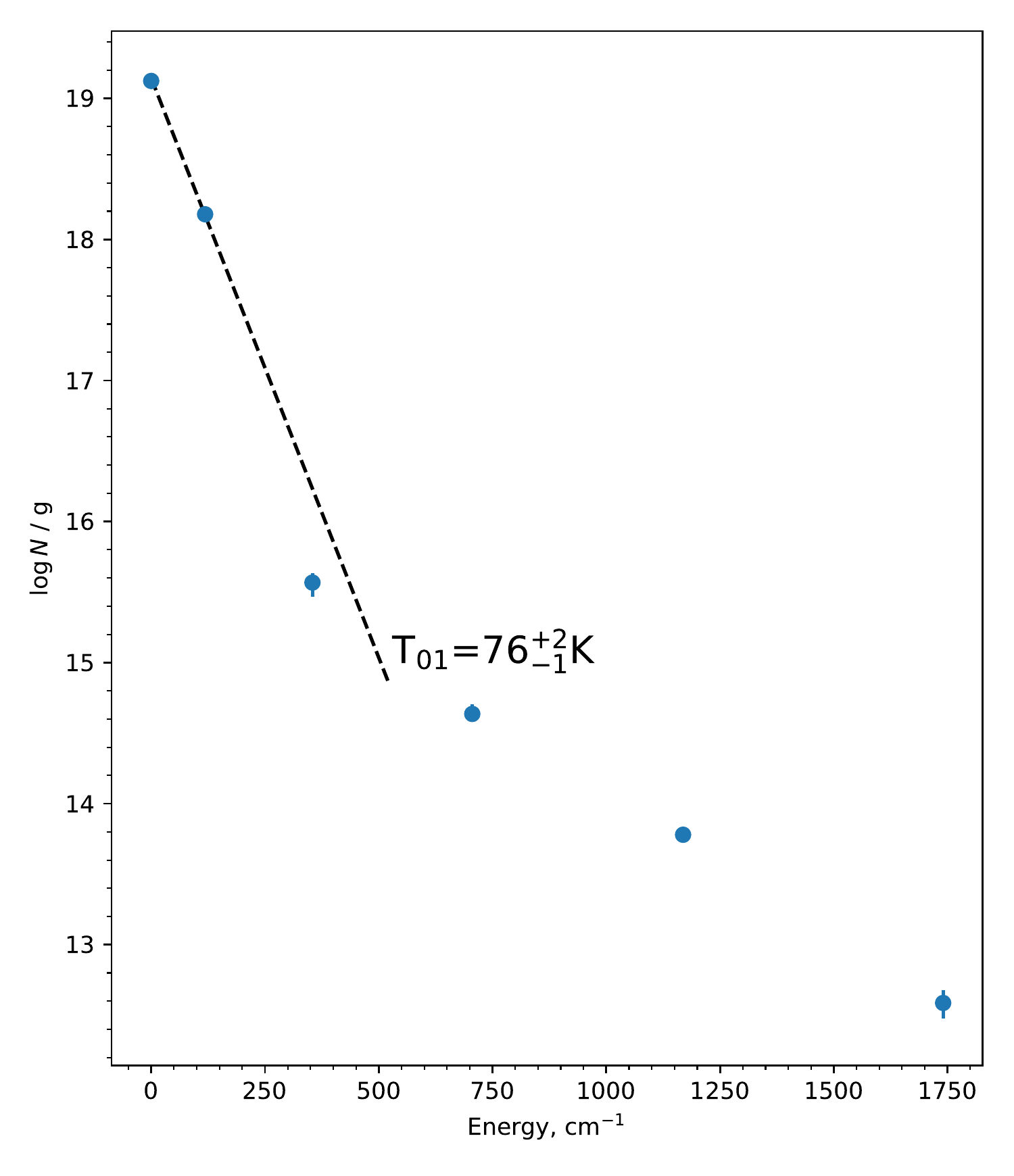}
		\caption{Excitation diagram of H$_2$ at $z=2.588$ towards \Jsevens.
		\label{fig:J2347_H2_exc}}
\end{figure}

We fitted the H$_2$ absorption lines towards \Jsevens\ using a one component model. The result of the fit is given in Table~\ref{table:J2347_H2} and H$_2$ line profiles are shown in Fig.~\ref{fig:J2347_H2}. The excitation diagram of H$_2$ rotation levels is shown in Fig.~\ref{fig:J2347_H2_exc} and we derived $T_{01} = 76^{+2}_{-1}$~K. 

\begin{figure}
    \centering
    \includegraphics[trim={0.0cm 0.0cm 0.0cm 0.0cm},clip,width=\hsize]{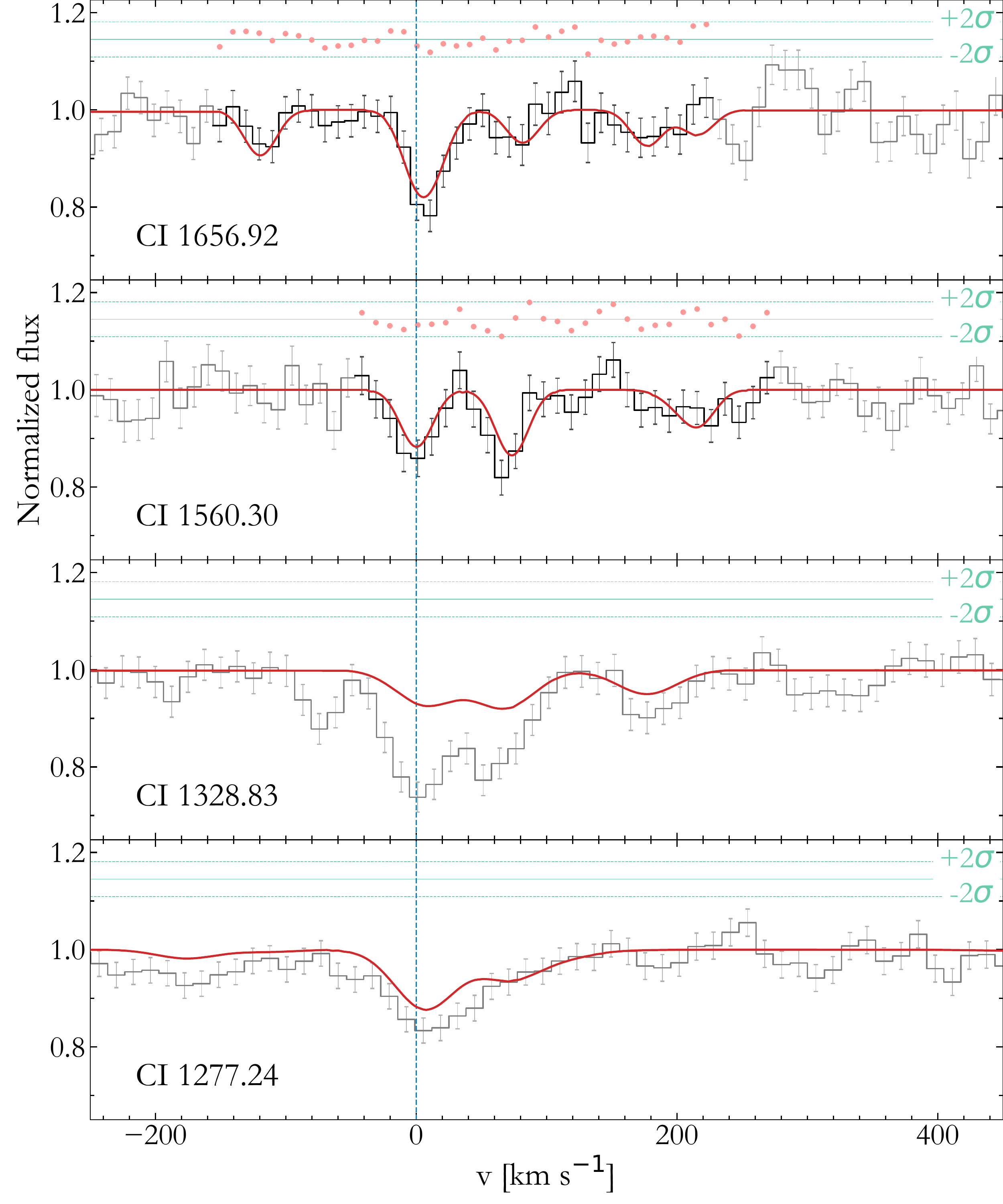}
	\caption{Fit to the \CI\ absorption lines at $z=2.588$ in \Jsevens. \CI\ lines near \CI$\lambda1328$ transition are blended with \ion{C}{IV}\ doublet at $z\approx2.0744$.}
	\label{fig:J2347_CI}
\end{figure}

We detected \CI\ absorption lines in three fine structure levels in a single component corresponding to H$_2$. 
The result from fitting these lines is given in Table~\ref{table:J2347_CI} and the corresponding profiles are shown in Fig.~\ref{fig:J2347_CI}. \CI$\lambda1565$\AA\ line is located at the wing of \ion{C}{IV} emission line of quasar. Therefore we included partial coverage of at this line into consideration. We estimated the the covering factor at this line to be $0.93^{+0.07}_{-0.11}$ which is consistent with the total coverage of emission region of quasar by \CI-bearing absorption system. However X-shooter resolution do not allow us to resolve the velocity components of the system and therefore this case should be studied using high resolution spectrum.

\begin{figure}
\centering
\includegraphics[trim={0.0cm 0.0cm 0.0cm 0.0cm},clip,width=\hsize]{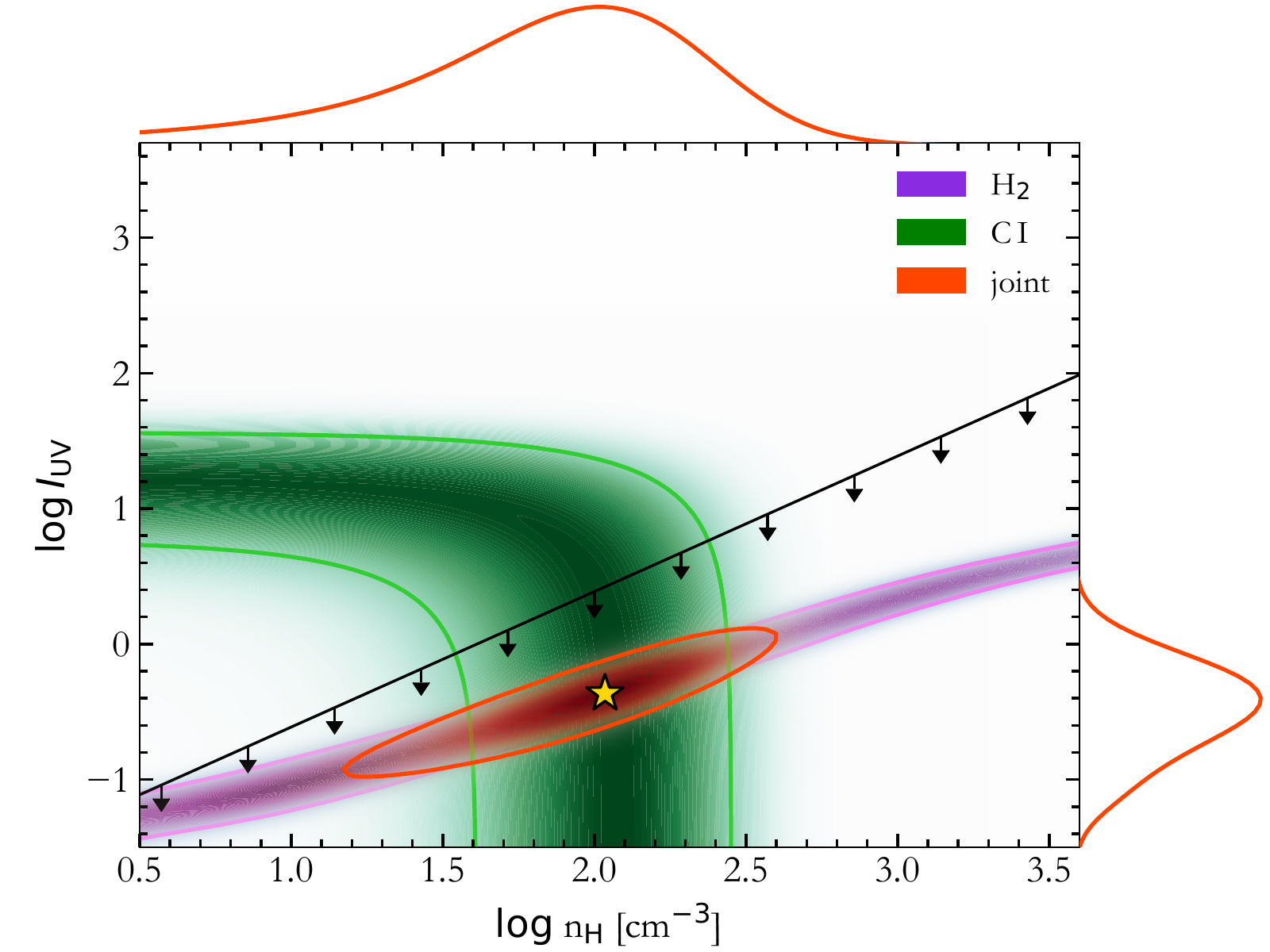}
		\caption{Constraints on the number density and UV field in H$_2$-bearing medium at $z=2.588$ towards \Jsevens\ from the excitation of \CI\ fine-structure and H$_2$ rotational levels. The colors and lines are the same as in Fig.~2.
		}
		\label{fig:J2347_phys}
\end{figure}

The constrained values of the hydrogen number density and UV field in the H$_2$/\CI-bearing medium are shown in Fig.~\ref{fig:J2347_phys} and the comparison between observed and model column densities at the best fit value (yellow star in Fig.~\ref{fig:J2347_phys}) is provided in Table~\ref{table:J2347_phys}. 

\begin{figure*}
\centering
\includegraphics[trim={0.0cm 0.0cm 0.0cm 0.0cm},clip,width=\hsize]{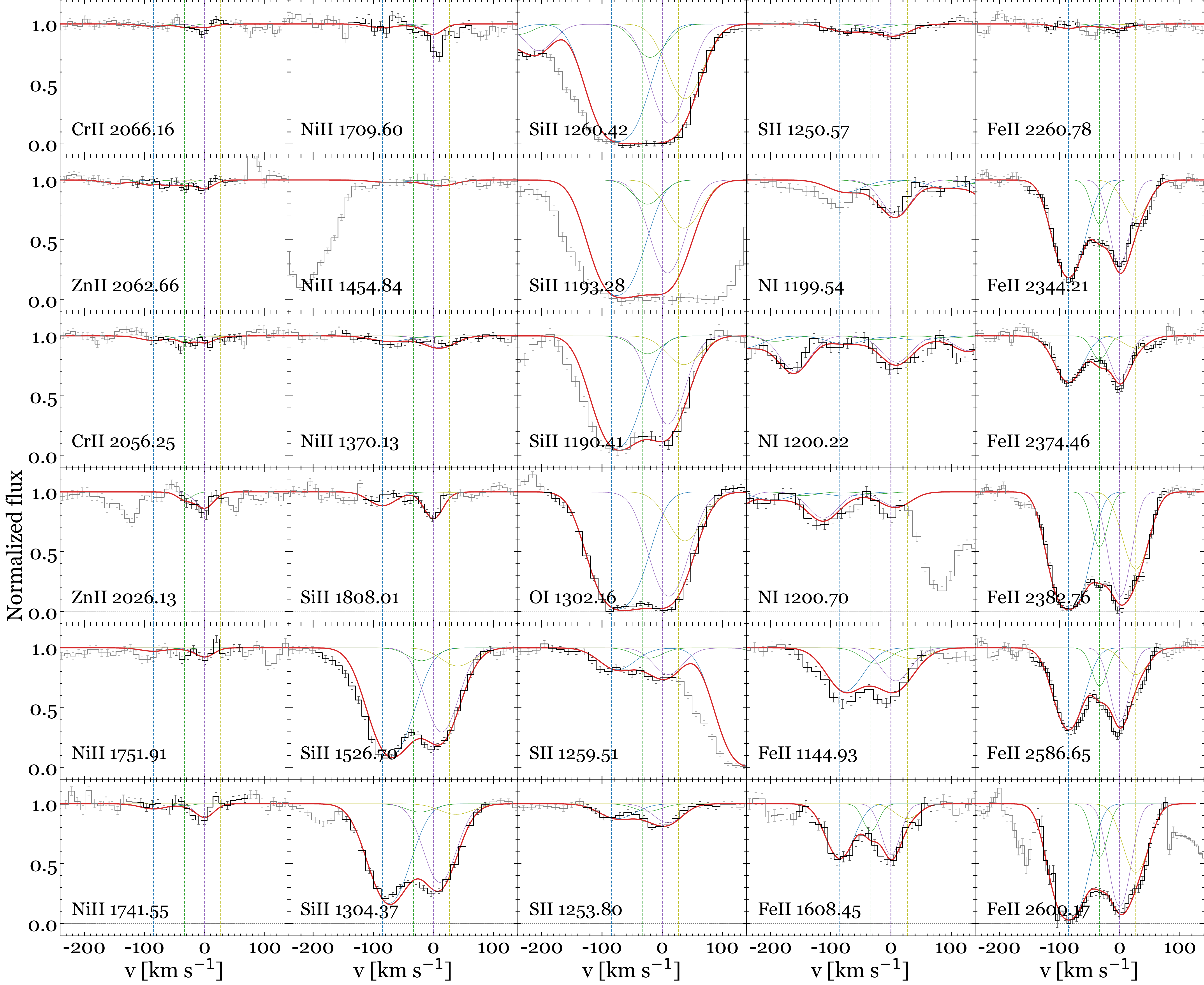}
		\caption{Fit to the metal absorption lines at $z=2.588$ towards \Jsevens.}
		\label{fig:J2347_metals}
\end{figure*}

Metal lines are fitted using a four components model. The third component of metals corresponds to H$_2$/\CI-bearing component of the DLA. The results of the fit are given in Table~\ref{table:J2347_metals} and the fit profiles are shown in Fig.~\ref{fig:J2347_metals}.

\begin{table}
\caption{Fit results of H$_2$ towards \Jsevens.
\label{table:J2347_H2}
}
\begin{tabular}{cc}
\hline
z & $2.587969(^{+10}_{-9})$ \\
b, km\,s$^{-1}$ & $6.1^{+0.2}_{-0.1}$ \\
$\log N$(H$_2$ J0) & $19.14^{+0.01}_{-0.01}$ \\
$\log N$(H$_2$ J1) & $19.12^{+0.01}_{-0.01}$ \\
$\log N$(H$_2$ J2) & $16.23^{+0.07}_{-0.10}$ \\
$\log N$(H$_2$ J3) & $15.92^{+0.07}_{-0.06}$ \\
$\log N$(H$_2$ J4) & $14.77^{+0.04}_{-0.05}$ \\
$\log N$(H$_2$ J5) & $14.10^{+0.09}_{-0.11}$ \\
$\log N_{\rm tot}$ & $19.44^{+0.01}_{-0.01}$ \\
\hline
\end{tabular}
\end{table}

\begin{table}
\caption{Fit results of \CI\ towards \Jsevens.
\label{table:J2347_CI}
}
\begin{tabular}{cc}
\hline
z & $2.587986(^{+11}_{-13})$ \\
b, km\,s$^{-1}$ & $1.1^{+0.3}_{-0.2}$ \\
$\log N$(CI) & $14.04^{+0.40}_{-0.54}$ \\
$\log N$(CI*) & $13.31^{+0.12}_{-0.10}$ \\
$\log N$(CI**) & $12.99^{+0.12}_{-0.14}$ \\
$\log N_{\rm tot}$ & $14.16^{+0.28}_{-0.44}$ \\
\hline
\end{tabular}
\end{table}

\begin{table}
\caption{Comparison of the observed column densities in the DLA towards \Jsevens\ with the model ones, calculated at the best fit value of number density and UV field from Fig.~\ref{fig:J2347_phys}.
\label{table:J2347_phys}
}
\begin{tabular}{ccc}
\hline
species & \multicolumn{2}{c}{$\log N$}\\ 
& observed & model \\
\hline
CI & $14.28^{+0.36}_{-0.37}$ & 14.24$^a$ \\
CI* & $13.70^{+0.11}_{-0.08}$ & 13.87$^a$ \\
CI** & $13.18^{+0.11}_{-0.16}$ & 12.96$^a$ \\
H$_2$ J0 & $19.14^{+0.01}_{-0.01}$ & 19.24 \\
H$_2$ J1 & $19.12^{+0.01}_{-0.01}$ & 18.97 \\
H$_2$ J2 & $16.23^{+0.07}_{-0.10}$ & 16.09 \\
\hline
H$_2$ J3$^b$ &  & 14.93 \\
H$_2$ J4$^b$ &  & 13.59 \\
H$_2$ J5$^b$ &  & 13.14 \\
\hline
\end{tabular}
\begin{tablenotes}
    \item (a) These values were obtained under the additional assumption that the total \CI\ column density equates the observed one.
    \item (b) We did not use these levels to compare with observed values, since the column densities measurements for these levels are subject to a large systematic uncertainty (see text).
\end{tablenotes}
\end{table}

\begin{table*}
\caption{Fit results of metals towards \Jsevens.
\label{table:J2347_metals}
}
\begin{tabular}{cccccccc}
\hline
comp & 1 & 2 & 3 & 4 & $\log N_{\rm tot}$ & $\rm [X/H]$ & $\rm [X/SII]$ \\
\hline
z & $2.587008(^{+9}_{-5})$ & $2.587633(^{+12}_{-11})$ & $2.5879841(^{+26}_{-4})$ & $2.5883292(^{+4}_{-70})$ &  &  &  \\
$\Delta$v, km\,s$^{-1}$ & -81.2 & -29.3 & 0.0 & 28.4 &  &  &  \\
b, km\,s$^{-1}$ & $24.98^{+0.88}_{-0.47}$ & $6.17^{+0.52}_{-0.83}$ & $8.00^{+0.75}_{-0.65}$ & $19.38^{+0.74}_{-1.04}$ &  &  &  \\
$\log N$(SiII) & $14.64^{+0.03}_{-0.02}$ & $13.96^{+0.11}_{-0.11}$ & $14.89^{+0.07}_{-0.07}$ & $14.05^{+0.03}_{-0.04}$ & $15.14^{+0.05}_{-0.03}$ & $-0.91^{+0.05}_{-0.03}$ & $-0.09^{+0.06}_{-0.06}$ \\
$\log N$(FeII) & $14.15^{+0.01}_{-0.01}$ & $13.59^{+0.04}_{-0.04}$ & $14.06^{+0.04}_{-0.05}$ & $13.50^{+0.03}_{-0.02}$ & $14.51^{+0.01}_{-0.02}$ & $-1.53^{+0.02}_{-0.02}$ & $-0.71^{+0.03}_{-0.05}$ \\
$\log N$(NiII) & $13.05^{+0.13}_{-0.26}$ & $12.67^{+0.28}_{-0.84}$ & $13.38^{+0.06}_{-0.09}$ & $<12.8$ & $13.55^{+0.07}_{-0.06}$ & $-1.21^{+0.07}_{-0.06}$ & $-0.40^{+0.08}_{-0.08}$ \\
$\log N$(SII) & $14.41^{+0.05}_{-0.03}$ & $13.85^{+0.25}_{-0.39}$ & $14.63^{+0.07}_{-0.11}$ & $<14.2$ & $14.84^{+0.05}_{-0.03}$ & $-0.82^{+0.05}_{-0.03}$ & 0.0 \\
$\log N$(CrII) & $12.63^{+0.23}_{-0.48}$ & $12.49^{+0.16}_{-0.29}$ & $12.57^{+0.14}_{-0.18}$ & $<12.4$ & $12.94^{+0.16}_{-0.08}$ & $-1.24^{+0.16}_{-0.08}$ & $-0.42^{+0.16}_{-0.09}$ \\
$\log N$(ZnII) & $<11.4$ & $11.97^{+0.16}_{-0.27}$ & $12.31^{+0.08}_{-0.11}$ & $<11.4$ & $12.50^{+0.05}_{-0.09}$ & $-0.60^{+0.05}_{-0.09}$ & $0.22^{+0.06}_{-0.11}$ \\
$\log N$(NI) & $13.43^{+0.13}_{-0.21}$ & $<12.3$ & $13.91^{+0.11}_{-0.07}$ & $<13$ & $14.05^{+0.07}_{-0.08}$ & $-2.32^{+0.07}_{-0.08}$ & $-1.50^{+0.08}_{-0.10}$ \\
\hline
\end{tabular}
\end{table*}

\bsp	
\label{lastpage}
\end{document}